\documentclass[sigconf,screen,nonacm]{./texmf/tex/latex/acmart/acmart}

\usepackage{stfloats}

\usepackage{minted}
\makeatletter
\let\orgdescriptionlabel\descriptionlabel
\renewcommand*{\descriptionlabel}[1]{%
  \let\orglabel\label
  \let\label\@gobble
  \phantomsection
  \edef\@currentlabel{#1}%
  \let\label\orglabel
  \orgdescriptionlabel{#1}%
}
\renewcommand\subsubsection{\def\@toclevel{3}%
  \@startsection{subsubsection}{3}{\z@}%
  {-.5\baselineskip \@plus -2\p@ \@minus -.2\p@}%
  {.25\baselineskip}%
  {\ACM@NRadjust{\@subsubsecfont}}}
\def\@subsubsecfont{\sffamily\section@raggedright}
\makeatother

\copyrightyear{2025}
\setcopyright{cc}
\setcctype{by}

\begin{document}

\title{Nagare Media Engine: A System for Cloud- and Edge-Native Network-based Multimedia Workflows}

\author{Matthias Neugebauer}
\orcid{0000-0002-1363-0373}
\affiliation{%
  \institution{University of M\"unster}
  \city{M\"unster}
  \country{Germany}
}
\email{matthias.neugebauer@uni-muenster.de}

\begin{abstract}

  Before media playback is possible, live and video-on-demand content alike usually undergoes various operations described as tasks within a multimedia workflow. Where previously ingest, transcode, packaging and delivery tasks might have run on a single machine, today's workflows are significantly more complex distributed systems. Describing and implementing multimedia workflows is challenging and requires new approaches.

  A standards-based multimedia workflow system is described in ISO/IEC~23090-8 Network-Based Media Processing~(NBMP) developed by MPEG. This technical report discusses details of \texttt{nagare media engine}, our open source research prototype implementation of NBMP. Built upon the Kubernetes platform, \texttt{nagare media engine} provides a cloud- and edge-native solution that meets today's requirements for multimedia workflow systems.

\end{abstract}

% Generated with http://dl.acm.org/ccs.cfm


\maketitle

\section{Introduction}
\label{sec:introduction}

Modern multimedia workflows are challenging due to their structural and dynamic nature. With the move to cloud- and edge-infrastructure, multimedia workflows are structured as complex distributed systems that might span across multiple regionally separated data centers. Workflow authors therefore need to design, implement and operate multimedia workflows holistically. Product differences between various cloud- and edge-providers have to be overcome, workloads have to be deployed in selected environments as well as securely connected over the network and observability data has to be aggregated to monitor the complete workflow.

Additional complexity is introduced with the dynamic nature of today's workflows. On the one hand, this is a consequence of the distributed structure which is more error-prone. Workflows now have to react to new failure modes such as server, network or data center outages. On the other hand, workflows are subject to business objectives and therefore require constant adaptation to changed circumstances. For instance, energy efficiency and environmental targets might require rescheduling workloads to other data centers based on the current electricity mix.

For these reasons, describing and implementing multimedia workflows can be hard. To meet these challenges, the Moving Picture Experts Group~(MPEG) recently published ISO/IEC~23090-8 Network-Based Media Processing~(NBMP)~\citep{isoiec_isoiec230908_2020,isoiec_isoiec230908_2025}, a standard that defines a multimedia workflow system together with data models, common Application Programming Interfaces~(APIs) and a reference architecture. In NBMP, workflows are described declaratively as JavaScript Object Notation~(JSON) documents and submitted to the workflow system via a Representational State Transfer~(REST) API. The NBMP workflow manager will then first select appropriate multimedia functions to fulfill the defined constraints. Afterwards, it schedules the functions to media processing entities~(MPEs) that represent different infrastructure environments. An instantiated function runs as a task within one MPE and receives, processes and outputs media and metadata streams over the network. Thus, the collection of network-connected tasks forms the workflow. The NBMP workflow manager observes tasks in all MPEs and ensures a proper execution.

This technical report outlines the requirements, design decisions and implementation of the \texttt{nagare media engine} research prototype, an open source NBMP implementation based on the Kubernetes platform~\citep{neugebauer_nagaremediaengine_2023,neugebauer_nagaremediaengine_2024,neugebauer_nagaremediaengine_2025}. Over the last decade, Kubernetes has established itself as an orchestrator for containerized workloads in on-premises, cloud and edge environments. As such, it provides a common platform regardless of infrastructure provider differences. Kubernetes clusters are therefore good candidates for MPEs. At the same time, Kubernetes has extension points for custom solutions. The NBMP workflow manager could hence integrate directly with Kubernetes. We thus designed \texttt{nagare media engine} as a Kubernetes-native solution. In \citep{neugebauer_nagaremediaengine_2023,neugebauer_nagaremediaengine_2024,neugebauer_nagaremediaengine_2025}, we previously showed how \texttt{nagare media engine} can meet the structural and dynamic challenges of multimedia workflows.

The rest of this technical report is structured as follows. In Section~\ref{sec:related-work}, we discuss related work. Next, Sections~\ref{sec:overview-of-nagare-media-engine} and~\ref{sec:models} give an overview of \texttt{nagare media engine} and related (data) models, respectively. Section~\ref{sec:common-packages} describes commonly used packages before going into the details of the various components in the following sections. We outline the NBMP Gateway, Workflow Manager, Workflow Manager Helper, Task Shim and Functions components in Sections~\ref{sec:nbmp-gateway},~\ref{sec:workflow-manager},~\ref{sec:workflow-manager-helper},~\ref{sec:task-shim} and~\ref{sec:functions}, respectively. Finally, Section~\ref{sec:conclusion} concludes this report. Wherever appropriate, this report inlines figures with the text. Note, however, that we intentionally simplified and abbreviated some of the included UML diagrams to increase the readability and understandability. For instance, unrelevant types, attributes or methods might be omitted, and in some cases we left out the method argument names and only depicted the types. Full package names are also often abbreviated.

\section{Related Work}
\label{sec:related-work}

The first edition of NBMP standard has been published in late 2020~\cite{isoiec_isoiec230908_2020}. In the following years, work continued on the second edition. During that process, JSON schema files~\cite{wright_jsonschemamedia_2022} for the revised NBMP data model were published beforehand as supplementary files\footnote{see \url{https://standards.iso.org/iso-iec/23090/-8/ed-2/en/} (published together with the second edition) and \url{https://github.com/MPEGGroup/NBMP} (version repository publicly available before the release of the second edition)}. The second edition was finally released mid-2025~\cite{isoiec_isoiec230908_2025}. For \texttt{nagare media engine}, we already used the revised JSON schema files of the second edition. It will therefore correctly parse workflow descriptions. However, since the development of \texttt{nagare media engine} was carried out between the years 2022 and 2025~\citep{neugebauer_nagaremediaengine_2023,neugebauer_nagaremediaengine_2024,neugebauer_nagaremediaengine_2025}, our understanding of NBMP comes mainly from the first edition and changes of the second still need to be incorporated. This technical report will use the terminology of the first edition. As far as we know, \texttt{nagare media engine} is the only publicly available NBMP implementation.

NBMP is part of the ISO/IEC~23090 \mbox{MPEG-I} standard family for immersive media. \textsc{\citeauthor{wien_standardizationstatusimmersive_2019}} give an overview of \mbox{MPEG-I} including NBMP in~\cite{wien_standardizationstatusimmersive_2019}. A more detailed introduction to NBMP specifically is provided by \textsc{\citeauthor{xu_mediaproductionusing_2022}} in~\cite{xu_mediaproductionusing_2022}. They also showcase potential applications for NBMP.

The standard has already been used in research. In~\cite{ramoschavez_mpegnbmptestbed_2021}, \textsc{\citeauthor{ramoschavez_mpegnbmptestbed_2021}} developed an NBMP testbed that was used to evaluate and optimize scalable workflows for MPEG Dynamic Adaptive Streaming over HTTP~(DASH)~\cite{isoiec_isoiec230091_2022} and HTTP Live Streaming~(HLS)~\cite{pantos_httplivestreaming_2017} streaming. Similarly, \textsc{\citeauthor{mueller_contextawarevideoencoding_2022}} proposed NBMP for an encoding parameters optimization problem~\cite{mueller_contextawarevideoencoding_2022}. \textsc{\citeauthor{you_omaf4cloudstandardsenabled360_2020}} used NBMP to describe workflows in the context of the Omnidirectional Media Format~(OMAF)~\citep{isoiec_isoiec230902_2023,you_omaf4cloudstandardsenabled360_2020} and augmented reality~\cite{you_nbmpstandarduse_2021}. Lastly, \textsc{\citeauthor{bassbouss_5gvictorioptimizingmedia_2021}} proposed NBMP for a media caching workflow in~\cite{bassbouss_5gvictorioptimizingmedia_2021}.

\section{Overview of \texttt{nagare media engine}}
\label{sec:overview-of-nagare-media-engine}

This section gives a first overview of \texttt{nagare media engine}. We start by discussing the requirements and design decisions in Section~\ref{subsec:design}. Afterwards, Section~\ref{subsec:implementation} goes over our implementation. Section~\ref{subsec:usage} concludes this section with a description of how \texttt{nagare media engine} can be used. Throughout, we will point to later sections for further details.

\subsection{Design}
\label{subsec:design}

The high-level goal of \texttt{nagare media engine} is to provide a state-of-the-art cloud- and edge-native multimedia workflow solution. For this purpose, we followed a standards-based approach and implemented \texttt{nagare media engine} based on ISO/IEC~23090-8 NBMP~\cite{isoiec_isoiec230908_2020}. This standard defines options for many use cases. As research prototype, \texttt{nagare media engine} currently only implements selected features relevant for a basic NBMP workflow system. The implementation of \texttt{nagare media engine} was conducted iteratively with each of our previous works~\citep{neugebauer_nagaremediaengine_2023,neugebauer_nagaremediaengine_2024,neugebauer_nagaremediaengine_2025} expanding the feature set.

NBMP defines data models, REST APIs and a reference architecture (see Section~\ref{subsec:network-based-media-processing} for a detailed discussion). Users of the workflow system submit workflows by sending a request with a JSON formatted workflow description to the workflow API. Workflows consist of tasks that are scheduled to MPEs. Tasks are instantiations of multimedia functions. The workflow system uses the task API and sends a request with a JSON-formatted task description to configure the multimedia function. A task has input and output ports that may be bound to ports of other tasks or inputs and outputs of the whole workflow. Ports thus abstractly describe network connections between tasks. In this way, tasks in the NBMP model logically form a directed acyclic graph~(DAG).

We target the following roles: \emph{Users} that want to execute workflows, \emph{administrators} that configure and provide the workflow system and \emph{developers} that implement multimedia functions. With that, we formulate the following high-level functional and non-functional requirements for \texttt{nagare media engine}:
\begin{description}
  \item[FR1\label{req:nbmp-workflow-api}] A user should be able to submit an NBMP compatible workflow description to an NBMP compatible workflow API.
  \item[FR2\label{req:nbmp-task-api}] A developer should be able to implement an NBMP compatible multimedia function that provides an NBMP compatible task API to be used by the workflow system to control task instantiations of this multimedia function.
  \item[FR3\label{req:non-nbmp-functions}] An administrator should be able to easily adapt existing non-NBMP multimedia functions.
  \item[FR4\label{req:function-repo}] An administrator should be able to define multimedia functions and make them available for selected users. A user should be able to use available multimedia functions.
  \item[FR5\label{req:mpe}] An administrator should be able to define MPEs and make them available for selected users. A user should be able to submit workflows whose tasks run distributed across multiple available MPEs.
  \item[FR6\label{req:report-api}] A multimedia function should be able to report occurring events, i.e.~the workflow system keeps a record of task state changes.
  \item[NFR1\label{req:scalable}] The system should be scalable, i.e.~it should be possible to easily deploy it to a single machine or to many clusters of machines.
  \item[NFR2\label{req:reliable}] The system should be reliable, i.e.~failures should be detected and corrected automatically.
\end{description}

\begin{figure}[h]
  \centering
  \includegraphics[width=0.9\columnwidth]{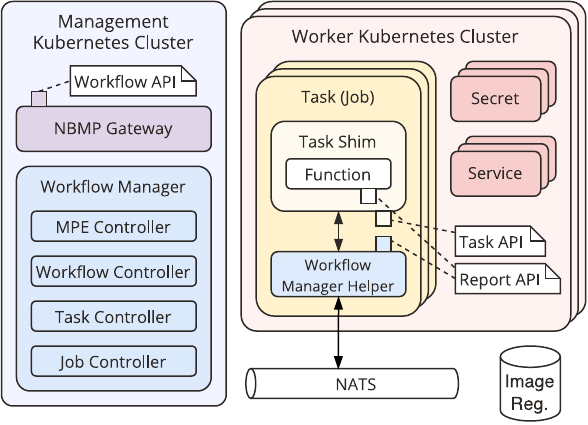}
  \caption{\texttt{nagare media engine} architecture, compare~\cite{neugebauer_nagaremediaengine_2025}.}
  \label{fig:nme-architecture}
  \Description{nagare media engine architecture, compare~\cite{neugebauer_nagaremediaengine_2025}}
\end{figure}

As a result of these requirements, we derived a design for \texttt{nagare media engine} that is illustrated in Figure~\ref{fig:nme-architecture}. Our decisions are explained in the following paragraphs.

\texttt{nagare media engine} should provide the NBMP workflow API as the main outward facing interface for users~(\ref{req:nbmp-workflow-api}). Although currently not all fields of the NBMP data model have corresponding effects, all requests with valid workflow definitions should be accepted first. NBMP then allows signaling unsupported fields by responding with an error and an accompanying \texttt{acknowledge} object that further details unsupported and partially supported fields as well as any other failure that occurred when trying to fulfill the request. In this way, new features can be integrated gradually.

Various computational models could be used to implement an NBMP workflow system. The NBMP standard describes a hypothetical implementation in Annex~B that uses Amazon Web Services~(AWS)\footnote{\url{https://aws.amazon.com/}} virtual machines~(VMs) as the computing environment. Alternatively, a serverless / Function as a Service~(FaaS) approach could be taken. While both approaches are viable, the interfaces for managing VMs and functions are not homogeneous across infrastructure providers. We, therefore, propose using Kubernetes\footnote{\url{https://kubernetes.io/}} instead. Kubernetes is an open source orchestration system for containerized applications. It runs in many environments and various providers offer managed Kubernetes solutions. As such, a universal interface is available across providers. Moreover, the container image format in combination with image registries allows packaging and shipping any application in a common way. Simultaneously, Kubernetes is extendable with custom resource types and corresponding controllers. \texttt{nagare media engine} thus not only uses Kubernetes clusters for MPEs but directly integrates the workflow system in it.

In our design, we called the cluster that hosts the control plane, i.e.~all workflow-managing components, the \emph{Management Kubernetes Cluster}. Here, we extend the Kubernetes API to implement our design. Tasks, on the other hand, are deployed to \emph{Worker Kubernetes Clusters} that do not need these extensions. Note that in a simple case, both cluster roles can fall to a single cluster, i.e.~one cluster acts as management and worker cluster. In an even simpler case, the Kubernetes cluster could consist of a single node. Projects like K3s\footnote{\url{https://k3s.io/}}, Microk8s\footnote{\url{https://microk8s.io/}} or KubeEdge\footnote{\url{https://kubeedge.io/}} allow provisioning Kubernetes even to smaller edge devices. Our design therefore scales from single machines to many clusters~(\ref{req:scalable}).

Furthermore, building on Kubernetes eases the implementation of a reliable system~(\ref{req:reliable}). As an orchestrator, Kubernetes monitors running workloads by performing regular health checks. It then tries to automatically remediate any unhealthy application, e.g.~by restarting it. Availability can be increased by running workloads replicated across failure domains, e.g.~on different machines or racks of machines.

Between \texttt{nagare media engine} components, a custom data model is used. First, this is a consequence of the Kubernetes-native design and, second, this allows the data model to be tailored to the needs of our implementation. We therefore introduced the \emph{NBMP Gateway} component that implements the NBMP workflow API and translates between the two data representations. Internally, the state is persisted in the form of custom Kubernetes resources. Administrators also manage available MPEs and multimedia functions through custom resources~(\ref{req:function-repo} and~\ref{req:mpe}). We will detail our custom data model in Section~\ref{subsec:custom-kubernetes-resources} and NBMP Gateway in Section~\ref{sec:nbmp-gateway}.

In the NBMP reference architecture (see Section~\ref{subsec:network-based-media-processing}), the NBMP workflow manager is responsible for controlling the system. In our design, we have a correspondingly named \emph{Workflow Manager} component. It consists of multiple Kubernetes controllers, each responsible for one aspect of the system. The \emph{MPE Controller} manages the connection to configured MPEs. The \emph{Workflow} and \emph{Task Controllers} manage the lifecycle of workflows and tasks. Internally, tasks are executed as Kubernetes jobs. The \emph{Job Controller} is responsible for observing the lifecycle of these jobs. These controllers work in tandem, which results in the desired effect. We will detail the Workflow Manager in Section~\ref{sec:workflow-manager}.

Multimedia functions are packaged as container images. We can therefore easily deploy tasks to Worker Kubernetes Clusters. For that, a \texttt{Job}, \texttt{Secret} and \texttt{Service} resource is created for each task. These are built-in Kubernetes resources available in any cluster. The \texttt{Job} resource describes the task that should be executed. The \texttt{Service} ensures that other tasks have a consistent endpoint when sending to input or receiving from output ports. Lastly, the \texttt{Secret} contains necessary configuration for the instantiation of the task.

Each job consists of two containers that are scheduled and deployed together as a Kubernetes pod, i.e.~the smallest deployable unit in a Kubernetes cluster. The \emph{Workflow Manager Helper} is an extension of the Workflow Manager within the worker cluster. It reads the \texttt{Secret} and instantiates the task accordingly using the NBMP task API~(\ref{req:nbmp-task-api}). Next to it runs the multimedia function that implements the task API. It starts processing inputs after the instantiation. Developers thus need to bundle the multimedia processing logic together with the NBMP task API. Implementing this API repeatedly can be error-prone. As an alternative, we therefore introduce a small layer called \emph{Task Shim} that implements this API once. Through a configuration file, the Task Shim is instructed what actions should be taken when certain API requests are handled. Ultimately, the Task Shim will start the multimedia function as a subprocess. In this way, even non-NBMP functions can be adapted to our workflow system~(\ref{req:non-nbmp-functions}). We will detail the Workflow Manager Helper and the Task Shim in Sections~\ref{sec:workflow-manager-helper} and~\ref{sec:task-shim}, respectively. \texttt{nagare media ingest} already implements various multimedia functions. These will be described in Section~\ref{sec:functions}.

NBMP allows tasks to report events and thus give an understanding of the inner workings. The standard only defines that HTTP \texttt{POST} requests should be used, but otherwise leaves the data format unspecified. Instead, the workflow manager should signal the desired format in the \texttt{report-type} field during task initialization. In our design, the Workflow Manager Helper implements this report API and makes it available to the multimedia function~(\ref{req:report-api}). The Task Shim already reports generic events that indicate task state changes (e.g.~creation, update, deletion, start, stop, etc.). Developers can extend this and emit custom events. The Workflow Manager Helper persists events in the event sourcing system NATS\footnote{\url{https://nats.io/}} JetStream. In previous work, we used this event record to implement multimedia functions that were able to perform error recovery~\cite{neugebauer_nagaremediaengine_2024}. Failed and restarted tasks would replay events from earlier executions and thus recover the previous state. For this purpose, we also incorporated the reports API into these functions as a special metadata input port. The Workflow Manager Helper then replays existing events to this input port.

\subsection{Implementation}
\label{subsec:implementation}

We implemented our design as the open source software \texttt{nagare media engine} and published it to \url{https://github.com/nagare-media/engine} under the Apache~2.0 license. \texttt{nagare media engine} is implemented in Go, which is common for Kubernetes-native applications. Kubernetes itself is written in Go and supporting libraries and frameworks for Kubernetes are most mature in this programming language.

Each component in our design is compiled as its own executable. The included \texttt{Makefile} helps with building binaries and container images for various operating systems and hardware architectures. Additionally, a continuous integration pipeline automatically builds and pushes container images to the GitHub container registry\footnote{\url{https://github.com/orgs/nagare-media/packages?repo_name=engine}}. Having all components in the same repository allows easily sharing code in common packages (see Section~\ref{sec:common-packages}).

We use a number of software libraries and frameworks to implement \texttt{nagare media engine}. An overview of the most important ones is given here while libraries specific to one component are introduced in the respective section. For the Workflow Manager, we use the Kubebuilder framework\footnote{\url{https://github.com/kubernetes-sigs/kubebuilder}}. It bundles relevant libraries for creating custom Kubernetes controllers. Moreover, code generation tools are provided that help to keep code and configuration files in sync with our custom resource definitions. We also use some Kubernetes libraries directly. \texttt{k8s.io/api}\footnote{\url{https://github.com/kubernetes/api}} and \texttt{k8s.io/apimachinery}\footnote{\url{https://github.com/kubernetes/apimachinery}} provide definitions of built-in Kubernetes resources, while \texttt{k8s.io/client-go}\footnote{\url{https://github.com/kubernetes/client-go}} offers client libraries for interacting with the Kubernetes API. Finally, \texttt{k8s.io/utils}\footnote{\url{https://github.com/kubernetes/utils}} implements supporting functionality.

Various components interact via HTTP. To implement the HTTP servers, we use the \texttt{github.com/gofiber/fiber/v2} (Fiber)\footnote{\url{https://github.com/gofiber/fiber}} \linebreak{}framework in combination with the \texttt{github.com/valyala/\linebreak{}fasthttp} (fasthttp)\footnote{\url{https://github.com/valyala/fasthttp}} HTTP implementation. As a distributed system, \texttt{nagare media engine} might temporarily experience failed network calls. Throughout our codebase we therefore use retries with exponential backoff and timeouts as offered by \texttt{github.com/\linebreak{}cenkalti/backoff/v5}\footnote{\url{https://github.com/cenkalti/backoff}}. For interactions with NATS JetStream, we use the official Go client library \texttt{github.com/nats-io/\linebreak{}nats.go}\footnote{\url{https://github.com/nats-io/nats.go}}. Lastly, we extracted the Go implementation of the NBMP data model in its own library called \texttt{github.com/nagare-media/\linebreak{}models.go}\footnote{\url{https://github.com/nagare-media/models.go}}. This way, other developers can reuse our implementation.

We further rely on existing applications both during development and also at runtime. We already described \texttt{controller-gen}\footnote{\url{https://github.com/kubernetes-sigs/controller-tools}}, a tool bundled with Kubebuilder for generating code and configuration files. Additionally, we use Kustomize\footnote{\url{https://kustomize.io/}} and \texttt{yq}\footnote{\url{https://mikefarah.gitbook.io/yq/}} to generate and patch YAML Ain't Markup Language~(YAML)~\cite{benkiki_yamlaintmarkup_2004} files. YAML is a data exchange format often used in the Kubernetes ecosystem, e.g.~for configuration or as a representation of Kubernetes resources. For our multimedia functions, we build upon FFmpeg\footnote{\url{https://ffmpeg.org/}}, a command line interface~(CLI) tool and library for working with multimedia data.

\subsection{Usage}
\label{subsec:usage}

\texttt{nagare media engine} is a Kubernetes-native system. As such, it requires a working Kubernetes cluster as its execution environment. Setting up a Kubernetes cluster can be non-trivial. Managed offerings from infrastructure providers make this easy in cloud and edge environments. For local test and development clusters, we automated the provisioning with the Kubernetes in Docker~(\texttt{kind})\footnote{\url{https://kind.sigs.k8s.io/}} tool. It starts a management and worker cluster as containers running in Docker\footnote{\url{https://www.docker.com/}}. The management cluster also hosts required dependencies such as NATS. Although \texttt{kind} Kubernetes clusters run nested in containers, they behave in the same way as regular clusters. As such, we also provide an easy way to create a local environment for \texttt{nagare media engine}. The included \texttt{Makefile} has the \texttt{kind-up} target to execute the provisioning process.

Deploying \texttt{nagare media engine} to the management cluster means creating the necessary Kubernetes resources. With \texttt{output-\linebreak{}crds} and \texttt{output-deployment} we provide two \texttt{make}-targets for generating the Custom Resource Definitions~(CRDs) and deployment resources, respectively. They can also be directly applied to the Kubernetes cluster using the \texttt{install} and \texttt{deploy} targets. For local clusters, the Skaffold\footnote{\url{https://skaffold.dev/}} tool can be used alternatively. It automates the container image build and deployment process allowing for quicker development loops. Moreover, Skaffold simplifies the configuration of remote debugging, allowing to halt and step through the code that is running within a container in the cluster. In future work, we plan to release Helm\footnote{\url{https://helm.sh/}} charts for \texttt{nagare media engine}. Helm is a package manager for Kubernetes often used to deploy and update applications.

\section{Models}
\label{sec:models}

This section discusses (data) models as well as reference architectures and concepts that are necessary to understand \texttt{nagare media engine}. In Section~\ref{subsec:network-based-media-processing}, we start with models that are defined in the NBMP standard. Afterwards, Section~\ref{subsec:custom-kubernetes-resources} details our custom models as used in Kubernetes. Finally, Section~\ref{subsec:data-transformations} describes how we transform data between different representations.

\subsection{Network-Based Media Processing}
\label{subsec:network-based-media-processing}

\begin{figure}[h]
  \centering
  \includegraphics[width=\columnwidth]{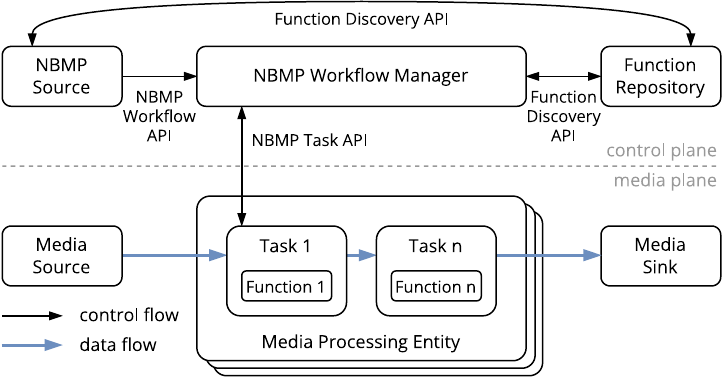}
  \caption{First edition of the NBMP reference architecture, compare~\citep{isoiec_isoiec230908_2020,neugebauer_nagaremediaengine_2025}.}
  \label{fig:nbmp-reference-architecture}
  \Description{First edition of the NBMP reference architecture}
\end{figure}

In this section, we will discuss NBMP in more detail. Throughout this section, we will refer to NBMP notions as depicted in the reference architecture in Figure~\ref{fig:nbmp-reference-architecture}.

NBMP considers workflows as a set of tasks that logically form a DAG. In this graph, nodes represent tasks while edges are network connections between input and output ports. Connected ports are said to be bound to a stream. Next to streams originating from tasks, the workflow as a whole also has input and output streams. Streams can be push- or pull-based, i.e. either the sender or the receiver facilitates the transport. NBMP further differentiates between media and metadata streams but is otherwise indifferent regarding specific data and media formats as well as network protocols. It is also open regarding supported functions. The second edition includes reference function templates in Annex~F and implementations can signal compatibility with these templates. However, developers are free to implement custom functions. Workflows and tasks follow a lifecycle model that consists of the following steps. Both start out in the ``instantiated'' state. After the configuration, they move to the ``idle'' state. In the ``running'' state, streams are processed by functions. If an error occurs, they move into the ``error'' state. Here, a state transition to earlier states is still allowed. Terminated workflows or tasks are in the ``destroyed'' state from where no further transitions are possible.

This logical model is specified through NBMP descriptions. The first edition defines workflow, task and function descriptions. The second edition extends this model with MPE capabilities descriptions that represent the current state of MPEs. Descriptions can share certain descriptors and parameters. For instance, the \texttt{general} descriptor contains general properties such as an ID or name. The standard also defines descriptions, descriptors and properties \linebreak{} through a JSON schema~\cite{wright_jsonschemamedia_2022} specification. Represented as JSON objects, they are called workflow, task, function and MPE capabilities description documents~(WDD, TDD, FDD and~MDD), respectively. The JSON representation is then the basis for the REST API specifications. All documents are declarative, i.e.~they specify intent and constraints rather than instructions.

The NBMP reference architecture consists of various entities running within the control or media plane. While control plane components manage the workflow execution, media plane components implement the actual flow and processing of data. The architecture starts with an \emph{NBMP source} (in the second edition renamed to \emph{NBMP client}) that submits a WDD via the workflow API. The \emph{NBMP workflow manager} receives the request and constructs a logical DAG based on the workflow description. For that it may consult the \emph{function repository} with the function discovery API. Given the constraints included in the WDD, suitable functions are thus selected for the DAG. Alternatively, the NBMP source could already specify known or discovered functions directly in the WDD. Next, the NBMP workflow manager schedules tasks onto MPEs. The communication between the NBMP workflow manager and MPEs is intentionally unspecified allowing for competing implementations. After the tasks are instantiated, the NBMP workflow manager initializes the task by sending a TDD via the task API. Among others, the TDD contains descriptors about input and output streams bound to ports as well as function configuration. Tasks thus consume from the \emph{media source}, i.e.~the workflow inputs, or send to the \emph{media sink}, i.e.~the workflow outputs. In case the media source or sink should initialize the data flow, the NBMP workflow manager will update the WDD with corresponding descriptors in its response to the NBMP source. The NBMP source is then responsible to instruct the media source and sink. MPEs, tasks as well as media source and sink together form the media plane.

The workflow and task APIs follow the typical create, read, update and delete~(CRUD) pattern of REST APIs by mapping these operations to the HTTP verbs \texttt{POST}, \texttt{GET}, \texttt{PATCH} and \texttt{DELETE}. In addition to \texttt{PATCH}, we also allow \texttt{PUT} for updates in \texttt{nagare media engine}. The function discovery API only allows read, i.e.~\texttt{GET}, operations. Multiple query parameters are standardized to filter the returned list of functions. Finally, the MPE API of the second edition allows reading (\texttt{GET}) the capabilities of an MPE and updating some descriptors using a \texttt{PATCH} request.

In Section~\ref{sec:overview-of-nagare-media-engine}, we already introduced the design of \texttt{nagare media engine}. Note that we currently implement components corresponding to all elements of the first edition of the NBMP reference architecture except for the function discovery API. The standard allows NBMP sources to include a \texttt{repository} descriptor in the WDD thus referencing arbitrary function repositories. We regard this as an unnecessary security risk and intentionally do not support this descriptor. Instead, only functions made explicitly available by administrators of the workflow system are allowed (see Section~\ref{subsec:custom-kubernetes-resources}). Our workflow manager will therefore not consult the function discovery API. However, in future work we could implement this API to help NBMP sources find available functions. Additionally, future work should investigate how the MDD API of the second edition can be integrated into \texttt{nagare media engine}.

\subsection{Custom Kubernetes Resources}
\label{subsec:custom-kubernetes-resources}

At a high level, Kubernetes has a simple orchestration model consisting of \emph{resources} and \emph{controllers} as the two central elements. Users express declaratively what the system should look like, i.e.~the \emph{desired} state. For this purpose, users submit resources of various types with a state specification to Kubernetes. Controllers then try to take the necessary steps towards reaching this state. Along this way, the resource may get updated by controllers with a description of the \emph{actual} state. Logically, controllers are implemented in the form of a reconciliation loop that constantly identifies and reconciles state differences. The actual state thus converges towards the desired state. In practice, reconciliation loop implementations are optimized and only run when triggered by certain events. Multiple controllers can work in tandem to achieve some behavior. For instance, a \texttt{CronJob} controller could regularly create \texttt{Job} resources based on a \texttt{CronJob} specification. In turn, the \texttt{Job} controller would create \texttt{Pod} resources that would be scheduled to and executed on Kubernetes nodes as a collection of containers. Although this model is fairly simple to understand, it leads to a highly fault-tolerant system because state deviations are automatically detected and corrected.

\begin{listing}[b]
\begin{minted}{yaml}
apiVersion: batch/v1
kind: Job
metadata:
  name: encode-01
  namespace: workflow-system
spec:
  template:
    metadata:
      labels:
        app.kubernetes.io/component: workflow-system
        app.kubernetes.io/name: encode-01
    spec:
      containers:
        - name: encode-container
          image: encode-image:v1.0.0
          resources:
            requests:
              cpu: 250m
              memory: 512Mi
            limits:
              cpu: 1000m
              memory: 512Mi
status:
  ready: 0
  succeeded: 1
  terminating: 0
  startTime: "2025-08-15T10:00:00Z"
  completionTime: "2025-08-15T11:00:00Z"
  conditions:
  - type: SuccessCriteriaMet
    status: "True"
    reason: CompletionsReached
    message: Reached expected number of succeeded pods
    lastProbeTime: "2025-08-15T11:00:00Z"
    lastTransitionTime: "2025-08-15T11:00:00Z"
  - type: Complete
    status: "True"
    message: Reached expected number of succeeded pods
    reason: CompletionsReached
    lastProbeTime: "2025-08-15T11:00:00Z"
    lastTransitionTime: "2025-08-15T11:00:00Z"
\end{minted}
  \caption{Example Kubernetes \texttt{Job} resource in YAML representation.}
  \label{lst:job}
  \Description{Example Kubernetes \texttt{Job} resource in YAML representation.}
\end{listing}

Resources follow a common structure and are usually represented as YAML. Listing~\ref{lst:job} depicts an example \texttt{Job} resource. All resources are versioned and part of an API group both identified by the \texttt{apiVersion} field (line~1). The resource type is determined by the \texttt{kind} field (line~2). API groups bundle different resource types, e.g.~\texttt{Job} and \texttt{CronJob} belong to the same API group ``\texttt{batch}''. The last mandatory field is \texttt{metadata} that contains common subfields (lines~3--5). All resources are identified by a \texttt{name}. Resource types can further be cluster- or namespace-scoped. If a resource type is namespace-scoped, the \texttt{namespace} field must be specified; a resource is then identified by \texttt{namespace} and \texttt{name}.

Next to these three mandatory fields, most resources also have \texttt{spec} and \texttt{status} fields (lines~6--22 and~23--41). While the former describes the desired state, the latter shows the actual state. Note that \texttt{status} is normally only changed by a corresponding controller and therefore requires additional permissions. As such, it represents the state as seen by the last iteration of the reconciliation loop. The available subfields depend on the resource type, yet some patterns emerged.

Resource types that will eventually result in the creation of further resources (e.g.~a~\texttt{Job} will result in one or more \texttt{Pod}s) often have a \texttt{template} field for that dependent resource type (lines~7--22). In fact, relationships between resources are so common in Kubernetes that references can be expressed directly in the \texttt{metadata} subfield \texttt{ownerReferences}. Doing that has the additional benefit that Kubernetes will automatically delete dependent resources once the ``owner'' resource is deleted (e.g.~\texttt{Pod}s resulting from a \texttt{Job} will be garbage collected when the \texttt{Job} is deleted).

A common pattern in \texttt{status} fields is the \texttt{conditions} subfield (lines~29--41). It allows controllers to convey in what conditions the resource is at the moment. The example \texttt{Job} in Listing~\ref{lst:job} is in the \texttt{SuccessCriteriaMet} and \texttt{Complete} conditions. Additional subfields give a reason, a human-readable message and a timestamp when the condition change was observed. Other systems can then react when certain conditions are met.

Kubernetes already includes many built-in resources\footnote{For an up-to-date list of built-in resources see \url{https://kubernetes.io/docs/reference/kubernetes-api/}.} for generically describing containerized workloads. Beyond that, third-party systems can extend Kubernetes with custom resources and utilize the same model for orchestrating specialized workloads. Kubernetes offers two ways to add custom resources. API Aggregation requires developers to write a custom API server that runs beside the main Kubernetes API server and handles additional HTTP paths. This option is complex but offers great flexibility. However, in many cases it is enough to extend the Kubernetes API through \texttt{CustomResourceDefinition}~(CRD) resources. The CRD is a built-in resource type for describing custom resources including the API group, versioning as well as structure and field validations. This second option is easier to implement and only requires developers to submit a description of the custom resource. The Kubernetes API server will then automatically add additional HTTP paths itself to handle requests for these resources. If additional logic is necessary (e.g.~for complex validating, defaulting or altering logic), webhooks can be configured that are then called by the Kubernetes API server.

Kubebuilder supports the implementation of CRD-based API extensions as well as corresponding controllers. Here, custom resources are based on Go \texttt{struct} definitions annotated with \texttt{struct} field tags as well as additional Kubebuilder markers. Based on the \texttt{struct} definition, Kubebuilder's \texttt{controller-gen} tool will then automatically generate common \texttt{struct} methods, CRDs and other related Kubernetes resources.
\begin{listing}[t]
\begin{minted}{go}
import metav1 "k8s.io/apimachinery/pkg/apis/meta/v1"

// +kubebuilder:object:root=true

type ExampleList struct {
  metav1.TypeMeta `json:",inline"`
  metav1.ListMeta `json:"metadata,omitempty"`

  Items []Example `json:"items"`
}

// +kubebuilder:object:root=true
// +kubebuilder:subresource:status
// +kubebuilder:resource:categories={exmpl}
// +kubebuilder:printcolumn:name="Config",type="string",JSONPath=`.spec.myConfig`
// +kubebuilder:printcolumn:name="Age",type="date",JSONPath=`.metadata.creationTimestamp`
// +kubebuilder:printcolumn:name="Start",type="date",JSONPath=`.status.startTime`
// +kubebuilder:printcolumn:name="End",type="date",JSONPath=`.status.endTime`

type Example struct {
  metav1.TypeMeta   `json:",inline"`
  metav1.ObjectMeta `json:"metadata,omitempty"`

  Spec   ExampleSpec   `json:"spec,omitempty"`
  Status ExampleStatus `json:"status,omitempty"`
}

type ExampleSpec struct {
  // +kubebuilder:validation:Pattern="^[a-zA-Z0-9]$"
  MyConfig string `json:"myConfig"`
}

type ExampleStatus struct {
  // +optional
  StartTime *metav1.Time `json:"startTime,omitempty"`

  // +optional
  EndTime *metav1.Time `json:"endTime,omitempty"`

  // +optional
  // +patchMergeKey=type
  // +patchStrategy=merge
  // +listType=atomic
  Conditions []Condition `json:"conditions,omitempty" patchStrategy:"merge" patchMergeKey:"type"`
}
\end{minted}
  \caption{Go \texttt{struct} definition of the \texttt{Example} resource with Kubebuilder annotations.}
  \label{lst:configuration}
  \Description{Go \texttt{struct} definition of the \texttt{Example} resource with Kubebuilder annotations.}
\end{listing}
Listing~\ref{lst:configuration} shows how an \texttt{Example} resource could be implemented. Each resource has a list and object type definition (lines~5--10 and~20--26). The \texttt{+kubebuilder:object:\linebreak{}root} marker specifies that the following \texttt{struct} is a Kubernetes resource (lines~3 and~12). Further \texttt{struct} markers indicate the availability of a \texttt{status} field (line~13), the association to a custom resource category (line~14) or what fields should be printed when this resource is retrieved with the Kubernetes CLI tool (lines~15--18).

List and object \texttt{struct}s must embed the mandatory metadata \texttt{struct}s from the Kubernetes library. \texttt{TypeMeta} adds the \linebreak{}\texttt{apiVersion} and \texttt{kind} fields, while \texttt{ListMeta} and \texttt{ObjectMeta} add the \texttt{metadata} field for lists and objects, respectively. Go \texttt{struct}s for the \texttt{spec} and \texttt{status} fields are custom implementations (lines~28--31 and~33--45). In addition to \texttt{struct} tags that govern how fields are serialized in a data exchange format, Kubebuilder \texttt{struct} field markers further configure how the CRD is generated. For instance, line~29 defines a pattern validation with a regular expression that is later enforced by the Kubernetes API server. By convention, optional fields are pointer types and have the \texttt{+optional} marker (e.g. lines~34--35). Moreover, there are markers that configure how the Kubernetes API server should merge lists when handling patch requests (lines~\mbox{41--44}).

After this introduction to Kubernetes, the following sections will introduce common types and custom resources defined by \texttt{nagare media engine}.

\subsubsection{Common Types}
\label{subsubsec:common-types}

\begin{figure}[h]
  \centering
  \includegraphics[width=0.7\columnwidth]{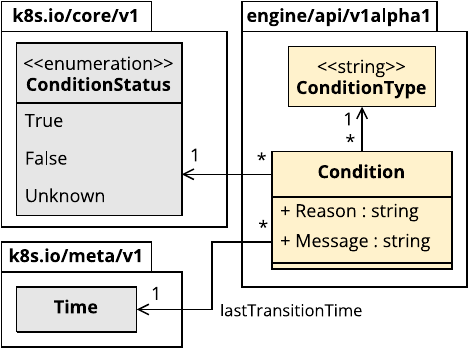}
  \caption{UML class diagram of \texttt{Condition} related types.}
  \label{fig:nme-api-condition-uml-class}
  \Description{UML class diagram of \texttt{Condition} related types}
\end{figure}

Throughout our custom data model, we use various common types that are detailed in this section. As mentioned earlier, the \linebreak{}\texttt{conditions} subfield is a common pattern in \texttt{status} fields. We also employ this pattern and define the types as seen in Figure~\ref{fig:nme-api-condition-uml-class}. We implemented our own types, as Kubernetes only defines resource-dependent condition types (e.g.~the \texttt{JobCondition} for \texttt{Job} \linebreak{}resources). Nevertheless, we follow the typical structure with \texttt{type}, \texttt{status}, \texttt{reason}, \texttt{message} and \texttt{lastTransitionTime} fields.

\begin{figure}[t]
  \centering
  \includegraphics[width=0.8\columnwidth]{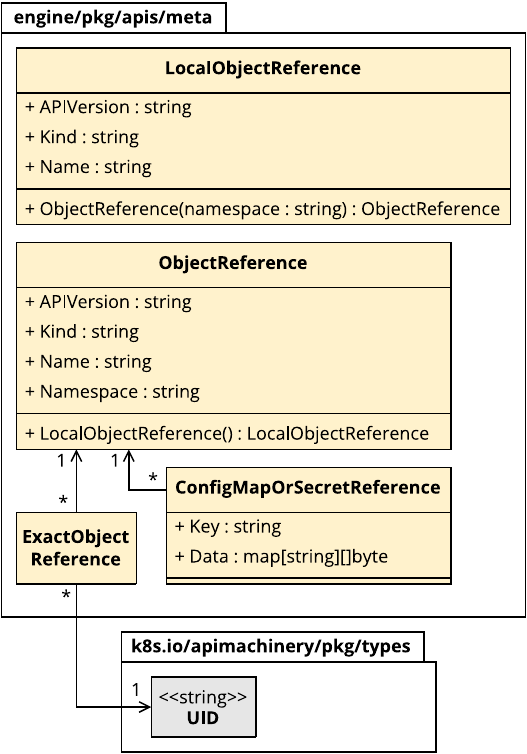}
  \caption{UML class diagram of the types in the \texttt{.../meta} package.}
  \label{fig:nme-apis-meta-uml-class}
  \Description{UML class diagram of the types in the \texttt{.../meta} package}
\end{figure}

Next, we define types for referencing other resources. This is used, for instance, when a \texttt{Task} references a \texttt{Workflow} (see Sections~\ref{subsubsec:task} and~\ref{subsubsec:workflow}). Figure~\ref{fig:nme-apis-meta-uml-class} depicts the types in an UML class diagram. Note that we define different reference types. \texttt{LocalObject\linebreak{}Reference}s can be used when referring to a resource in the same namespace. As such, it is only identified by the \texttt{APIVersion}, \texttt{Kind} and \texttt{Name}. An \texttt{ObjectReference}, on the other hand, additionally includes the \texttt{Namespace} in order to refer to resources in arbitrary namespaces. Sometimes it is necessary to reference an exact resource and identify cases where a resource was first deleted and later recreated under the same name. For this, \texttt{ExactObject\linebreak{}Reference} includes the \texttt{UID}, which is a unique identifier for each resource. We provide methods to convert between these types. Lastly, the \texttt{ConfigMapOrSecretReference} type is a special case for references to the built-in resource types \texttt{ConfigMap} and \texttt{Secret}. It combines an \texttt{ObjectReference} with a \texttt{Key} that indexes to data within the \texttt{ConfigMap} or \texttt{Secret}. The \texttt{Data} field is only used internally by the controllers and contains an in-memory copy of the content of the resolved \texttt{ConfigMap} or \texttt{Secret}. It is not exposed in the custom resources.

\begin{figure}[h]
  \centering
  \includegraphics[width=\columnwidth]{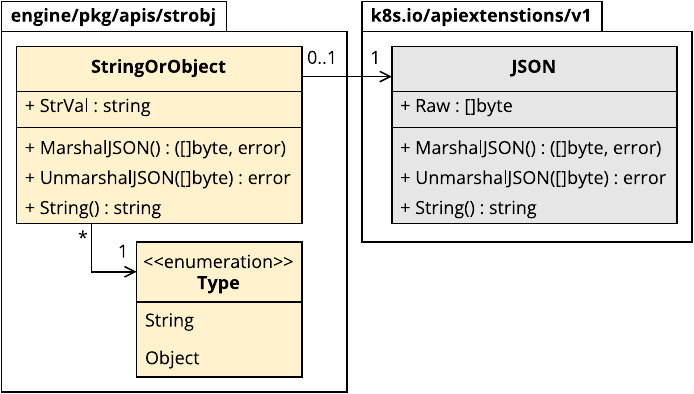}
  \caption{UML class diagram of the types in the \texttt{.../strobj} package.}
  \label{fig:nme-apis-strobj-uml-class}
  \Description{UML class diagram of the types in the \texttt{.../strobj} package}
\end{figure}
A special data type is \texttt{StringOrObject}. We currently use this in some configuration files for fields that can either be a string or an object, i.e.~a nested structure with arbitrary subfields. Figure~\ref{fig:nme-apis-strobj-uml-class} depicts the corresponding UML class diagram. Note that we implement the \texttt{MarshalJSON} and \texttt{UnmarshalJSON} methods required by the JSON (and YAML) en- and decoders. \texttt{Type} is used to differentiate between strings and objects. This design is derived from the \texttt{IntOrString} type that is used in Kubernetes.

\begin{figure}[h]
  \centering
  \includegraphics[width=0.75\columnwidth]{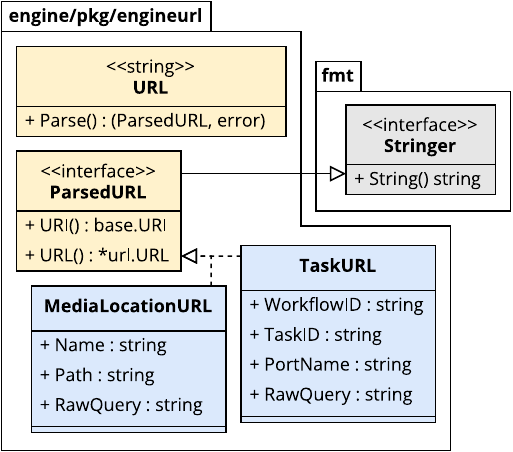}
  \caption{UML class diagram of the types in the \texttt{.../engineurl} package.}
  \label{fig:nme-engineurl-uml-class}
  \Description{UML class diagram of the types in the \texttt{.../engineurl} package}
\end{figure}

We use a custom URL schema to identify and locate multimedia streams. Here, we differentiate between streams originating from tasks and streams from pre-configured media locations (see Section~\ref{subsubsec:medialocation-and-clustermedialocation}). Both \texttt{TaskURL} and \texttt{MediaLocationURL} have corresponding types to easily access parsed URL components as illustrated in Figure~\ref{fig:nme-engineurl-uml-class}.

\begin{figure*}[t]
  \centering
  \includegraphics[width=0.7\textwidth]{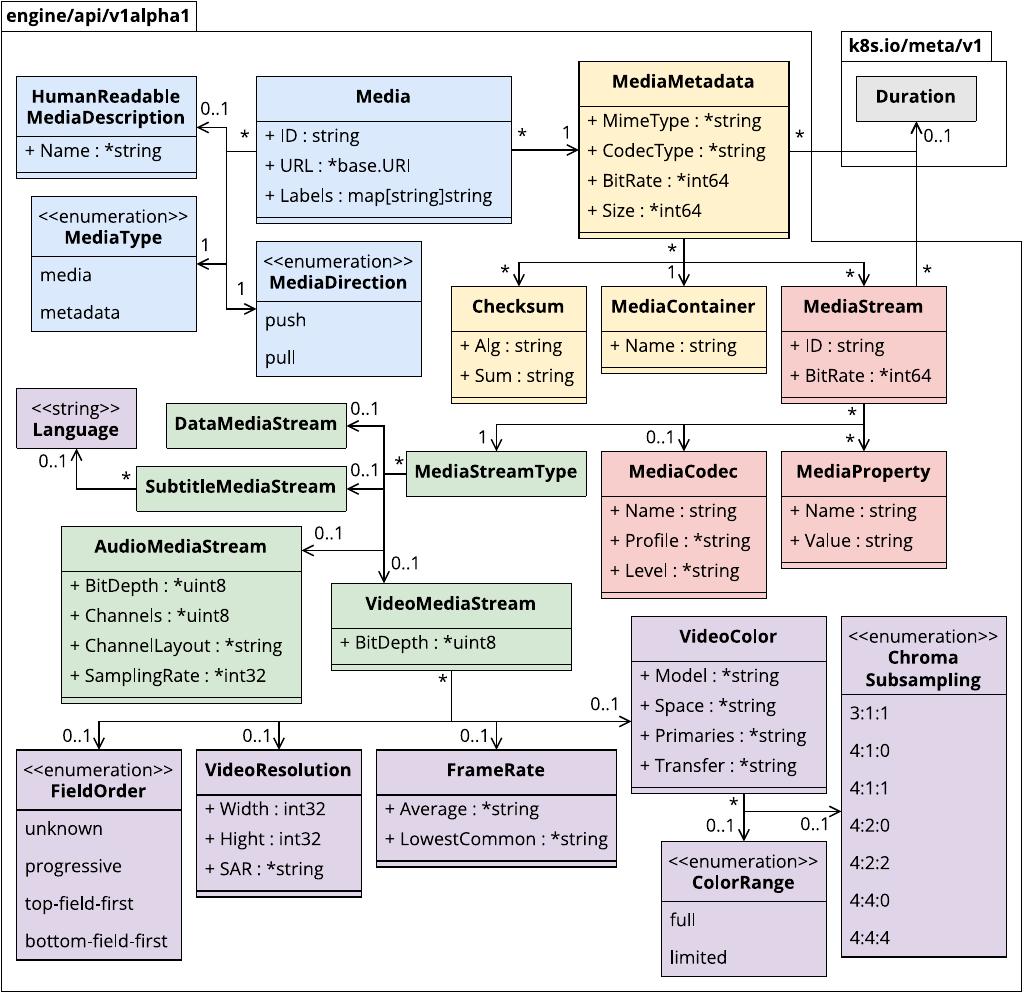}
  \caption{UML class diagram of \texttt{Media} related types.}
  \label{fig:nme-api-media-uml-class}
  \Description{UML class diagram of \texttt{Media} related types}
\end{figure*}

We model multimedia content itself with the \texttt{Media} type. Figure~\ref{fig:nme-api-media-uml-class} depicts a class diagram of \texttt{Media} and related types. Note that some properties directly map to NBMP properties, while others can only be represented as generic key-value pairs in NBMP. \texttt{MediaType} (media and metadata) and \texttt{MediaDirection} (push and pull) refer to the same NBMP notions. Moreover, all \texttt{Media} instances are identified by an \texttt{ID} that corresponds to the NBMP stream ID. The \texttt{URL} is optional, e.g.~when it is to be set later by the workflow manager. It can be a URL representing a common streaming format as well as our custom URL schema representing an internal task or media location stream.

\texttt{Media} instances then have additional container- and stream-level metadata properties. \texttt{MediaMetadata} captures general properties such as the media type (\texttt{MimeType})~\cite{freed_mediatypespecifications_2013} including \texttt{codecs} and \texttt{profiles} parameters (\texttt{CodecType})~\cite{frojdh_codecsprofilesparameters_2011}, the overall bit rate and size, the duration, any checksums as well as media container metadata. \texttt{Media} instances then can have arbitrary many \texttt{MediaStream}s. Note that the term ``stream'' in this context does not refer to an NBMP stream that would include the whole \texttt{Media} instance. Rather, \texttt{MediaStream} refers to elementary streams, i.e.~individual video, audio, subtitle and data streams, that are muxed together in a media container. For instance, the ISO Base Media File Format~(ISO BMFF)~\cite{isoiec_isoiec1449612_2022}~container format could contain one video stream and two corresponding audio streams in different languages. General stream-level metadata properties are modelled in the \texttt{MediaStream} and \texttt{MediaCodec} types. Furthermore, \texttt{MediaProperty} allows capturing general key-value pairs. The \texttt{Video-}, \texttt{Audio-}, \texttt{Subtitle-} and \texttt{DataMediaStream} types then model properties of the corresponding stream types. Here, we included the most prominent properties such as the video resolution or audio sampling rate.

\subsubsection{\texttt{MediaProcessingEntity} and \texttt{ClusterMediaProcessingEntity}}
\label{subsubsec:mediaprocessingentity-and-clustermediaprocessingentity}

\begin{figure*}[t]
  \centering
  \includegraphics[width=0.85\textwidth]{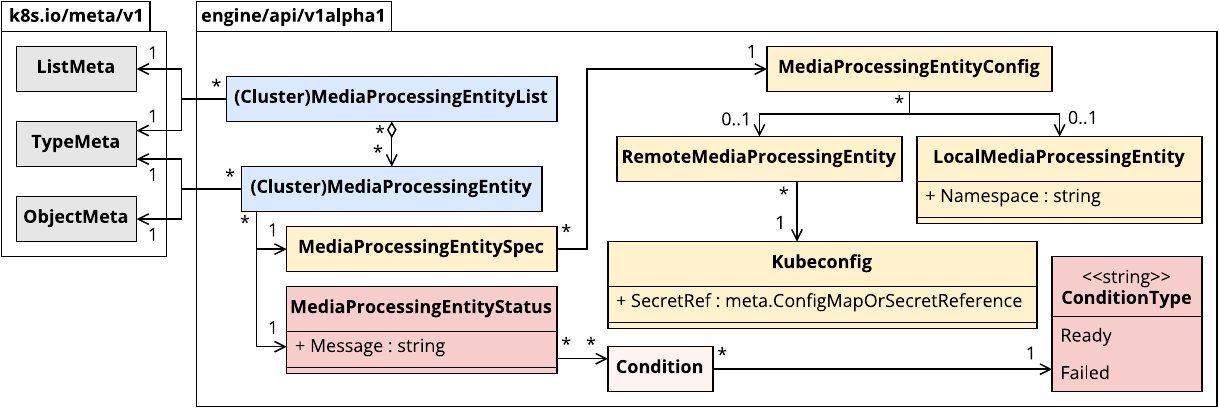}
  \caption{UML class diagram of \texttt{MediaProcessingEntity} related types.}
  \label{fig:nme-api-mediaprocessingentities-uml-class}
  \Description{UML class diagram of \texttt{MediaProcessingEntity} related types}
\end{figure*}

Administrators can manage the available MPEs through the \texttt{Media\linebreak{}ProcessingEntity} and \texttt{ClusterMediaProcessingEntity} custom resource types. \texttt{MediaProcessingEntity} is namespace-scoped. As such, it is only available to workflows running in that Kubernetes namespace. \texttt{ClusterMediaProcessingEntity}, on the other hand, configures an MPE that is available to workflows in all namespaces. Administrators can thus control which MPEs are available to which users assuming that Kubernetes namespaces are created around organizational boundaries. Apart from the scope, both resource types have the same structure as depicted in Figure~\ref{fig:nme-api-mediaprocessingentities-uml-class}.

We differentiate between local and remote MPEs. As MPEs in \texttt{nagare media engine} correspond to Kubernetes clusters, a local MPE refers to the cluster local to the control plane, i.e.~the management cluster (see Section~\ref{subsec:design}). Each local MPE has a configured \texttt{Namespace} where tasks should be scheduled. In contrast, remote MPEs refer to other Kubernetes clusters. Administrators therefore need to define a kubeconfig file, i.e.~a configuration file that includes the necessary details to connect to a Kubernetes cluster. The structure of kubeconfig files is standardized by Kubernetes and already sets a default namespace. Because kubeconfig files contain secrets for the authentication, the content is not directly associated with our custom resources. Instead, administrators need to create a \texttt{Secret} resource and set a corresponding reference.

\texttt{(Cluster)MediaProcessingEntity} resources have a \texttt{status} field that includes the observed condition. Either they are in the \texttt{Ready} condition and thus new tasks can be scheduled, or they are in the \texttt{Failed} condition, e.g. because a connection to the remote cluster could not be established. In the latter case, the \texttt{Message} field will give further details about the error.

\subsubsection{\texttt{Function} and \texttt{ClusterFunction}}
\label{subsubsec:function-and-clusterfunction}

\begin{figure}[h]
  \centering
  \includegraphics[width=0.9\columnwidth]{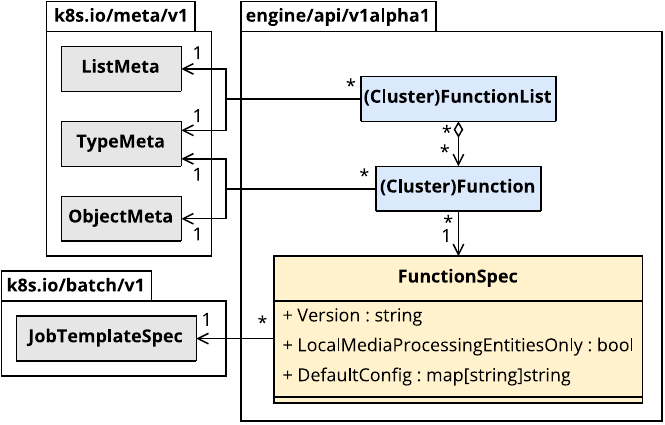}
  \caption{UML class diagram of \texttt{Function} related types.}
  \label{fig:nme-api-function-uml-class}
  \Description{UML class diagram of \texttt{Function} related types}
\end{figure}
Analogous to the MPE configuration, the \texttt{Function} and \texttt{Cluster\linebreak{}Function} custom resources offer a way for administrators to configure multimedia functions on a namespace or cluster level. Again, the scope allows administrators to control the availability for users. Figure~\ref{fig:nme-api-function-uml-class} shows a UML class diagram of these types.

Each custom resource includes a version and optionally a default configuration in the form of key-value pairs. Functions can further be restricted to only run on a local MPE, e.g.~if they need access to control plane components. Most importantly, \texttt{(Cluster)Function} resources have a \texttt{template} field with a \texttt{Job} resource specification. Instantiated functions will run as Kubernetes \texttt{Job}s that are constructed based on this field. \texttt{(Cluster)Function} resources only configure available multimedia functions and therefore have no \texttt{status} field.

\subsubsection{\texttt{MediaLocation} and \texttt{ClusterMediaLocation}}
\label{subsubsec:medialocation-and-clustermedialocation}

\begin{figure*}[t]
  \centering
  \includegraphics[width=0.9\textwidth]{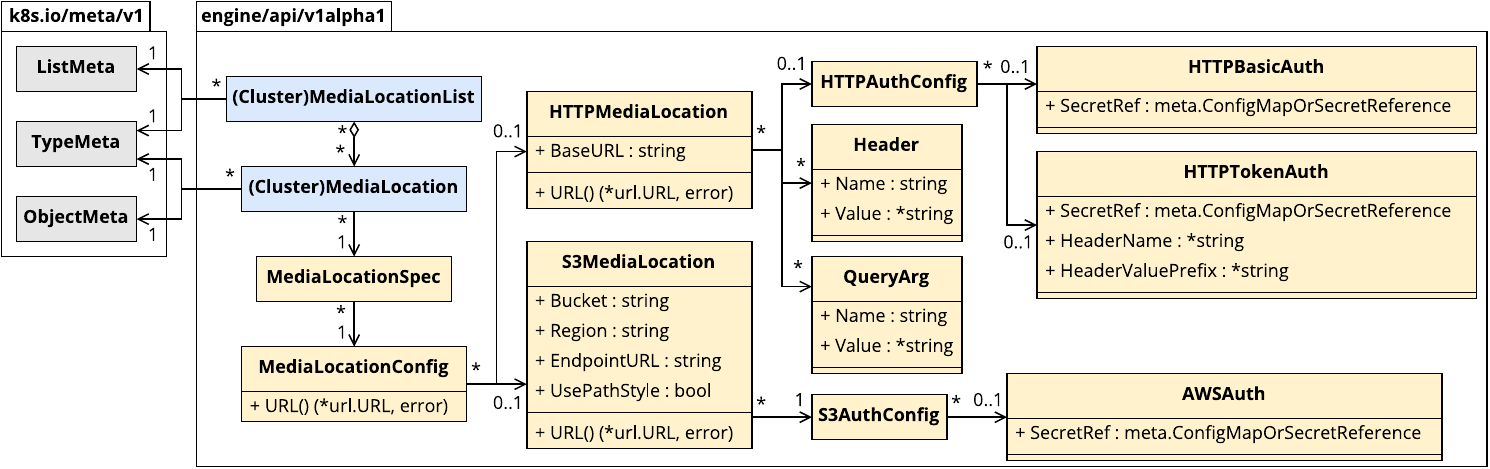}
  \caption{UML class diagram of \texttt{MediaLocation} related types.}
  \label{fig:nme-api-medialocation-uml-class}
  \Description{UML class diagram of \texttt{MediaLocation} related types}
\end{figure*}

Multiple workflows potentially use the same storage system for inputs and outputs. For instance, the same S3 bucket (an~object storage system) could serve as the input and output destination for file-based workflows. With \texttt{MediaLocation} and \texttt{ClusterMedia\linebreak{}Location} we provide a way for administrators to pre-configure media locations for users. A special URL schema can then be used in the input or output description to reference a media location (see Section~\ref{subsubsec:common-types}). The types are depicted in Figure~\ref{fig:nme-api-medialocation-uml-class}.

Currently, administrators can describe HTTP and S3 media locations. \texttt{HTTPMediaLocation} includes a \texttt{BaseURL} that is extended with additional path segments when used in the workflow. Moreover, administrators can define headers and query parameters. Finally, HTTP media locations secured by HTTP Basic Authentication~~\cite{reschke_basichttpauthentication_2015} or tokens can also be described. An \texttt{S3MediaLocation} requires the typical S3 parameters, i.e.~the bucket name and region as well as the URL endpoint and authentication information.

When a media location is used in a workflow, \texttt{nagare media engine} will encode the pre-configured parameters as URL that then needs to be interpreted by the multimedia function. The URL generation is implemented by the \texttt{URL} methods.

\texttt{(Cluster)MediaLocation} resources do not have a \texttt{status} field. Future work could implement further media locations, e.g. for Real-Time Messaging Protocol~(RTMP)~\cite{hparmar_adobesrealtime_2012} or Media over QUIC Transport~(MOQT)~\cite{nandakumar_mediaquictransport_2025} servers.

\subsubsection{\texttt{TaskTemplate} and \texttt{ClusterTaskTemplate}}
\label{subsubsec:tasktemplate-and-clustertasktemplate}

In NBMP workflows, multiple tasks can share a function instantiation. For instance, a workflow could consist of two encoding tasks that are configured in the same way and only differ in the input and output streams. We implement this idea with the \texttt{TaskTemplate} and \texttt{ClusterTaskTemplate} custom resources. Apart from the missing \linebreak{}
\begin{figure}[H]
  \vspace{-1pt}
  \centering
  \includegraphics[width=\columnwidth]{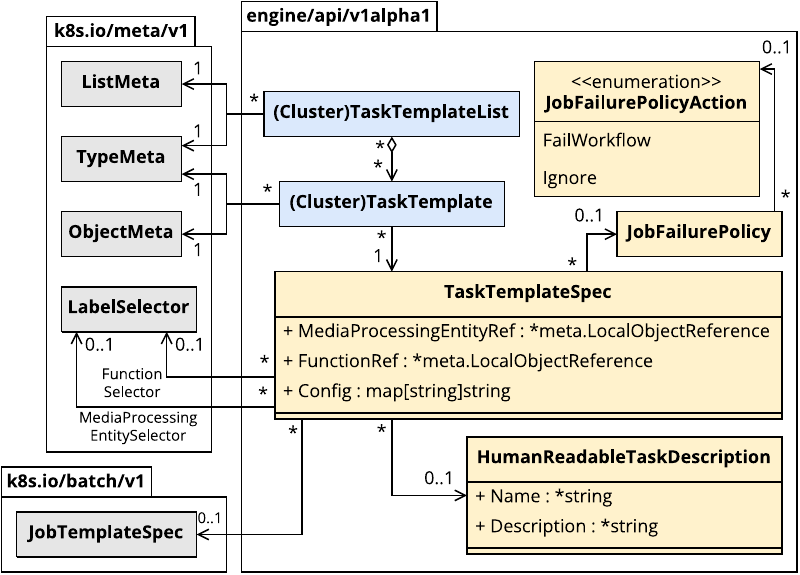}
  \caption{UML class diagram of \texttt{TaskTemplate} related types.}
  \label{fig:nme-api-tasktemplate-uml-class}
  \Description{UML class diagram of \texttt{TaskTemplate} related types}
\end{figure}
\noindent\texttt{status} field, they follow a structure similar to \texttt{Task} resources (see Section~\ref{subsubsec:task}). Figure~\ref{fig:nme-api-tasktemplate-uml-class} shows a UML class diagram of the relevant types.

All fields are optional and can later be overwritten by the \texttt{Task} resource. The MPE and multimedia function are either referenced directly with a \texttt{LocalObjectReference} or indirectly with a \texttt{Label\linebreak{}Selector}. In Kubernetes, labels are key-value pairs set within the \texttt{metadata} field of any resource. Many built-in resources use a \texttt{LabelSelector} to model a loosely coupled reference. Next, the \texttt{Config} field contains key-value pairs that are passed to the function as configuration. For this, it is merged with the \texttt{DefaultConfig} field from the referenced \texttt{(Cluster)Function}. Similarly, \texttt{(Cluster)\linebreak{}TaskTemplate} resources can overwrite fields from the function's \texttt{Job} template. The execution of a task can be unsuccessful. How \texttt{nagare media engine} should handle task failures is determined by the \texttt{JobFailurePolicy}. By default, failed tasks will fail the complete workflow and thus halt other tasks as well. Alternatively, if a task is non-critical, the failure can be ignored and other tasks will continue normally. Note that the \texttt{Job} template can contain a \texttt{PodFailurePolicy} that handles errors on the \texttt{Pod} level. Therefore, before the \texttt{Job} is considered unsuccessful, Kubernetes potentially already restarted \texttt{Pod}s multiple times. Finally, the \texttt{(Cluster)\linebreak{}TaskTemplate} can contain a human-readable name and description that can be used in user interfaces~(UIs) or the Kubernetes CLI tool.

\begin{figure*}[hb]
  \centering
  \includegraphics[width=0.6\textwidth]{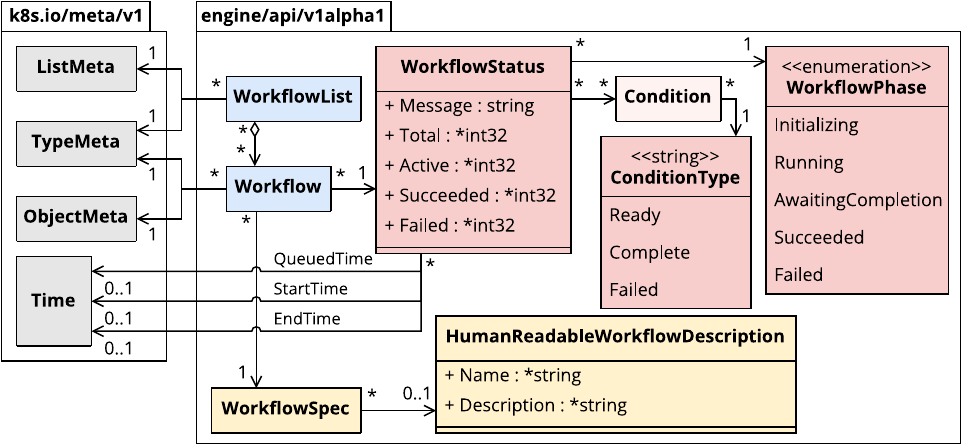}
  \caption{UML class diagram of \texttt{Workflow} related types.}
  \label{fig:nme-api-workflow-uml-class}
  \Description{UML class diagram of \texttt{Workflow} related types}
\end{figure*}
\subsubsection{\texttt{Workflow}}
\label{subsubsec:workflow}

Workflows are modelled in the \texttt{Workflow} custom resource. There is no cluster-scoped variant as workflows always execute within a namespace. Figure~\ref{fig:nme-api-workflow-uml-class} illustrates the structure in a UML class diagram.

The \texttt{spec} field only contains a human-readable name and description. The \texttt{status} field, on the other hand, gives an overview of the whole workflow execution. First, the \texttt{Queued-}, \texttt{Start-} and \texttt{EndTime} give information about when the \texttt{Workflow} was first seen by the workflow manager, when tasks first started executing and when the \texttt{Workflow} terminated, respectively. Next, the \texttt{Total} field reports the number of tasks that belong to this workflow while \texttt{Active}, \texttt{Succeeded} and \texttt{Failed} count the number of tasks in these states. \texttt{Workflow} resources follow a lifecycle analogous to the NBMP lifecycle model. The current state is indicated by the \texttt{WorkflowPhase}. We will detail each phase later in Section~\ref{subsec:workflow-controller}. Lastly, \texttt{Workflow} resources can be in three conditions that are commonly used in Kubernetes. \texttt{Ready} indicates that the workflow is in a normal state and actively running. \texttt{Complete} and \texttt{Failed} are termination conditions for successful or unsuccessful executions.

\subsubsection{\texttt{Task}}
\label{subsubsec:task}

The last custom resource in the \texttt{nagare media engine} model is \texttt{Task} that describes an NBMP task execution. As mentioned in Section~\ref{subsubsec:tasktemplate-and-clustertasktemplate}, \texttt{(Cluster)TaskTemplate} and \texttt{Task} share some \texttt{spec} fields. \texttt{Task} resources can thus inherit fields by defining a reference to the \texttt{(Cluster)TaskTemplate}. MPE and function references as well as the \texttt{Config}, \texttt{JobFailurePolicy} and the human-readable name and description fields are adopted if they are defined in the template. Otherwise or if no \texttt{(Cluster)TaskTemplate} is used at all, these fields need to be set in the \texttt{Task} resource. Additionally, the \texttt{WorkflowRef} determines to which \texttt{Workflow} this \texttt{Task} belongs. Furthermore, \texttt{Input-} and \texttt{OutputPortBinding}s connect a \texttt{Media} stream to function ports.

\begin{figure*}[t]
  \centering
  \includegraphics[width=0.75\textwidth]{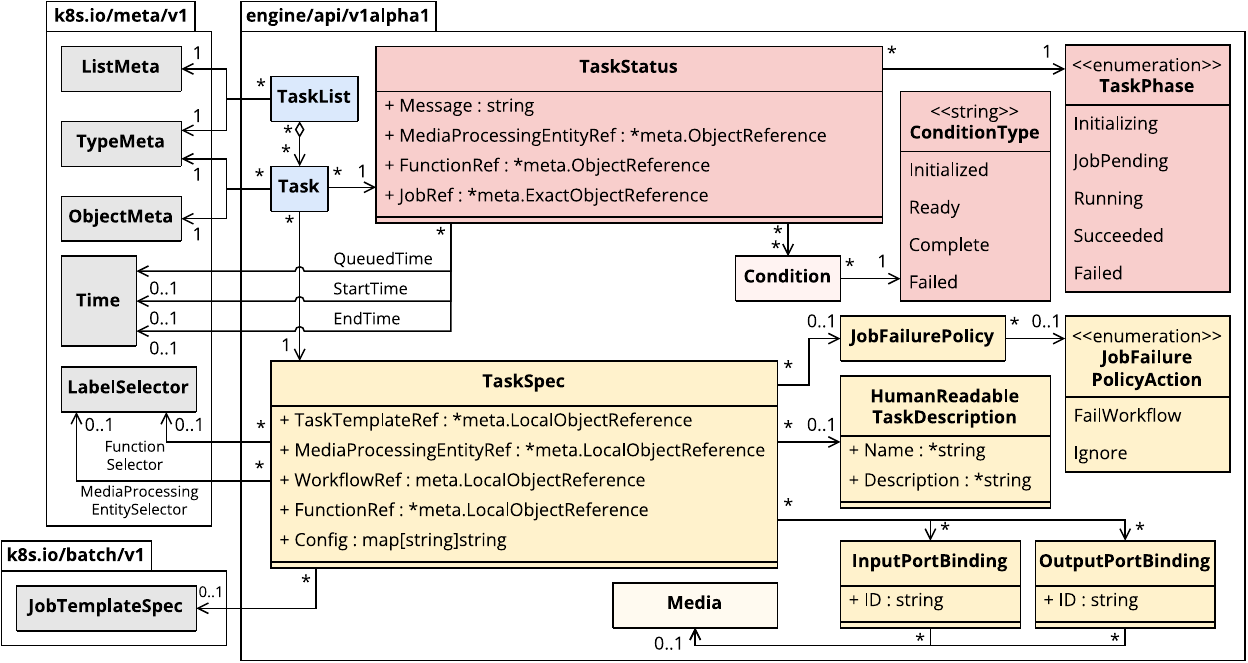}
  \caption{UML class diagram of \texttt{Task} related types.}
  \label{fig:nme-api-task-uml-class}
  \Description{UML class diagram of \texttt{Task} related types}
\end{figure*}

Similar to \texttt{Workflow} resources, the \texttt{Task} \texttt{status} field contains the \texttt{Queued-}, \texttt{Start-} and \texttt{EndTime}. We also follow a lifecycle model as indicated by the \texttt{TaskPhase}. Section~\ref{subsec:task-controller} will detail each phase. \texttt{Task} resources are in the \texttt{Initialized} condition before a \texttt{Job} is running, in the \texttt{Ready} condition during the \texttt{Job} execution and finally in either the \texttt{Complete} or \texttt{Failed} condition depending on the \texttt{Job} outcome. The workflow manager will determine the used MPE and function in various ways depending on the \texttt{Task} \texttt{spec} fields used (see Section~\ref{subsec:task-controller}). The \texttt{status} field contains a reference to the selected MPE and function. It will also reference the associated \texttt{Job} resource. Figure~\ref{fig:nme-api-task-uml-class} shows the \texttt{Task} resource and related types.

\subsection{Data Transformations}
\label{subsec:data-transformations}

\begin{figure}[b]
  \centering
  \includegraphics[width=\columnwidth]{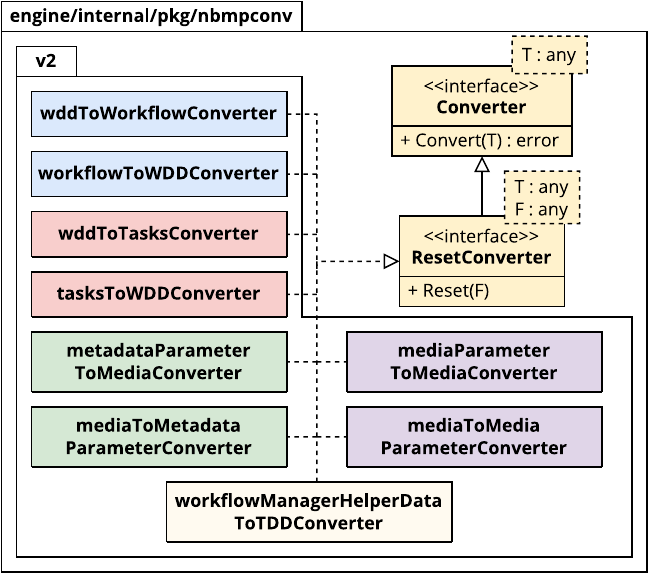}
  \caption{UML class diagram of the types in the \texttt{.../nbmpconv} package.}
  \label{fig:nme-nbmpconv-uml-class}
  \Description{UML class diagram of the types in the \texttt{.../nbmpconv} package}
\end{figure}

\texttt{nagare media engine} needs to translate between the NBMP (data) model and its own. Especially the NBMP Gateway and Workflow Manager Helper components perform data transformations. For this purpose, we implemented converters between various types. Figure~\ref{fig:nme-nbmpconv-uml-class} gives an overview of available converters.

The \texttt{Converter} interface describes how a type that performs data transformations looks like. We utilize Go's generics feature to implement type-safe \texttt{Converter} implementations that target any \texttt{T}~instance. The \texttt{ResetConverter} extends this interface and enables the reuse of \texttt{Converter}s by resetting the instance \texttt{F} from where the transformation should take place.

We generally implemented converters for both directions. For instance, NBMP Gateway uses \texttt{wddToWorkflowConverter} to transform incoming requests that contain a WDD to a \texttt{Workflow} resource. When responding, the \texttt{workflowToWDDConverter} is used for the reversed transformation. We implemented converters between the NBMP WDD and TDD data models including parameters for media and metadata streams on the one hand and the \texttt{nagare media engine} data models \texttt{Workflow}, \texttt{Task}, \texttt{Media} and Workflow Manager Helper data (see Section~\ref{sec:workflow-manager-helper}) on the other hand.

\section{Common Packages}
\label{sec:common-packages}

Before discussing each component of the \texttt{nagare media engine} design, this section will first outline shared functionality that was extracted into common packages. Section~\ref{subsec:updatable} starts with the Updatable package. In Section~\ref{subsec:starter-and-manager} we detail how we manage multiple Goroutines, Go's multi-threading primitive. Next, Section~\ref{subsec:http-server} outlines how the underlying HTTP server for the various APIs is implemented. Sections~\ref{subsec:context-middleware},~\ref{subsec:request-id-middleware} and~\ref{subsec:telemetry-middleware} discuss common HTTP middleware layers. Finally, in the last three sections, we explain the health, event as well as NBMP workflow and task APIs, respectively.

\subsection{Updatable}
\label{subsec:updatable}

Some \texttt{nagare media engine} components need to watch for configuration file changes. In Kubernetes, configuration files are often stored in \texttt{ConfigMap} or \texttt{Secret} resources. These are mounted as special volumes in a \texttt{Pod}, i.e.~Kubernetes creates corresponding folders and files within the container filesystem with the contents of the \texttt{ConfigMap} or \texttt{Secret}. Kubernetes will automatically update these files if the resources are changed. Applications can then react appropriately.

\begin{figure}[b]
  \centering
  \includegraphics[width=\columnwidth]{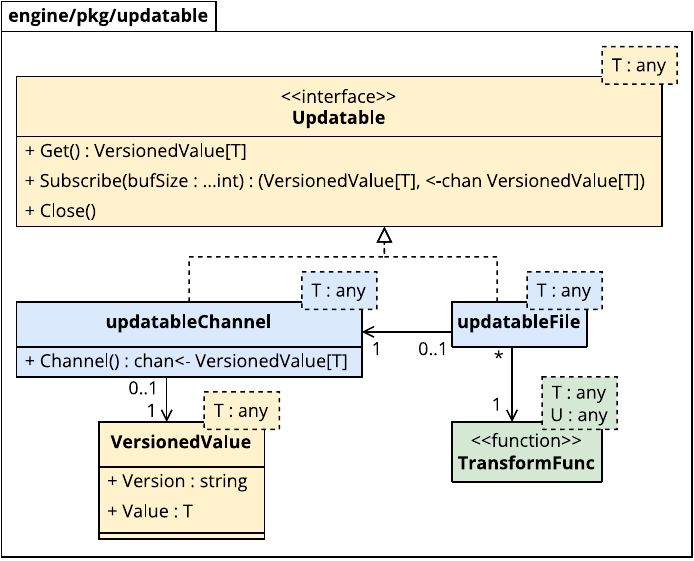}
  \caption{UML class diagram of the types in the \texttt{.../updatable} package.}
  \label{fig:nme-updatable-uml-class}
  \Description{UML class diagram of the types in the \texttt{.../updatable} package}
\end{figure}

We abstracted this process with generic types as depicted in Figure~\ref{fig:nme-updatable-uml-class}. The \texttt{Updatable} interface describes a way to get notified when a value was updated. Here, \texttt{VersionedValue} associates a \texttt{Version} attribute with a wrapped \texttt{Value}. \texttt{Updatable}'s \texttt{Get} method returns the current value, while \texttt{Subscribe} allows registering for changes. It also returns the current value and additionally a Go channel. Subscribers can read from that channel to be notified when the value has changed. \texttt{Close} stops further updates and closes all subscriber channels.

We provide two implementations of this interface. \texttt{updatable\linebreak{}Channel} is an implementation that also uses a channel to receive updated values. The \texttt{Channel} method therefore returns a send-only channel. \texttt{updatableFile} uses \texttt{updatableChannel} to implement \texttt{Updatable} for file changes. Here, the \texttt{github.com/fsnotify/fs\linebreak{}notify}\footnote{\url{https://github.com/fsnotify/fsnotify}} library is used to be notified by the operating system when a specific file was changed. The content of the file is then read and passed as an argument to the associated \texttt{TransformFunc}. This function may transform the content and return a different type, e.g.~it could parse the content of a configuration file and return a type with an in-memory representation. \texttt{updatableFile} creates a new \texttt{VersionedValue} with the return value of \texttt{TransformFunc} and sends it to the associated \texttt{updatableChannel}. It uses the \mbox{SHA-256} hash of the file content as \texttt{Version}. In this way, file changes unrelated to the content, e.g.~permission changes, are detected and will not notify subscribers.

\subsection{Starter and Manager}
\label{subsec:starter-and-manager}

\begin{figure}[h]
  \centering
  \includegraphics[width=0.7\columnwidth]{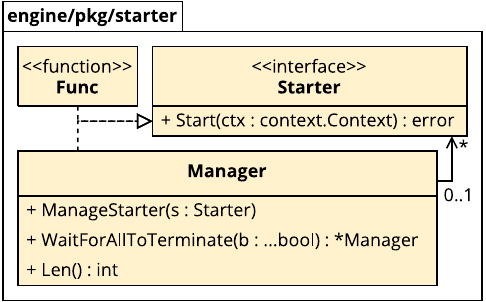}
  \caption{UML class diagram of the types in the \texttt{.../starter} package.}
  \label{fig:nme-starter-uml-class}
  \Description{UML class diagram of the types in the \texttt{.../starter} package}
\end{figure}

Support for concurrency is built directly into the Go language. Developers can easily spawn a new concurrently running Goroutine using the \texttt{go} keyword. The standard library also offers primitives such as \texttt{Mutex} and \texttt{WaitGroup} to coordinate multiple threads. We add additional types for \texttt{nagare media engine}. An overview is given in Figure~\ref{fig:nme-starter-uml-class}.

The \texttt{Starter} interface describes a type with a \texttt{Start} method that should execute in its own Goroutine. A \texttt{Context} is the only argument and it may return an error. The use of \texttt{Context} as the first argument is a common pattern in Go. \texttt{Context}s are mainly used to pass down cancellation signals. It is therefore expected that a canceled \texttt{Context} will lead to a swift termination of the \texttt{Start} method. In this way, \texttt{Starter} does not need a stop method. We use \texttt{Context}s right from the beginning in the main function and then throughout all of \texttt{nagare media engine}.

\texttt{Func} adapts an existing function to the \texttt{Starter} interface. In simple cases, the interface can thus be implemented simply by defining a function.

The \texttt{Manager} type helps with the orchestration. Developers can register \texttt{Starter}s using the \texttt{ManageStarter} method. The \texttt{Manager} will then make sure to start all registered \texttt{Starter}s at the same time. Developers can further control how termination should be handled. By default, the termination of one \texttt{Starter} will trigger the termination of all others. Alternatively, the \texttt{WaitForAllToTerminate} method can be used to configure the \texttt{Manager} to wait for all registered \texttt{Starter}s. Note that \texttt{Manager} itself implements the \texttt{Starter} interface. It is thus possible to construct hierarchies of \texttt{Manager}s with different behaviors.

\subsection{HTTP Server}
\label{subsec:http-server}

\begin{figure}[h]
  \centering
  \includegraphics[width=0.7\columnwidth]{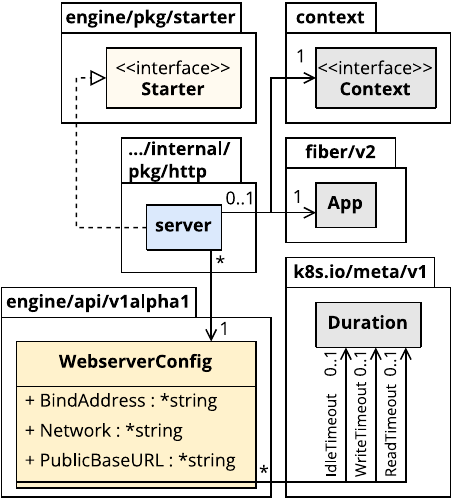}
  \caption{UML class diagram of HTTP Server related types.}
  \label{fig:nme-http-server-uml-class}
  \Description{UML class diagram of HTTP Server related types}
\end{figure}

We use the Fiber framework as the basis for the HTTP server in \texttt{nagare media engine}. Because various components implement HTTP-based APIs, a common implementation has been extracted into a separate package as depicted in Figure~\ref{fig:nme-http-server-uml-class}.

The \texttt{server} type bundles the logic to instantiate an HTTP server in \texttt{nagare media engine}. It implements the \texttt{Starter} interface in order to run concurrently to other threads. The \texttt{WebserverConfig} struct holds necessary configuration options and is passed to the \texttt{server} during the instantiation. The other components will typically embed this struct and allow users to set these fields in a configuration file. The \texttt{BindAddress} sets the network port and possibly the interfaces on which the HTTP server should listen for requests. \texttt{Network} controls whether Transmission Control Protocol~(TCP) with Internet Protocol Version~4~(IPv4), Internet Protocol Version~6~(IPv6) or both should be used. Go also supports UNIX domain sockets for inter-process communication, but because components in \texttt{nagare media engine} are distributed on multiple nodes, this is not supported. The \texttt{PublicBaseURL} defines the base URL that clients can use to reach this server. Some APIs report this address in responses, e.g.~the NBMP standard requires responses that created a resource to include a \texttt{Location} header with a URL to the created resource. Lastly, timeouts for idle connections as well as read and write operations can be configured. The \texttt{server} type wraps a Fiber \texttt{App} and thus allows its users to register request handlers. Additionally, it adds global HTTP middlewares that are discussed in the next sections.

\subsection{Context Middleware}
\label{subsec:context-middleware}

The context middleware is added globally and executes as the first request handler in the chain. It associates the Go \texttt{Context} passed to the \texttt{server} with the Fiber request. All subsequent handlers can thus retrieve the \texttt{Context}, e.g.~to receive cancellation signals.

\subsection{Request ID Middleware}
\label{subsec:request-id-middleware}

The request ID middleware is added globally and executes as the second request handler. Its purpose is to give each request a unique ID that can later be used in traces or for debugging. For this purpose, it first checks if the request already sets an ID in the commonly used \texttt{X-Request-ID} header. If that is not the case, a Universally Unique Identifier~(UUID) version~4~\cite{davis_universallyuniqueidentifiers_2024} will be generated and associated with the request. In any case, \texttt{X-Request-ID} will be included as a header in the response.

\subsection{Telemetry Middleware}
\label{subsec:telemetry-middleware}

The last common middleware is the telemetry middleware. It is not added globally by the \texttt{server} and needs to be included explicitly by the components. This is because it should only handle specific requests, e.g.~it should generally ignore health check requests (see Section~\ref{subsec:health-api}).

The telemetry middleware gathers information from the request and response in order to print it as a log message. For instance, it will log the remote address and user agent from the request or the status code and response time from the response. Future work could extend this middleware beyond logging and provide data for metric or request tracing systems.

\subsection{Health API}
\label{subsec:health-api}

Kubernetes defines the notion of healthy and unhealthy \texttt{Pod}s. Administrators can define health checks that are performed automatically by Kubernetes in regular intervals. If a \texttt{Pod} is considered unhealthy, this may have further consequences. For instance, Kubernetes may no longer route requests to this \texttt{Pod} (readiness check) or it may get restarted after some time (liveness check).

Kubernetes implements various health check mechanisms, but HTTP is commonly used. Here, Kubernetes sends a \texttt{GET} request to a configured endpoint and the application is expected to answer with a success status code (e.g.~``200~OK''). Error status codes or a general failure during the request, on the other hand, would indicate an unhealthy \texttt{Pod}.

\begin{figure}[b]
  \centering
  \includegraphics[width=0.45\columnwidth]{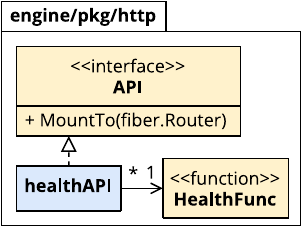}
  \caption{UML class diagram of \texttt{healthAPI} related types.}
  \label{fig:nme-http-api-health-uml-class}
  \Description{UML class diagram of \texttt{healthAPI} related types}
\end{figure}
With the \texttt{healthAPI} type we implemented a generic Health API. It handles requests sent to the \texttt{/readyz} and \texttt{/healthz} paths that are often used for readiness and liveness checks, respectively. The \texttt{healthAPI} has an associated \texttt{HealthFunc} function that is used to perform the health check. If this function returns an error, \texttt{health\linebreak{}API} will also return an HTTP error. Figure~\ref{fig:nme-http-api-health-uml-class} illustrates these types in a UML class diagram.

\subsection{Event API}
\label{subsec:event-api}

\begin{figure}[h]
  \centering
  \includegraphics[width=\columnwidth]{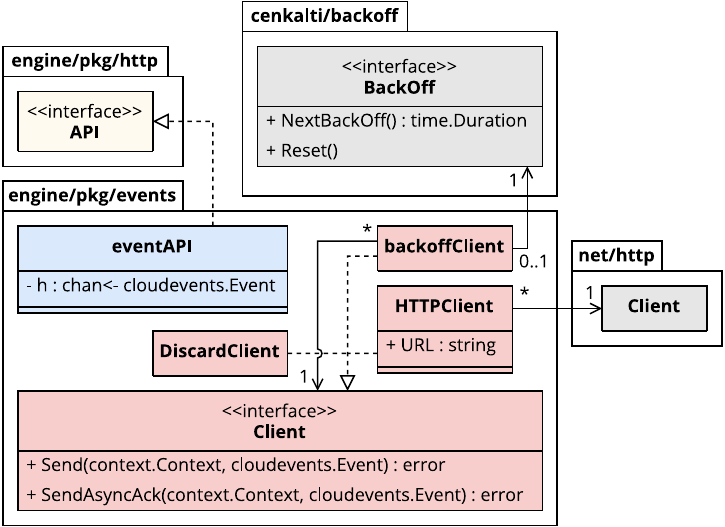}
  \caption{UML class diagram of the types in the \texttt{.../events} package.}
  \label{fig:nme-events-uml-class}
  \Description{UML class diagram of the types in the \texttt{.../events} package}
\end{figure}

NBMP standardizes a \texttt{report} descriptor that defines a URL to which events, e.g.~that occurred during the task execution, should be reported. It does not further specify event types or formats and only describes HTTP \texttt{POST} requests as the delivery mechanism. We implemented general server and client types for event reporting as depicted in Figure~\ref{fig:nme-events-uml-class}.

We use CloudEvents as the event format~\cite{cncf_cloudeventsversion102_2022}. It standardizes the event structure with common fields and simultaneously carries custom payloads. We use the official Go library \texttt{github.com/\linebreak{}cloudevents/sdk-go/v2}\footnote{\url{https://github.com/cloudevents/sdk-go}} for encoding to and decoding from JSON objects. CloudEvents defines three request formats for HTTP. In ``Binary Content Mode'', HTTP headers are used for transferring the common event fields while the request body contains the custom payload. In ``Structured Content Mode'', both the common event fields and the custom payload are contained in the request body represented as one JSON object. Finally, the ``Batched Content Mode'' extends ``Structured Content Mode'' and transfers multiple events in a single request. In this case, the request body contains a JSON array with multiple objects. In \texttt{nagare media engine}, we currently implement the last two modes.

\texttt{eventAPI} provides the server side as HTTP API. It accepts \texttt{POST} requests sent to the \texttt{/events} path. Decoded events are pushed to the associated channel \texttt{h} from where they can be received and handled by another Goroutine. Note that sending events to this channel is performed within the request context. The use of an unbuffered channel will therefore block \texttt{eventAPI} from answering until the receiver reads from the channel. This might be desired as the client can be sure the event was received. In other cases, the client might not want to be blocked. Buffered channels can alleviate this problem as long as the buffer is large enough. Alternatively, clients can set the \texttt{async} query parameter to \texttt{true}. Here, \texttt{eventAPI} will respond immediately with a ``202~Accepted'' status code after decoding the request and the events continue to be sent to the channel outside the request context.

With the \texttt{Send} and \texttt{SendAsyncAck} methods, these two delivery modes are also represented in the \texttt{Client} interface. We provide three implementations. \texttt{DiscardClient} is a dummy client useful for testing purposes. It will simply discard any event. \texttt{HTTPClient}, on the other hand, sends events to the configured URL using the Go HTTP client library. In practice, this might be subject to temporary network failures. We therefore provide a \texttt{backoffClient} that wraps another client and performs retries with exponential backoff if the wrapped client returns an error.

\subsection{NBMP}
\label{subsec:nbmp}

\begin{figure*}[t]
  \centering
  \includegraphics[width=\textwidth]{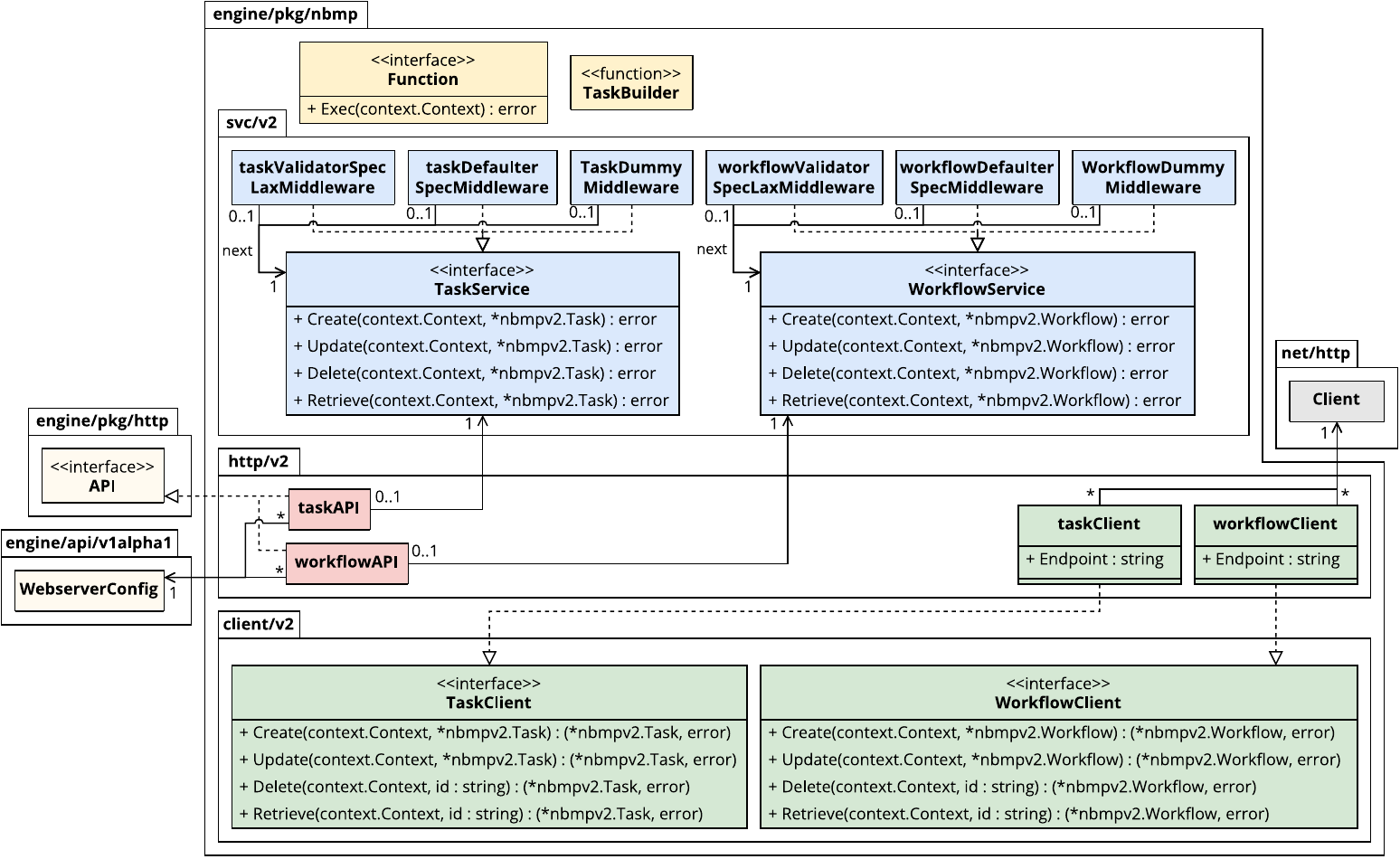}
  \caption{UML class diagram of the types in the \texttt{.../nbmp} package.}
  \label{fig:nme-nbmp-uml-class}
  \Description{UML class diagram of the types in the \texttt{.../nbmp} package}
\end{figure*}

The last common package discussed in this section contains NBMP-related types. Other packages use and extend these types for a working NBMP implementation. Figure~\ref{fig:nme-nbmp-uml-class} gives an overview in a UML class diagram.

The \texttt{Function} interface describes a multimedia function with an \texttt{Exec} method to start the execution. All multimedia functions we include in \texttt{nagare media engine} implement this interface. Additionally, each implementation is accompanied by a \texttt{TaskBuilder} function that takes a TDD as an argument and returns an initialized \texttt{Function} that is ready to be executed as a task.

Next, we implemented server and client types for the NBMP workflow and task APIs. The \texttt{WorkflowClient} and \texttt{TaskClient} interfaces describe the CRUD operations specified in the NBMP standard. \texttt{workflowClient} and \texttt{taskClient} implement compliant clients that send requests to the configured NBMP \texttt{Endpoint} using the Go HTTP client. Note that we currently do not provide a client that generally retries failed requests with exponential backoff and leave that as a choice for the client user.

The \texttt{workflowAPI} and \texttt{taskAPI} types implement the NBMP workflow and task API, respectively. We extract all the HTTP request and response handling into these types and leave the business logic to service types. As such, \texttt{workflowAPI} and \texttt{taskAPI} have an associated \texttt{WorkflowService} and \texttt{TaskService} implementation. Developers can thus concentrate on the core business logic. Moreover, we implemented service middleware types that already handle general aspects and then pass the call to the next service type. \texttt{workflowValidatorSpecLaxMiddleware} and \texttt{taskValidator\linebreak{}SpecLaxMiddleware} perform validation checks that are prescribed by the NBMP specification. However, they are not overly strict and forgive some deviations. Similarly, \texttt{workflowDefaulterSpec\linebreak{}Middleware} and \texttt{taskDefaulterSpecMiddleware} set default values defined in the standard. Finally, the \texttt{WorkflowDummyMiddleware} and \texttt{TaskDummyMiddleware} are dummy implementations that simply forward the call to the next service type. Developers can extend these types for custom middlewares and only overwrite the operations they are interested in.

\section{NBMP Gateway}
\label{sec:nbmp-gateway}

In this section we outline the NBMP Gateway, the first \texttt{nagare media engine} component. Its main purpose is to translate between the NBMP and the internal \texttt{nagare media engine} data model. Additionally, it provides the NBMP workflow API for NBMP sources.

\begin{figure}[b]
  \centering
  \includegraphics[width=\columnwidth]{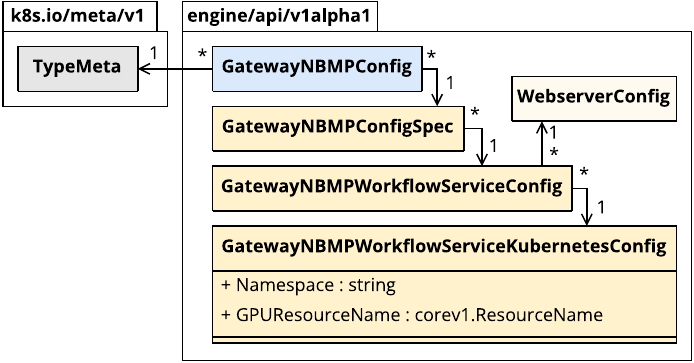}
  \caption{UML class diagram of \texttt{GatewayNBMPConfig} related types.}
  \label{fig:nme-api-config-gateway-nbmp-uml-class}
  \Description{UML class diagram of \texttt{GatewayNBMPConfig} related types}
\end{figure}
Because the NBMP Gateway is a control plane component, it runs as a Kubernetes \texttt{Deployment} in the management cluster. Administrators need to provide a configuration file that governs its behavior. All configuration files in \texttt{nagare media engine} are represented as YAML data that follows a structure similar to Kubernetes resources. We adopted the \texttt{apiVersion} and \texttt{kind} fields. They have the same purpose and identify the type of configuration represented in a YAML file. Unlike in Kubernetes resources, configuration files do not need a \texttt{metadata} field. Figure~\ref{fig:nme-api-config-gateway-nbmp-uml-class} illustrates the configuration types for the NBMP Gateway.

We use the \texttt{WebserverConfig} type to allow administrators to configure the HTTP server for the NBMP workflow API (see Section~\ref{subsec:http-server}). Next, the \texttt{Namespace} field defines in which Kubernetes namespace the NBMP Gateway should operate in. Resources are created, updated, deleted in and read from this namespace. If administrators want to use multiple namespaces, they would need to run the NBMP Gateway multiple times. Future work could improve support for multiple namespaces, e.g.~by mapping from virtual hosts to namespaces. At the same time, the NBMP Gateway is a lightweight application that only consumes a few computing resources. We therefore regard running multiple instances as a reasonable approach. The last field, \texttt{GPUResourceName}, determines the GPU resource name that should be used in Kubernetes. Both the NBMP and the Kuberntes data model allow defining limits on computing resource usage such as disk space, memory, CPU cores and GPU cores. While NBMP only allows to generally limit the number of GPU cores, Kubernetes differentiates between GPU resources from different vendors. For instance, the \texttt{amd.com/gpu} and \texttt{nvidia.com/gpu} resource names limit the usage of GPUs from the vendors AMD and NVIDIA, respectively. The \texttt{GPUResourceName} therefore instructs the NBMP Gateway which resource name it should use.

\begin{figure}[H]
  \centering
  \includegraphics[width=0.75\columnwidth]{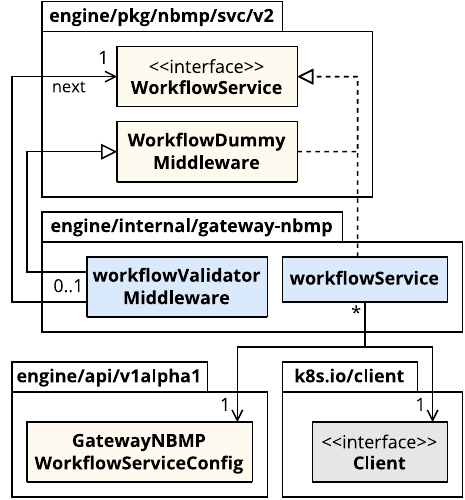}
  \caption{UML class diagram of the types in the \texttt{.../gateway-nbmp} package.}
  \label{fig:nme-cmd-gateway-nbmp-uml-class}
  \Description{UML class diagram of the types in the \texttt{.../gateway-nbmp} package}
\end{figure}

\begin{figure*}[b]
  \centering
  \includegraphics[width=0.8\textwidth]{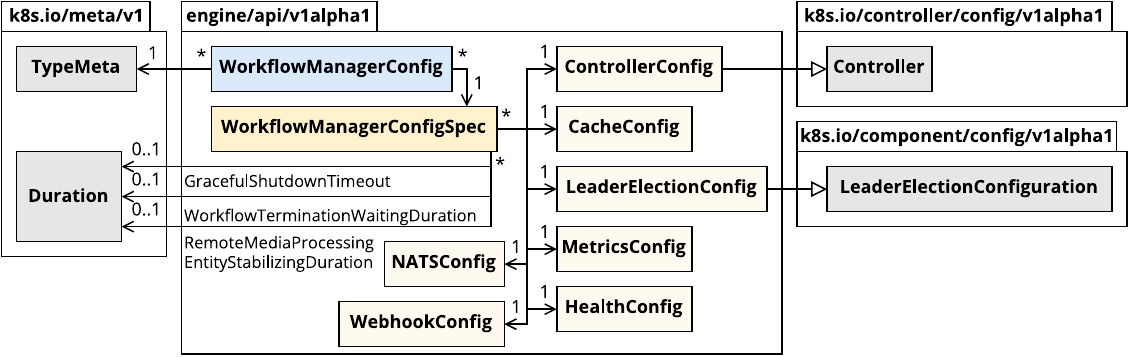}
  \caption{UML class diagram of \texttt{WorkflowManagerConfig} related types.}
  \label{fig:nme-api-config-workflowmanager-uml-class}
  \Description{UML class diagram of \texttt{WorkflowManagerConfig} related types}
\end{figure*}

After successfully decoding the configuration, an HTTP server (see Section~\ref{subsec:http-server}) is constructed. We mount the health and NBMP workflow APIs (see Sections~\ref{subsec:health-api} and~\ref{subsec:nbmp}) to the HTTP server. Moreover, the telemetry middleware (see Section~\ref{subsec:telemetry-middleware}) is added for NBMP workflow API requests. We use the common workflow service middlewares \texttt{workflowDefaulterSpecMiddleware} and \texttt{workflow\linebreak{}ValidatorSpecLaxMiddleware} (see Section~\ref{subsec:nbmp}). Further NBMP Gateway types are depicted in Figure~\ref{fig:nme-cmd-gateway-nbmp-uml-class}. \texttt{workflowValidator\linebreak{}Middleware} is an additional workflow service middleware that performs validations specific to \texttt{nagare media engine}. For instance, it will reject requests that use unsupported NBMP descriptors and properties. Finally, the \texttt{workflowService} type implements our business logic. Using the model converters (see Section~\ref{subsec:data-transformations}), a given WDD is transformed into \texttt{Workflow} and \texttt{Task} resources for create and update requests. The reverse transformation is applied when formulating the response. With the help of the Kubernetes client library, the NBMP CRUD operations lead to the equivalent operations on \texttt{Workflow} and \texttt{Task} resources in the configured Kubernetes namespace. The Kubernetes client is configured to use caching for read operations. This reduces the burden on the Kubernetes API server and simultaneously accelerates the response times of the NBMP Gateway.

We currently do not implement any special security mechanisms. The NBMP standard recommends the use of state-of-the-art approaches for data transport, authentication and authorization. Note that in cloud environments these functions are often implemented in common networking layers such as proxies, load balancers or single sign-on systems. Still, future work could add support directly in the NBMP Gateway.

\section{Workflow Manager}
\label{sec:workflow-manager}

The Workflow Manager is the central control plane application and implements the NBMP component with the same name. It bundles multiple concurrently running controllers that are outlined in the following sections. First, Section~\ref{subsec:workflow-manager-overview} gives a general overview. Next, Section~\ref{subsec:webhooks} discusses the bundled Kubernetes API webhooks. Finally, Sections~\ref{subsec:media-processing-entity-controller},~\ref{subsec:workflow-controller},~\ref{subsec:task-controller} and~\ref{subsec:job-controller} detail the controllers that reconcile \texttt{MediaProcessingEntity}, \texttt{Workflow}, \texttt{Task} and \texttt{Job} resources, respectively.

\subsection{Overview}
\label{subsec:workflow-manager-overview}

\texttt{nagare media engine} is a Kubernetes-native system. As such, it follows the Kubernetes orchestration model where the desired state is first defined through a Kubernetes resource and then materialized by an accompanying controller that implements a reconciliation loop. Systems such as \texttt{nagare media engine} that define more than one custom resource often bundle multiple controllers into a single application. Kubernetes itself bundles the controllers for built-in resources in the core component \texttt{kube-controller-manager}. This makes it easier for administrators because they just have to deploy a single application. However, splitting the system up gives more fine-grain control of what permissions are assigned to which application, resulting in a potential security advantage (principle of least privilege). We still decided to bundle all controllers in a single application. Hence, the Workflow Manager configuration file includes options for all controllers as depicted in Figure~\ref{fig:nme-api-config-workflowmanager-uml-class}.

Most of the types are derived from configuration for Kubebuilder types. \texttt{ControllerConfig} defines common Kubernetes controller settings. \texttt{CacheConfig} controls how the Kubernetes client builds an in-memory cache of resources. \texttt{MetricsConfig}, \texttt{HealthConfig} and \texttt{WebhookConfig} define settings for metric, health and webhook HTTP endpoints, respectively. For higher availability, Kubernetes controllers can run in a replicated manner. Nevertheless, at any given time, typically at most one replica is actively reconciling resources and thus considered the ``leader''. If the leader becomes unavailable, e.g.~the container or node terminates unexpectedly, one of the other replicas becomes the new leader and swiftly continues the reconciliation. Kubernetes facilitates this process by offering a distributed leader election system. The \texttt{LeaderElectionConfig} type controls if and how leader election should be used. Next, we also define configuration options specific to \texttt{nagare media engine}. The \texttt{NATSConfig} sets connection information to the NATS server. These are also passed on to the Workflow Manager Helper for persisting events sent to the report API (see Section~\ref{sec:workflow-manager-helper}). Most configuration options have default values, but a few must be provided by the administrator. The Workflow Manager validates the configuration file and then hands down the necessary settings to each controller. These are started in separate Goroutines and managed by the Kubebuilder framework.

In order to construct a Kubernetes controller, developers need to implement a reconciler type that is passed to a \texttt{Manager} type provided by the framework. Most controllers in \texttt{nagare media engine} follow a similar structure as shown in a generalized form in Listing~\ref{lst:general-reconciler}. The \texttt{Manager} implements the reconciliation loop with the help of a queue of resources that should be reconciled. By default, it calls the reconciler sequentially for each element in that queue, but users can instead also configure a maximum number of concurrently running reconciliations. When calling the reconciler, the \texttt{Manager} provides a Go \texttt{Context} and a reconciliation request that contains the namespace and name of the resource as arguments (line~14). The reconciler then uses the Kubernetes client to fetch the resource and start the reconciliation. This can result in various outcomes. If the reconciler returns an error, the \texttt{Manager} will requeue the resource for a retry while using exponential backoff for repeated errors. Alternatively, the reconciler can return a result with a custom requeue duration after which a reconciliation is tried again. This is useful for scenarios where a started process takes time to complete and the reconciliation loop should not be blocked by actively waiting. Lastly, the reconciler can return a successful result indicating the resource can be removed from the queue.
\begin{listing}[t]
\begin{minted}{go}
type ResourceReconciler struct {
  Client    client.Client
  APIReader client.Reader
}

func (r *ResourceReconciler) SetupWithManager(mgr ctrl.Manager) error {
  return ctrl.NewControllerManagedBy(mgr).
    Named("resource-controller").
    For(&v1.Resource{}).
    Owns(&v1.RelatedResource{}).
    Complete(r)
}

func (r *ResourceReconciler) Reconcile(ctx context.Context, req ctrl.Request) (_ ctrl.Result, reterr error) {
  // fetch resource
  resource := &v1.Resource{}
  err := r.Client.Get(ctx, req.NamespacedName, resource)
  if err != nil {
    if apierrors.IsNotFound(err) {
      return ctrl.Result{}, nil
    }
    return ctrl.Result{}, err
  }

  // patch resource when returning
  oldResource := resource.DeepCopy()
  defer func() {
    r.reconcileCondition(ctx, reterr, resource)
    _, err := utils.FullPatch(
      ctx, r.Client, resource, oldResource)
    apiErrs := kerrors.FilterOut(
      err, apierrors.IsNotFound)
    reterr = kerrors.Flatten(
      kerrors.NewAggregate([]error{apiErrs, reterr}))
  }()

  // init: add finalizer
  changed := controllerutil.AddFinalizer(
    resource, "engine.nagare.media/protection")
  if changed {
    return ctrl.Result{}, nil
  }

  // delete: reconcile deleted resource
  if !resource.DeletionTimestamp.IsZero() {
    return r.reconcileDelete(ctx, resource)
  }

  // normal: reconcile
  return r.reconcile(ctx, resource)
}
\end{minted}
  \caption{Generalized implementation of a reconciler type in \texttt{nagare media engine}.}
  \label{lst:general-reconciler}
  \Description{Generalized implementation of a reconciler type in \texttt{nagare media engine}.}
\end{listing}

Resources can be enqueued because of various events. By convention, a reconciler has a \texttt{SetupWithManager} method where it registers itself to the passed \texttt{Manager} and configures at least one event source (lines~6--12). Each controller reconciles one primary resource type for which the \texttt{Manager} typically defines a watch in order to enqueue changed resources, i.e.~upon create, update and delete events (line~9). Note that changing the resource during the reconciliation, e.g.~fields in its \texttt{status}, will result in new update events and thus requeue the resource again. Developers can furthermore easily add a watch for changes of related resource types. In this way, if a resource is changed that has an owner reference to the primary resource type, the owner can be enqueued (line~10). Alternatively, if owner references are not used, a mapping function can be implemented that maps from the watched related resource type to the primary type. Moreover, Kubebuilder allows enqueuing resources through a Go channel. Developers are thus able to react to any external event and trigger a reconciliation by sending a \texttt{GenericEvent} instance to the Go channel. Within its payload, a \texttt{GenericEvent} carries the namespace and name of the resource to reconcile. Before an event leads to an enqueued resource, developers can subject events to a list of filters. For instance, Kubebuilder provides filters to only consider events that affected the \texttt{spec} field.

By convention, the reconciler is initialized with a \texttt{Reader} and a \texttt{Client} instance that both provide access to the Kubernetes API (lines~2--3). Generally, the \texttt{Client} instance should be preferred. It provides read and write access in combination with an in-memory cache. The \texttt{Reader}, on the other hand, is read-only and directly talks to the Kubernetes API. It can be useful in situations where the \texttt{Client} could potentially return stale data. There are few places in \texttt{nagare media engine} where we used the \texttt{Reader}.

In \texttt{nagare media engine}, the \texttt{Reconcile} method then generally consists of the following steps. First, the \texttt{Client} is used to fetch the resource that should be reconciled (lines~16--23). If it cannot be found, i.e.~it was deleted, the reconciliation is finished (lines~19--21). Other errors are returned and result in a later retry (line~22).

On success, an in-memory copy of the fetched resource is created (line~26). For that, Kubebuilder's \texttt{controller-gen} tool automatically generates \texttt{DeepCopy} methods that return fully independent copies, including pointer fields or nested structures. We then register a function execution with the \texttt{defer} keyword. In Go, \texttt{defer} is used to execute a function right before returning from the current function. It is often used to clean up state, e.g.~closing a file. In this case, the function is used to persist the outcome of the reconciliation. First, the \texttt{reconcileCondition} helper method updates the \texttt{condition} field in the resource's \texttt{status} based on the returned error \texttt{reterr} (line~28). Next, the \texttt{FullPatch} helper function is called to compare the resource before and after the reconciliation (lines~29--30). At this point, any changes are in-memory only. If changes are detected, \texttt{FullPatch} thus sends the appropriate \texttt{PATCH} requests to the Kubernetes API. Lastly, any errors from the reconciliation and the API requests are combined to a final error for the \texttt{Reconcile} method (lines~31--34).

After registering the deferred function execution, the resource is ``initialized'' by adding a finalizer (lines~38--42). In Kubernetes, each resource has a \texttt{finalizers} field, a list of strings, in its \texttt{metadata}. If this list is not empty, the Kubernetes API server delays delete requests for that resource until the list is empty. Controllers that need to perform cleanup logic before resources are deleted therefore should add a finalizer to the list, i.e.~a string marker for the controller. After the cleanup logic, the controller removes the finalizer, thus allowing the Kubernetes API server to finish the removal. Note that multiple controllers can register finalizers. In this case, the Kubernetes API server will wait until all controllers finish the cleanup. Generally, the cleanup logic can omit deleting related resources if owner references are used. The removal of the owner resource automatically triggers a cascading removal of owned resources. We use the \texttt{AddFinalizer} helper function to add the finalizer. It returns whether the finalizer was already present. If it was newly added, we finish the reconciliation in order to immediately persist the change. This, in turn, results in a new update event and thus a new reconciliation.

In the final step, the reconciler checks whether the resource is marked for deletion (line~45). If that is the case, the cleanup logic is executed (line~46). Otherwise, the normal reconciliation specific to that resource is performed (line~50).

\subsection{Webhooks}
\label{subsec:webhooks}

\begin{figure*}[t]
  \centering
  \includegraphics[width=0.69\textwidth]{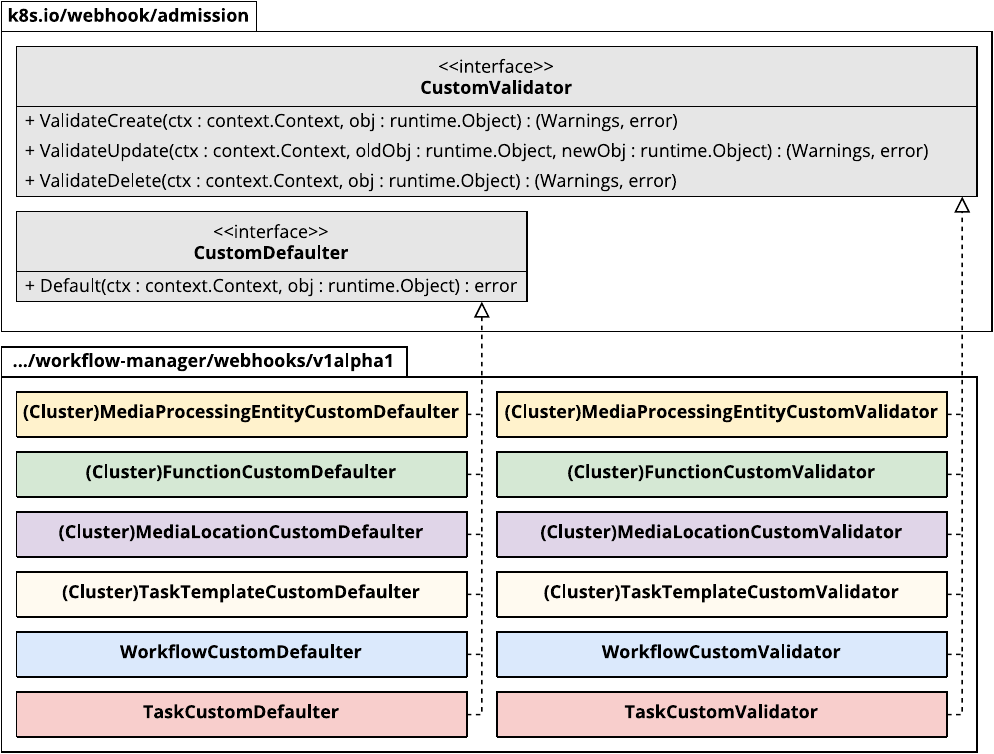}
  \caption{UML class diagram of the types in the \texttt{.../workflow-manager/webhooks/v1alpha1} package.}
  \label{fig:nme-cmd-workflow-manager-webhooks-uml-class}
  \Description{UML class diagram of the types in the \texttt{.../workflow-manager/webhooks/v1alpha1} package}
\end{figure*}

As already mentioned, the Kubernetes API server automatically adds new routes to its REST API for the typical operations, i.e.~CRUD and, depending on the configuration, some Kubernetes-specific operations when custom resources are registered through CRDs. Handling these operations primarily results in updates to the Kubernetes persistent store. The main purpose of the Kubernetes API server is then to inform interested controllers that defined a watch for these changes. Note that the Kubernetes API server first persists changes and only then informs interested controllers. In this scenario, validity checks can therefore only be performed relatively late. Ideally, users of the Kubernetes API are informed directly in the response about any validation errors. Besides general checks performed on all resources, a CRD can already describe basic constraints that are then maintained by the Kubernetes API server. However, more complex logic needs to be extracted into an admission controller that is registered as a webhook.

An admission controller is executed within the Kubernetes API request context and influences if and how the operation is executed. There are various built-in admission controllers administrators can enable. Custom admission controllers need to be implemented as webhooks, i.e.~the Kubernetes API server sends an HTTP request with details about the operation that is about to be performed to the admission controller. After an evaluation, the admission controller is then expected to answer with a standardized response that controls the further execution of the operation. Note that administrators can register multiple admission controllers that then run in sequence. Because all these interactions happen within the request context, custom admission controllers should run close to the Kubernetes API server and answer in a timely manner.

Kubernetes differentiates between mutating and validating webhooks. The former are allowed to alter the resource originally submitted by the API user, e.g.~in order to set default values. The latter only validate the resource and inform Kubernetes about any warnings and errors that are then passed on to the API user. Administrators use the built-in resources \texttt{MutatingWebhookConfiguration} and \texttt{ValidatingWebhookConfiguration} to register admission controllers for specific resources and operations.

The Kubebuilder framework simplifies the implementation of admission controllers that validate custom resources (\texttt{Custom\linebreak{}Validator} interface) or set default values (\texttt{CustomDefaulter} interface). Moreover, the \texttt{controller-gen} tool automatically generates the necessary Kubernetes resources for registering the admission controller webhooks. As can be seen in Figure~\ref{fig:nme-cmd-workflow-manager-webhooks-uml-class}, we define admission controllers for all our custom resources. The cluster-scoped resource variants generally reuse the logic of the namespace-scoped variants. Currently, we only implement basic checks. Future work should extend this logic to a more comprehensive validation.

Lastly, it should be mentioned that the Kubernetes API server also uses webhooks for converting between different resource versions. For instance, a CRD could define \texttt{v2} as the primary version and still support version \texttt{v1}. If users submit \texttt{v1} resources, the Kubernetes API server will first convert it to the primary version. For this purpose, a conversion webhook can be configured in the CRD. It receives the old version and responds with the new one. This simplifies the implementation of all further controllers as they only need to handle the primary version. In this way, systems are able to evolve the resource structure over time. As research prototype, \texttt{nagare media engine} currently only defines \texttt{v1alpha1} as the single version and therefore does not provide conversion~webhooks.

\subsection{Media Processing Entity Controller}
\label{subsec:media-processing-entity-controller}

\begin{figure*}[b]
  \centering
  \includegraphics[width=0.85\textwidth]{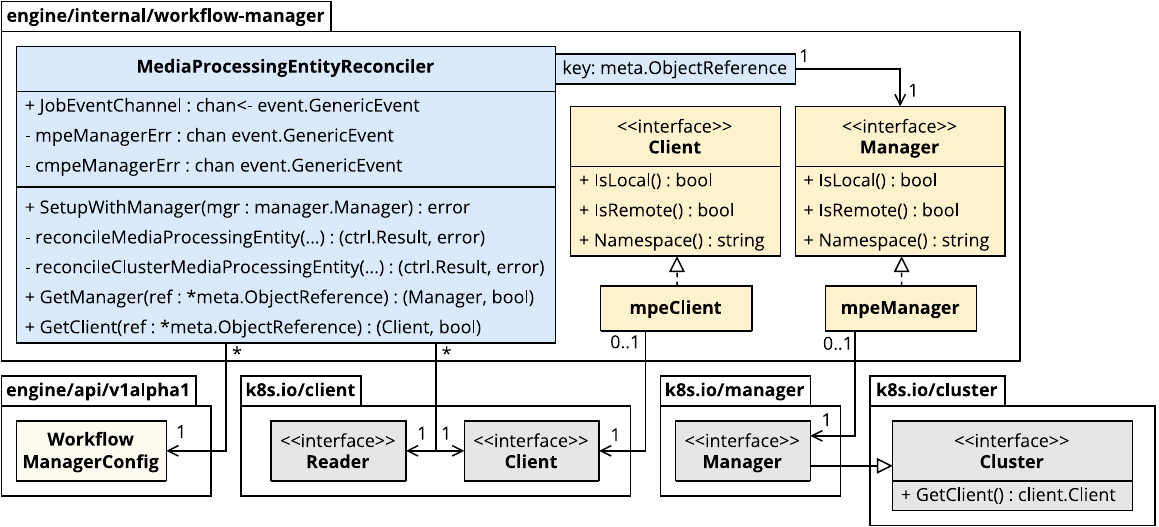}
  \caption{UML class diagram of \texttt{MediaProcessingEntityReconciler} related types.}
  \label{fig:nme-cmd-workflow-manager-media-processing-entity-reconciler-uml-class}
  \Description{UML class diagram of \texttt{MediaProcessingEntityReconciler} related types}
\end{figure*}

The first of four Kubernetes controllers is the MPE Controller. Its purpose is to reconcile \texttt{(Cluster)MediaProcessingEntity} resources. Figure~\ref{fig:nme-cmd-workflow-manager-media-processing-entity-reconciler-uml-class} gives an overview of the involved types.

As the reconciler, we implemented the \texttt{MediaProcessingEntity\linebreak{}Reconciler} type. The \texttt{reconcileMediaProcessingEntity} and \texttt{reconcileClusterMediaProcessingEntity} methods implement the reconciliation logic for \texttt{MediaProcessingEntity} and \texttt{Cluster\linebreak{}MediaProcessingEntity} resources, respectively. Consequently, two reconciliation loops are constructed with the help of a \texttt{Manager} instance. However, because the \texttt{spec} field has the same structure, the reconciliation logic is shared. Moreover, the \texttt{mpeManagerErr} and \texttt{cmpeManagerErr} channels are added as sources for \texttt{Generic\linebreak{}Event}s as will be explained further below. We did not add any filters.

The reconciliation follows the general structure laid out in Section~\ref{subsec:workflow-manager-overview}. In a normal reconciliation run, we try to establish and then maintain a connection to the MPE, i.e.~a Kubernetes cluster. If the reconciler encounters this MPE for the first time, a new \texttt{Manager} instance together with its own client and in-memory cache is created for that cluster. This is straightforward for a local MPE. However, for remote MPEs the referenced \texttt{Secret} with the kubeconfig file has to be read first. We create a \texttt{Manager} instance instead of a simple \texttt{Client} because we register a Job Controller for each MPE (see Section~\ref{subsec:job-controller}). Next, the \texttt{Manager} is started in its own Goroutine after which the reconciler waits for the \texttt{Manager} to ``stabilize''. We want to make sure that the \texttt{Manager} does not directly terminate with an error. For local MPEs, this is effectively skipped by setting the waiting duration to zero, but for remote MPEs, the reconciler waits by default for five seconds before declaring the reconciliation a success. Administrators can configure this duration with the \texttt{RemoteMediaProcessingEntityStabilizingDuration} option (cf.~Figure~\ref{fig:nme-api-config-workflowmanager-uml-class}). If the \texttt{Manager} terminates at a point after the reconciliation, the \texttt{Manager}'s Goroutine sends a \texttt{GenericEvent} to the \texttt{mpeManagerErr} or \texttt{cmpeManagerErr} channel to trigger a new reconciliation. The reconciler maintains an in-memory map of MPE references to \texttt{Manager} instances. In this way, other controllers can use the \texttt{GetManager} and \texttt{GetClient} methods to retrieve the \texttt{Manager} and \texttt{Client} instance for a particular MPE, respectively. When a \texttt{(Cluster)MediaProcessingEntity} resource is deleted, the \texttt{Manager} is stopped and removed from the in-memory map. Finally, the \texttt{JobEventChannel} is passed to the Job Controller and allows the Job Controller to trigger reconciliations of \texttt{Task} resources (see Sections~\ref{subsec:task-controller} and~\ref{subsec:job-controller}).

\subsection{Workflow Controller}
\label{subsec:workflow-controller}

\begin{figure}[b]
  \centering
  \includegraphics[width=0.9\columnwidth]{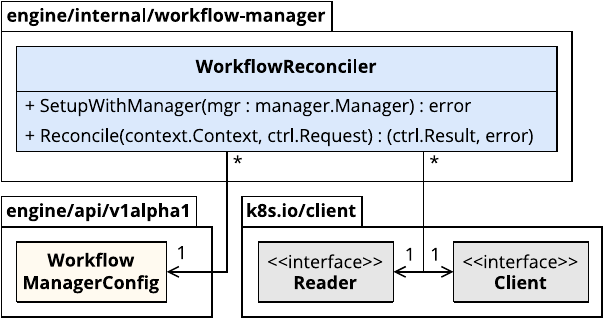}
  \caption{UML class diagram of \texttt{WorkflowReconciler} related types.}
  \label{fig:nme-cmd-workflow-manager-workflow-reconciler-uml-class}
  \Description{UML class diagram of \texttt{WorkflowReconciler} related types}
\end{figure}
The Workflow Controller reconciles \texttt{Workflow} resources. It again follows the general controller structure presented in Section~\ref{subsec:workflow-manager-overview} but additionally splits the reconciliation into multiple phases based on the workflow lifecycle. Figure~\ref{fig:nme-cmd-workflow-manager-workflow-reconciler-uml-class} depicts the \texttt{WorkflowReconciler} and related types. In addition to watching for \texttt{Workflow} resources, the \texttt{Manager} is configured to watch for \texttt{Task} resources with an owner reference to a \texttt{Workflow}. A reconciliation for the associated \texttt{Workflow} will therefore be triggered if a \texttt{Task} is changed.

\begin{figure}[h]
  \centering
  \includegraphics[width=0.7\columnwidth]{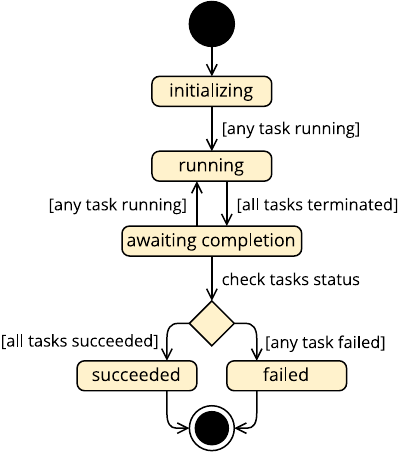}
  \caption{UML state diagram of the logical state changes of a workflow.}
  \label{fig:nme-cmd-workflow-manager-workflow-reconciler-uml-state-logically}
  \Description{UML state diagram of the logical state changes of a workflow}
\end{figure}

Logically, the workflow lifecycle in \texttt{nagare media engine} follows the state diagram shown in Figure~\ref{fig:nme-cmd-workflow-manager-workflow-reconciler-uml-state-logically}. This is an adaptation of the NBMP workflow lifecycle model (see Section~\ref{subsec:network-based-media-processing}).

It starts out in the ``initializing'' state and remains there until the first associated task starts executing. The workflow is then considered in the ``running'' state. If all tasks that belong to the workflow have terminated, the workflow moves to the ``awaiting completion'' state. After a configured duration (cf.~the \texttt{WorkflowTermination\linebreak{}WaitingDuration} option in Figure~\ref{fig:nme-api-config-workflowmanager-uml-class}), the workflow will either transition to the ``succeeded'' or ``failed'' state depending on the task execution. Both states are final, i.e.~a workflow in these states cannot start running again. We introduced the ``awaiting completion'' state to mitigate race conditions where tasks belonging to a workflow are still being created as others have already terminated. By default, the workflow will remain in that state for 20~seconds. In case new active tasks are detected, it will transition to the ``running'' state again.

\begin{figure}[t]
  \centering
  \includegraphics[width=0.75\columnwidth]{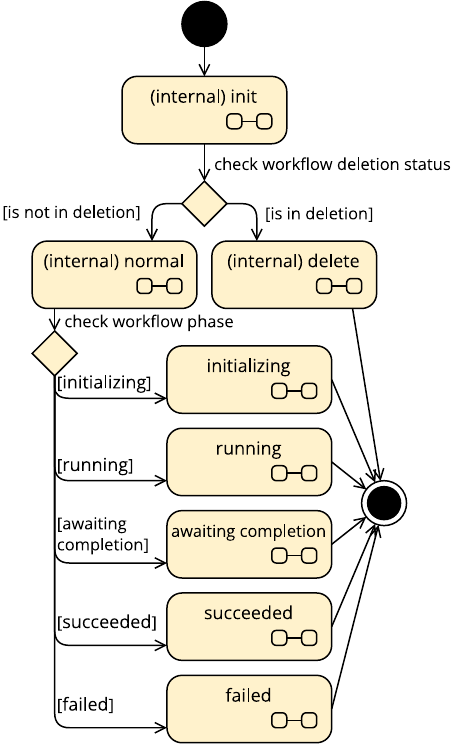}
  \caption{UML state diagram of the phase changes within one workflow reconciliation loop iteration.}
  \label{fig:nme-cmd-workflow-manager-workflow-reconciler-uml-state-reconciliation-loop}
  \Description{UML state diagram of the phase changes within one workflow reconciliation loop iteration}
\end{figure}

The current state of a \texttt{Workflow} resource is communicated via the \texttt{Phase} field in its \texttt{status} (see Section~\ref{subsubsec:workflow}). Since we only persist in-memory changes at the end of a reconciliation, we stop further reconciliations after changing the phase. Users will thus get an up-to-date view and the implementation can focus on one phase at a time. In practice, the phase changes during a single reconciliation therefore follow the state diagram in Figure~\ref{fig:nme-cmd-workflow-manager-workflow-reconciler-uml-state-reconciliation-loop}. ``init'', ``normal'' and ``delete'' are internal phases that are not reported to the user and correspond to the steps of the generalized \texttt{nagare media engine} controller. In the following, we will discuss the actions performed during each phase and visualize them in UML activity diagrams.

\begin{figure}[h]
  \centering
  \includegraphics[width=\columnwidth]{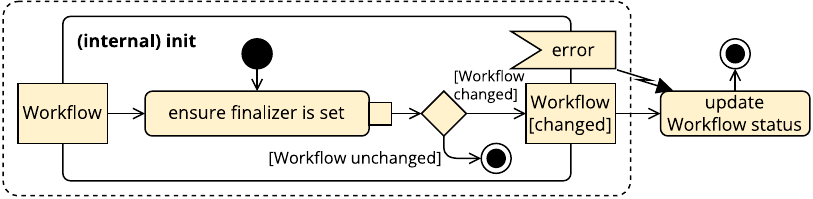}
  \caption{UML activity diagram of the ``(internal) init'' workflow phase.}
  \label{fig:nme-cmd-workflow-manager-workflow-reconciler-uml-activity-1-internal-init}
  \Description{UML activity diagram of the ``(internal) init'' workflow phase}
\end{figure}

In the ``(internal) init'' phase, a finalizer is added to the \texttt{Workflow} in order to trigger the cleanup logic upon deletion. This phase is illustrated in Figure~\ref{fig:nme-cmd-workflow-manager-workflow-reconciler-uml-activity-1-internal-init}. Note that at the end of each phase, the \texttt{Workflow} \texttt{status} field gets updated, which results in another reconciliation. Additionally, the \texttt{Workflow} is updated if an error occurs leading to a retry with exponential backoff. The actions in the ``(internal) init'' phase are executed during each reconciliation. Therefore, in case the finalizer is somehow lost, e.g.~a user removes it manually, the reconciler will automatically repair the error. We generally implement actions performed by reconcilers to be idempotent, i.e.~they can be executed multiple times with the same outcome. If no changes are required, the reconciler either moves to the ``(internal) normal'' or the ``(internal) delete'' phase.
\pagebreak

\begin{figure}[h]
  \centering
  \includegraphics[width=\columnwidth]{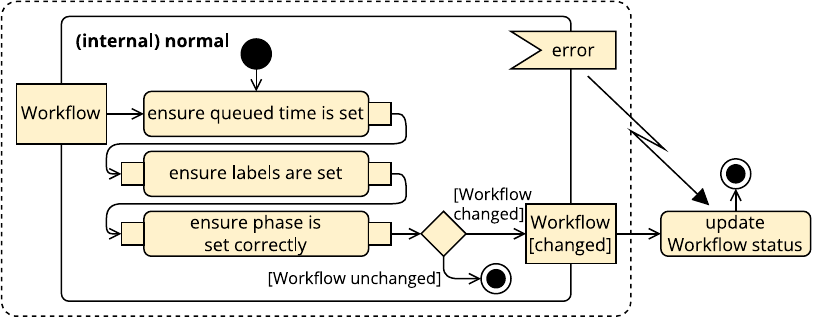}
  \caption{UML activity diagram of the ``(internal) normal'' workflow phase.}
  \label{fig:nme-cmd-workflow-manager-workflow-reconciler-uml-activity-2-internal-normal}
  \Description{UML activity diagram of the ``(internal) normal'' workflow phase}
\end{figure}

The ``(internal) normal'' phase is executed during each normal reconciliation. Figure~\ref{fig:nme-cmd-workflow-manager-workflow-reconciler-uml-activity-2-internal-normal} depicts the actions involved. First, the reconciler ensures that the \texttt{QueuedTime} field is set. It indicates to users when the Workflow Controller first started reconciling this \texttt{Workflow} resource. If this field is undefined, it is set to the current time. Next, common labels are added to the \texttt{Workflow}. This allows users and controllers to quickly filter \texttt{Workflows} by these labels. Finally, the reconciler checks whether the \texttt{Phase} field is set correctly and corresponds to the observed state. If not, it is corrected. If the \texttt{Phase} field is undefined, e.g.~it is a newly created \texttt{Workflow} resource, the \texttt{Phase} field is set to ``initializing''. The reconciler then either persists the \texttt{Workflow} in case there are changes or moves to the phase indicated by the \texttt{Phase} field.

\begin{figure}[h]
  \centering
  \includegraphics[width=\columnwidth]{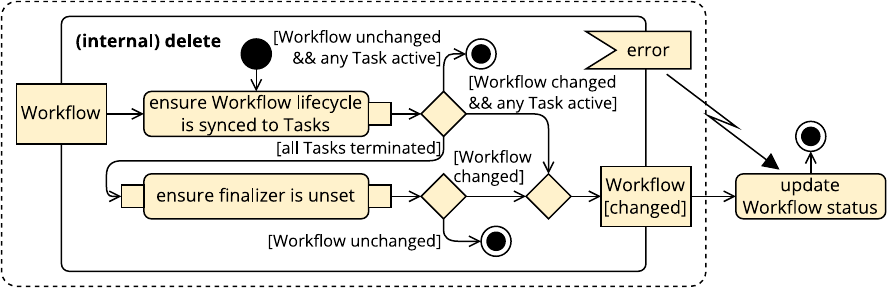}
  \caption{UML activity diagram of the ``(internal) delete'' workflow phase.}
  \label{fig:nme-cmd-workflow-manager-workflow-reconciler-uml-activity-3-internal-delete}
  \Description{UML activity diagram of the ``(internal) delete'' workflow phase}
\end{figure}

The ``(internal) delete'' phase implements the cleanup logic for \texttt{Workflow} resources. It is illustrated in Figure~\ref{fig:nme-cmd-workflow-manager-workflow-reconciler-uml-activity-3-internal-delete}. Before a \texttt{Workflow} resource can be removed by the Kubernetes API server, \texttt{nagare media engine} makes sure that all tasks belonging to that workflow have terminated. In the first step, this phase therefore checks the state of all associated tasks. Fetching \texttt{Task} resources is done efficiently with the help of labels that were added by the Task Controller (see Section~\ref{subsec:task-controller}). Depending on the \texttt{Task}'s \texttt{Phase} field, the \texttt{Total}, \texttt{Active}, \texttt{Succeeded} and \texttt{Failed} counters of the \texttt{Workflow} are calculated (see Section~\ref{subsubsec:workflow}). If there are any active tasks left, the reconciliation is stopped with a custom requeue duration. Otherwise, the finalizer is removed. In any case, changes to the \texttt{Workflow} resource are persisted. Note that the Workflow Controller does not remove \texttt{Task} resources. Because the Task Controller sets an owner reference to the associated \texttt{Workflow}, Kubernetes will automatically remove related \texttt{Task} resources.
\pagebreak

\addtocounter{figure}{2}
\begin{figure*}[t]
  \centering
  \includegraphics[width=0.66\textwidth]{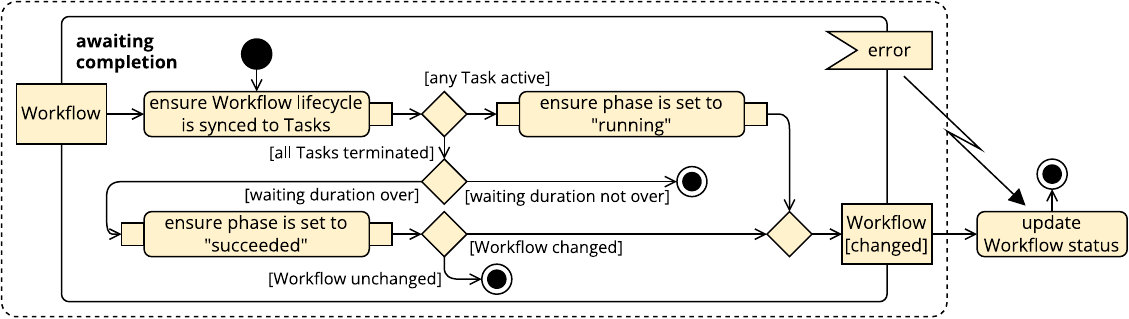}
  \caption{UML activity diagram of the ``awaiting completion'' workflow phase.}
  \label{fig:nme-cmd-workflow-manager-workflow-reconciler-uml-activity-6-awaiting-completion}
  \Description{UML activity diagram of the ``awaiting completion'' workflow phase}
\end{figure*}
\addtocounter{figure}{-3}

\begin{figure}[h]
  \centering
  \includegraphics[width=\columnwidth]{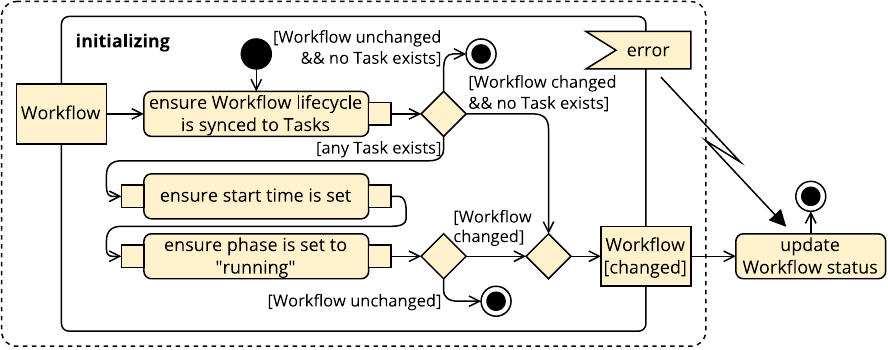}
  \caption{UML activity diagram of the ``initializing'' workflow phase.}
  \label{fig:nme-cmd-workflow-manager-workflow-reconciler-uml-activity-4-initializing}
  \Description{UML activity diagram of the ``initializing'' workflow phase}
\end{figure}

After the internal phases, a workflow starts out in the ``initializing'' phase. It is depicted in Figure~\ref{fig:nme-cmd-workflow-manager-workflow-reconciler-uml-activity-4-initializing}. Workflows remain in this phase until the first task belonging to that workflow is created. Consequently, the reconciler first updates the task counters of the \texttt{Workflow} resource. If there is any active task, the \texttt{Workflow} \texttt{Phase} is set to ``running''. Additionally, the \texttt{StartTime} field is set to the current time.

\begin{figure}[b]
  \centering
  \includegraphics[width=\columnwidth]{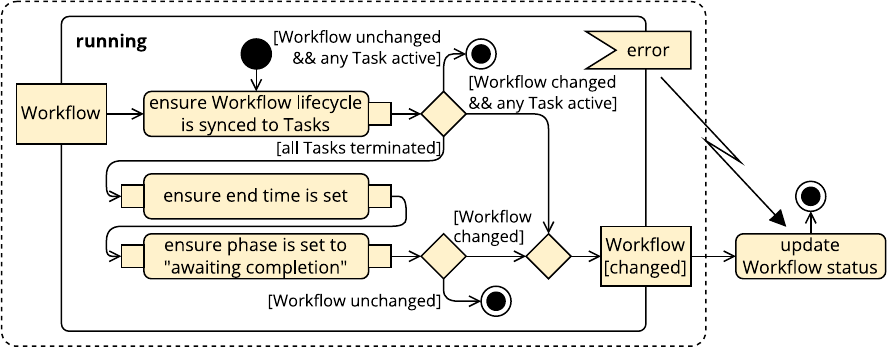}
  \caption{UML activity diagram of the ``running'' workflow phase.}
  \label{fig:nme-cmd-workflow-manager-workflow-reconciler-uml-activity-5-running}
  \Description{UML activity diagram of the ``running'' workflow phase}
\end{figure}
\addtocounter{figure}{1}
In the ``running'' phase, the reconciler continues to observe the execution of related tasks as shown in Figure~\ref{fig:nme-cmd-workflow-manager-workflow-reconciler-uml-activity-5-running}. Because we defined a watch for owned \texttt{Task} resources, the reconciler is called once a \texttt{Task} is updated. It can thus keep the counters in the \texttt{Workflow} \texttt{status} up to date. If a failed task is observed, the first step consults the \texttt{JobFailurePolicy} of the \texttt{Task} (see Section~\ref{subsubsec:task}). Only if it is set to \texttt{Ignore}, the workflow will continue. Otherwise, the \texttt{Phase} is set to ``failed'' which will ultimately lead to the termination of all other active tasks (see Section~\ref{subsec:task-controller}). Instead, if all tasks are observed to have terminated successfully, the \texttt{Phase} is set to ``awaiting completion''. Moreover, the current time is recorded in the \texttt{EndTime} field.

Similar to the previous phase, tasks continue to be observed in the ``awaiting completion'' phase. It is visualized in Figure~\ref{fig:nme-cmd-workflow-manager-workflow-reconciler-uml-activity-6-awaiting-completion}. If after updating the counters, a new active task is observed, the \texttt{Phase} is set to ``running'' again. Otherwise, the reconciler calculates the remaining time it should wait based on the \texttt{EndTime} field and the \texttt{WorkflowTerminationWaitingDuration} configuration option. In case the waiting duration is not over, it will stop the reconciliation with the requeue duration set to the remaining time. Otherwise, the \texttt{Phase} is set to ``succeeded''.

The ``succeeded'' and ``failed'' phases are final and no further actions are taken. The \texttt{Workflow} resource is still available and users can inspect its recorded \texttt{status} field. Future work could add an option to automatically delete terminated \texttt{Workflow} resources after some time.

\subsection{Task Controller}
\label{subsec:task-controller}

This section discusses the Task Controller that reconciles \texttt{Task} resources. Analogous to the Workflow Controller, it follows the general controller structure but splits the reconciliation into further phases. These correspond to the task lifecycle and are again based on the NBMP task lifecycle model (see Section~\ref{subsec:network-based-media-processing}).

\begin{figure}[b]
  \centering
  \includegraphics[width=0.9\columnwidth]{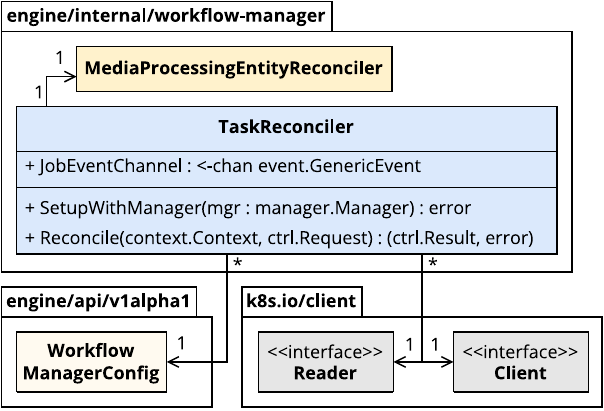}
  \caption{UML class diagram of \texttt{TaskReconciler} related types.}
  \label{fig:nme-cmd-workflow-manager-task-reconciler-uml-class}
  \Description{UML class diagram of \texttt{TaskReconciler} related types}
\end{figure}
The \texttt{TaskReconciler} type implements the reconciler. It is depicted in Figure~\ref{fig:nme-cmd-workflow-manager-task-reconciler-uml-class}. Three event sources are registered. First, a watch on \texttt{Task}s as the primary resource type is configured. Next, a watch for \texttt{Workflow} resources is added with a custom mapping function that adds all \texttt{Task} resources belonging to that \texttt{Workflow} to the reconciliation queue. The reconciler can thus react to workflow state changes, e.g.~when it is marked as failed. Finally, the \texttt{JobEventChannel} channel is included as a \texttt{GenericEvent} source. The Job Controller sits on the sending side of that channel and informs the Task Controller about changed \texttt{Job} resources that belong to \texttt{Task}s. We will detail this process and why this approach was taken in Section~\ref{subsec:job-controller}. No event filters are configured for the Task Controller.

\begin{figure}[h]
  \centering
  \includegraphics[width=0.55\columnwidth]{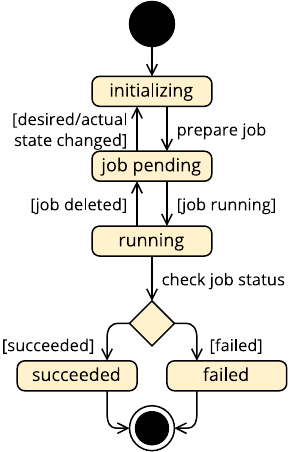}
  \caption{UML state diagram of the logical state changes of a task.}
  \label{fig:nme-cmd-workflow-manager-task-reconciler-uml-state-logically}
  \Description{UML state diagram of the logical state changes of a task}
\end{figure}

Figure~\ref{fig:nme-cmd-workflow-manager-task-reconciler-uml-state-logically} illustrates the logical task lifecycle in a UML state diagram. Newly created tasks are in the ``initializing'' state. After the Task Controller determines the MPE and multimedia function, the task transitions to the ``job pending'' state. Here the Task Controller proceeds to create a \texttt{Job} resource in the MPE and then sets the task state to ``running''. Alternatively, if the task was updated, e.g.~another MPE was selected, the task moves back to the ``initializing'' state. In the ``running'' state, the \texttt{Job} is continually observed. When it is deleted, the task transitions back to the ``job pending'' state. Instead, when the \texttt{Job} terminates, the task either moves to the ``succeeded'' or the ``failed'' state depending on the outcome of the \texttt{Job}.

\begin{figure}[t]
  \centering
  \includegraphics[width=0.7\columnwidth]{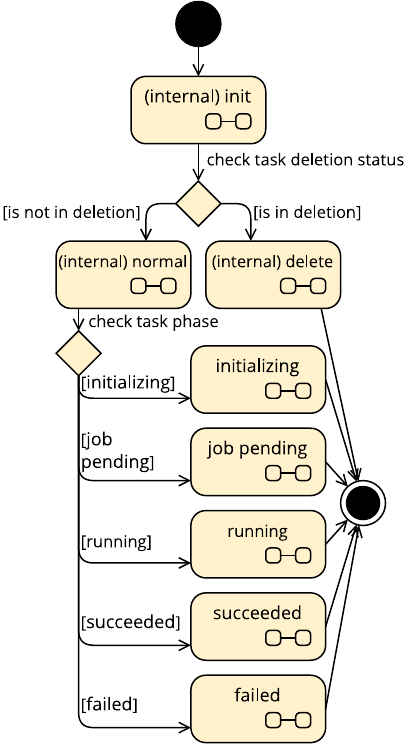}
  \caption{UML state diagram of the phase changes within one task reconciliation loop iteration.}
  \label{fig:nme-cmd-workflow-manager-task-reconciler-uml-state-reconciliation-loop}
  \Description{UML state diagram of the phase changes within one task reconciliation loop iteration}
\end{figure}

The current state is reported in the same way as for \texttt{Workflow} resources, i.e.~through the \texttt{Phase} field in the \texttt{Task}'s \texttt{status}. Similarly, the implementation of the reconciler stops the reconciliation if a phase change is observed. This results in the state model depicted in Figure~\ref{fig:nme-cmd-workflow-manager-task-reconciler-uml-state-reconciliation-loop} for a single iteration of the reconciliation loop. The following paragraphs will discuss each phase and illustrate the involved actions in UML activity diagrams.

\begin{figure}[b]
  \centering
  \includegraphics[width=\columnwidth]{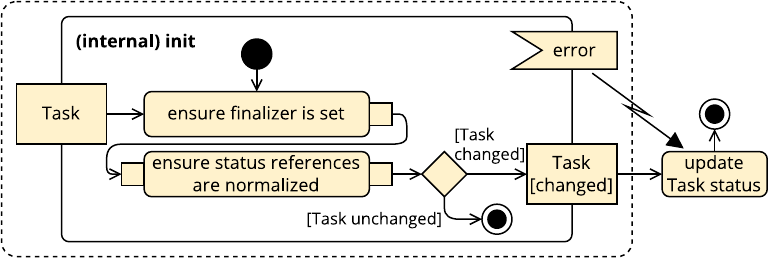}
  \caption{UML activity diagram of the ``(internal)~init'' task phase.}
  \label{fig:nme-cmd-workflow-manager-task-reconciler-uml-activity-1-internal-init}
  \Description{UML activity diagram of the ``(internal)~init'' task phase}
\end{figure}
The ``(internal) init'' phase is traversed during each reconciliation. Figure~\ref{fig:nme-cmd-workflow-manager-task-reconciler-uml-activity-1-internal-init} depicts the actions involved. First, a finalizer is added to the \texttt{Task} resource. In a second step, the references in the \texttt{status} field to the \texttt{(Cluster)MediaProcessingEntity}, the \texttt{(Cluster)Function} as well as the \texttt{Job} resources are normalized. This is only relevant for \texttt{Task} resources that have been previously reconciled. With the normalization we make sure that all fields of the references have been correctly set. Subsequent phases can therefore more easily resolve these references to the corresponding resources. When the \texttt{Task} was changed, it is persisted and the reconciliation stops. Otherwise, it either transitions to the ``(internal) normal'' or the ``(internal) delete'' phase.

\begin{figure}[h]
  \centering
  \includegraphics[width=\columnwidth]{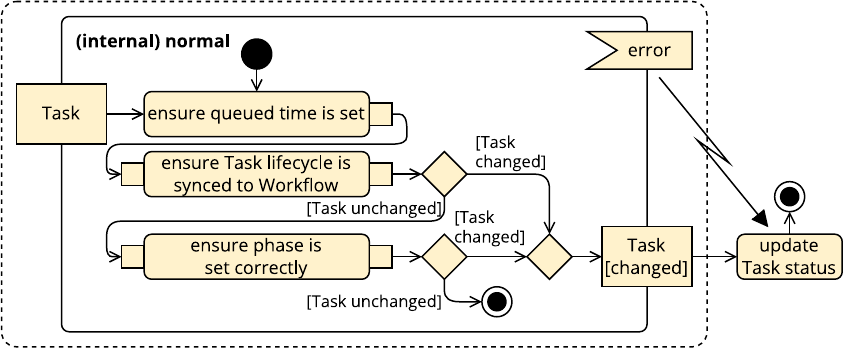}
  \caption{UML activity diagram of the ``(internal)~normal'' task phase.}
  \label{fig:nme-cmd-workflow-manager-task-reconciler-uml-activity-2-internal-normal}
  \Description{UML activity diagram of the ``(internal)~normal'' task phase}
\end{figure}

The ``(internal) normal'' phase is executed in each normal reconciliation. It is visualized in Figure~\ref{fig:nme-cmd-workflow-manager-task-reconciler-uml-activity-2-internal-normal}. It starts out by setting the \texttt{QueuedTime} field to the current time if it is undefined. The second action consists of multiple steps that make sure the task is synced to the workflow. First, it adds an owner reference in the \texttt{Task} to the \texttt{Workflow} resource. Additionally, common labels are set that help users and controllers to filter for specific \texttt{Task} resources. Lastly, it checks whether the workflow is in a final state while the task is still considered active. Typically, this is the case when another task failed, thus failing the whole workflow. Yet creating a new task for an already finished workflow leads to this condition, too. The reconciler will consequently set the \texttt{Task}'s \texttt{Phase} to ``failed''. Moreover, if the associated \texttt{Workflow} resource is in the process of being deleted, further reconciliations of the \texttt{Task} are stopped. In the final action, the reconciler ensures that the \texttt{Phase} field is set correctly, i.e.~consistent to the other \texttt{status} fields. If that is not the case, the error is corrected and the reconciliation stopped. Otherwise, the reconciler transitions to the phase indicated by the \texttt{Phase} field.

\begin{figure}[h]
  \centering
  \includegraphics[width=\columnwidth]{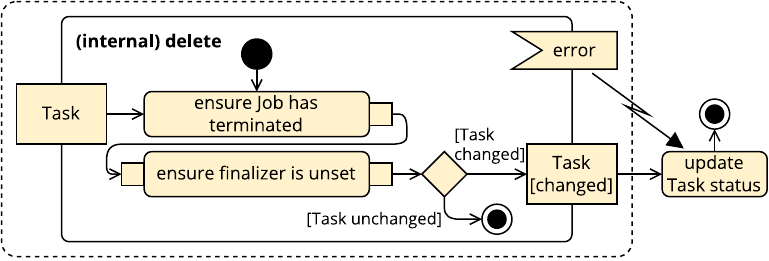}
  \caption{UML activity diagram of the ``(internal)~delete'' task phase.}
  \label{fig:nme-cmd-workflow-manager-task-reconciler-uml-activity-3-internal-delete}
  \Description{UML activity diagram of the ``(internal)~delete'' task phase}
\end{figure}

Before a \texttt{Task} resource is deleted, the ``(internal) delete'' phase is executed. It implements the necessary cleanup logic as depicted in Figure~\ref{fig:nme-cmd-workflow-manager-task-reconciler-uml-activity-3-internal-delete}. Most crucial is the termination and deletion of the associated \texttt{Job}. Because \texttt{Task} resources are always created in the management cluster and \texttt{Job} resources potentially live in another worker cluster, we cannot use owner references. Therefore, the reconciler deletes the \texttt{Job} resource manually. Only after that, the finalizer is removed.

\begin{figure}[h]
  \centering
  \includegraphics[width=\columnwidth]{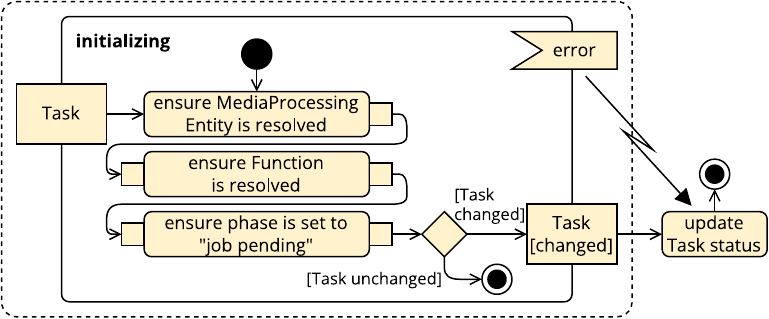}
  \caption{UML activity diagram of the ``initializing'' task phase.}
  \label{fig:nme-cmd-workflow-manager-task-reconciler-uml-activity-4-initializing}
  \Description{UML activity diagram of the ``initializing'' task phase}
\end{figure}

``Initializing'' is the first phase that is communicated to users. Here, the reconciler prepares for the construction of the \texttt{Job} resource as illustrated in Figure~\ref{fig:nme-cmd-workflow-manager-task-reconciler-uml-activity-4-initializing}. Consequently, the MPE and multimedia function have to be determined based on the task specification. Starting with the MPE, the reconciler first checks if the \texttt{MediaProcessingEntityRef} field is used and points to an existing \texttt{(Cluster)MediaProcessingEntity}. Alternatively, the \texttt{Media\linebreak{}ProcessingEntitySelector} field is consulted that selects based on labels. If both fields are undefined, this process is repeated for the referenced \texttt{(Cluster)TaskTemplate}. Lastly, if no \texttt{(Cluster)\linebreak{}TaskTemplate} is referenced, administrators can define a default MPE by annotating a \texttt{(Cluster)MediaProcessingEntity} re\-source with a special annotation, i.e.~a key-value pair. When even failing to resolve an MPE in this way, the reconciliation is aborted with an error. The user then has to supply a valid updated specification. This process is identical for the \texttt{(Cluster)Function} except that there is no default function administrators could configure. Once the MPE and the function have been determined, the resource references are persisted in the \texttt{status} fields. Subsequent phases can thus directly resolve these resources and users get a report about the outcome of this process. Finally, the \texttt{Phase} is set to ``job pending''.

\begin{figure}[b]
  \centering
  \includegraphics[width=\columnwidth]{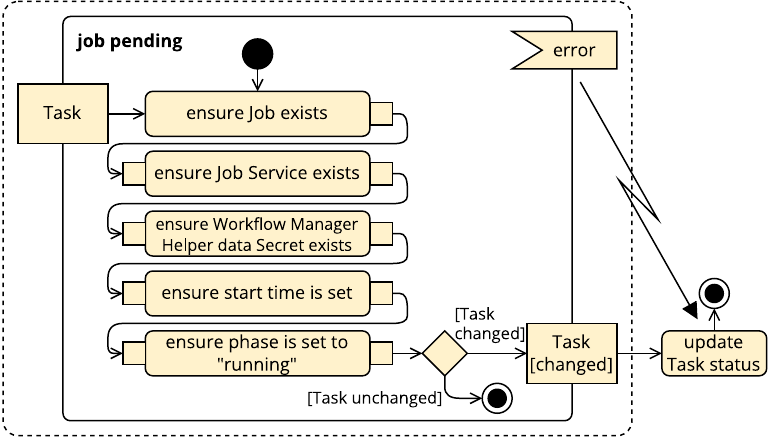}
  \caption{UML activity diagram of the ``job pending'' task phase.}
  \label{fig:nme-cmd-workflow-manager-task-reconciler-uml-activity-5-job-pending}
  \Description{UML activity diagram of the ``job pending'' task phase}
\end{figure}
In the ``job pending'' phase, the \texttt{Job} as well as related resources are created in the MPE. Figure~\ref{fig:nme-cmd-workflow-manager-task-reconciler-uml-activity-5-job-pending} gives an overview of the actions involved. The Task Controller uses the MPE reference to get a Kubernetes \texttt{Client} instance from the \texttt{MediaProcessingEntity\linebreak{}Reconciler} (see Section~\ref{subsec:media-processing-entity-controller}). The \texttt{Client} is then used to create the necessary resources. First, the \texttt{Job} resource is constructed. For this, we merge the \texttt{Job} template fields from the resolved \texttt{(Cluster)\linebreak{}Function}, \texttt{(Cluster)TaskTemplate} and \texttt{Task} resources. Furthermore, the \texttt{Job} is extended with common labels, a finalizer and a volume to a \texttt{Secret} resource. The \texttt{Job} is either created or patched if it already exists. Next, a \texttt{Service} resource is constructed. \texttt{Service}s provide a consistent IP address, port number and domain name within a Kubernetes cluster and thus simplify the communication between workloads. The reconciler sets the \texttt{Service} name and in this way the domain name based on the \texttt{Task}. Currently, we configure a port for HTTP-based traffic. Future work can extend the support for other streaming protocols or make this configurable. The last constructed resource is a \texttt{Secret} that is also referenced in the \texttt{Job} volume mount. It contains data for the Workflow Manager Helper component that runs as a sidecar container next to the main multimedia function container (see Section~\ref{sec:workflow-manager-helper}). The \texttt{Service} and \texttt{Secret} resources are configured with an owner reference to the \texttt{Job} resource. The cleanup logic therefore only needs to delete the \texttt{Job} to remove all created resources. After ensuring the \texttt{StartTime} field is set, the task transitions to the ``running'' phase.

\begin{figure}[h]
  \centering
  \includegraphics[width=\columnwidth]{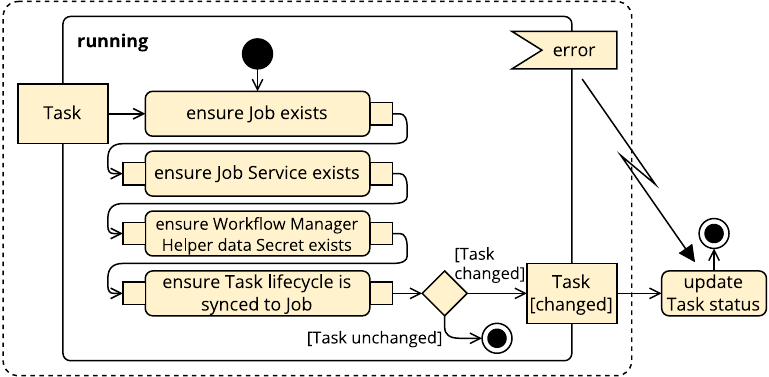}
  \caption{UML activity diagram of the ``running'' task phase.}
  \label{fig:nme-cmd-workflow-manager-task-reconciler-uml-activity-6-running}
  \Description{UML activity diagram of the ``running'' task phase}
\end{figure}

In the ``running'' phase, the \texttt{Job} is continually observed. As can be seen in Figure~\ref{fig:nme-cmd-workflow-manager-task-reconciler-uml-activity-6-running}, the first three actions are identical to the previous phase. The reconciler thus makes sure that the \texttt{Job}, \texttt{Service} and \texttt{Secret} resources exist and have the correct specifications. If one of these is deleted, e.g.~a user manually deletes a \texttt{Job}, it is recreated. Similarly, if the specification of the \texttt{Task} resource changes, e.g.~the function configuration is changed, it will also be reflected in updated specifications of the resources in the worker cluster. In the last action, the conditions of the \texttt{Job} are checked. If the \texttt{Job} is still running, no further action is taken and the reconciliation stops. If, on the other hand, the \texttt{Job} has terminated, the \texttt{Phase} of the \texttt{Task} is set to ``succeeded'' or ``failed'' depending on the outcome of the \texttt{Job}.

\begin{figure}[t]
  \centering
  \includegraphics[width=\columnwidth]{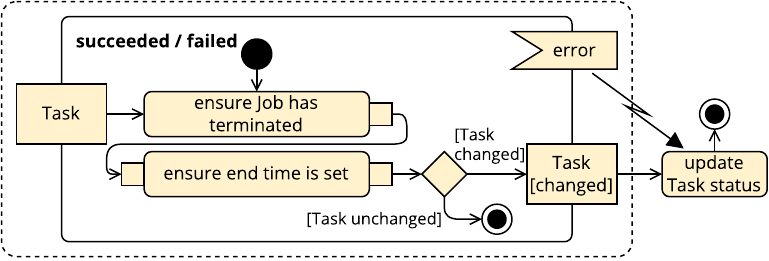}
  \caption{UML activity diagram of the ``succeeded'' and ``failed'' task phase.}
  \label{fig:nme-cmd-workflow-manager-task-reconciler-uml-activity-7-succeeded-failed}
  \Description{UML activity diagram of the ``succeeded'' and ``failed'' task phase}
\end{figure}
The ``succeeded'' and ``failed'' phases consist of the identical actions and are illustrated in Figure~\ref{fig:nme-cmd-workflow-manager-task-reconciler-uml-activity-7-succeeded-failed}. First, the reconciler makes sure that the \texttt{Job} has terminated. A discrepancy between the \texttt{Task} and \texttt{Job} states should only be possible when the \texttt{Task} was marked as ``failed'' during the ``(internal) normal'' phase. In any case, this action ensures that any differences are reconciled quickly. Finally, the \texttt{EndTime} field is set to the current time if it is undefined.

The created resources in the worker cluster are still available after the task has terminated. This gives users the option to inspect the outcome of a task more closely and additionally retrieve generated log output. The cleanup logic ensures that the resources are removed once the \texttt{Task} is deleted.

\subsection{Job Controller}
\label{subsec:job-controller}

\begin{figure}[h]
  \centering
  \includegraphics[width=\columnwidth]{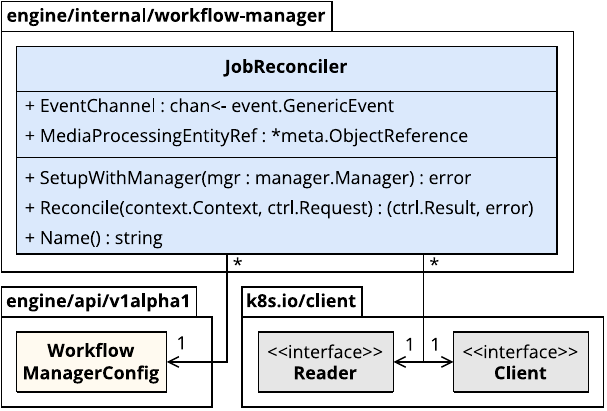}
  \caption{UML class diagram of \texttt{JobReconciler} related types.}
  \label{fig:nme-cmd-workflow-manager-job-reconciler-uml-class}
  \Description{UML class diagram of \texttt{JobReconciler} related types}
\end{figure}

The Job Controller is the last controller bundled in the Workflow Manager component. Unlike the other three controllers, it is rather simple and does not follow the general controller structure explained in Section~\ref{subsec:workflow-manager-overview}. This is because the reconciliation of \texttt{Job} resources is handled by a controller provided by Kubernetes. The \texttt{nagare media engine} Job Controller acts in-parallel to that and primarily informs the Task Controller about changes to \texttt{Job} resources. Figure~\ref{fig:nme-cmd-workflow-manager-job-reconciler-uml-class} gives an overview of the \texttt{JobReconciler} and related types.

One Job Controller is spawned by the MPE Controller for each reconciled \texttt{(Cluster)MediaProcessingEntity} (see Section~\ref{subsec:media-processing-entity-controller}). Because MPEs represent distinct Kubernetes clusters, individual Kubernetes \texttt{Client} and \texttt{Manager} instances must be created. Consequently, the Job Controller is instantiated multiple times with the help of the corresponding \texttt{Manager}. The MPE Controller sets the \texttt{MediaProcessingEntityRef} field to the MPE reference the reconciler is responsible for. Furthermore, the \texttt{EventChannel} field of all reconcilers points to the same Go channel.

The \texttt{SetupWithManager} method defines a watch for \texttt{Job} resources. However, we configure an event filter such that only events affecting \texttt{Job} resource created by the Task Controller are considered. For that, the labels set by the Task Controller are used. Each Job Controller is registered with a unique name based on the MPE.

The reconciliation starts out similarly by first fetching the \texttt{Job} resource with the Kubernetes \texttt{Client}. In the next step, however, the name and namespace of the corresponding \texttt{Task} resource are extracted from the \texttt{Job}. The Task Controller also uses labels to convey this information. With that, a \texttt{GenericEvent} is created and sent to the \texttt{EventChannel}. This, in turn, will trigger a reconciliation of the \texttt{Task} by the Task Controller. Finally, the reconciler examines whether the \texttt{Job} has terminated. If that is the case, it will remove the finalizer set by the Task Controller to allow deletions.

\section{Workflow Manager Helper}
\label{sec:workflow-manager-helper}

In this section we will outline the Workflow Manager Helper component. First, Section~\ref{subsec:workflow-manager-helper-overview} gives a general overview. Section~\ref{subsec:workflow-manager-helper-task-controller} and Section~\ref{subsec:reports-controller} then follow with more details about the Task and Reports Controllers, respectively.

\subsection{Overview}
\label{subsec:workflow-manager-helper-overview}

\begin{figure}[b]
  \centering
  \includegraphics[width=\columnwidth]{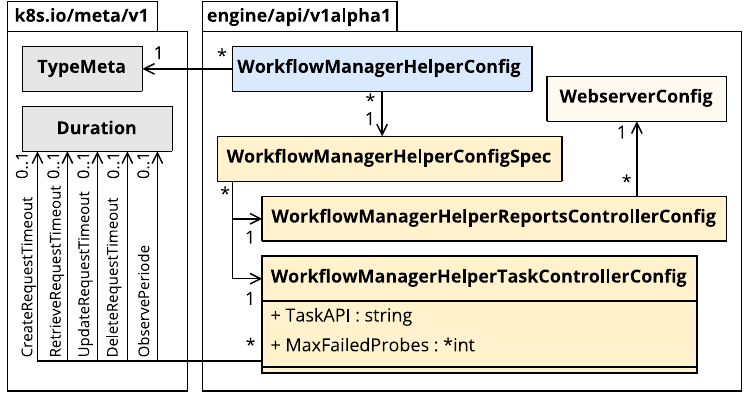}
  \caption{UML class diagram of \texttt{WorkflowManagerHelper\linebreak{}Config} related types.}
  \label{fig:nme-api-config-workflowmanagerhelper-uml-class}
  \Description{UML class diagram of \texttt{WorkflowManagerHelperConfig} related types}
\end{figure}
The Workflow Manager Helper component is an extension of the Workflow Manager in the worker Kubernetes clusters. It runs as a small sidecar container for each task. The Kubernetes \texttt{Job} hence always consists of at least two containers: the Workflow Manager Helper and the multimedia function itself. As a lightweight component, it consumes a small amount of computing resources and therefore only causes a small overhead. Its primary purpose is to control the task lifecycle through the NBMP task API. It thus mediates between the \texttt{nagare media engine} and the NBMP data models. Additionally, it implements an API for task event reporting. Because the Workflow Manager Helper and the multimedia function run side-by-side within one Kubernetes \texttt{Pod}, they share the same networking namespace. Thus, the communication through a local network interface is easily possible. This stands in contrast to an alternative design of the \texttt{nagare media engine} system where the functionality of the Workflow Manager Helper would be provided by a central component in the management Kubernetes cluster. Here, all tasks would need to provide access to their API through potentially multiple networking layers over various hops between the management and worker clusters. Moreover, limiting access to the NBMP task API to \texttt{Pod}-local requests reduces the attack surface and therefore increases the overall security level.

\begin{figure}[b]
  \centering
  \includegraphics[width=0.65\columnwidth]{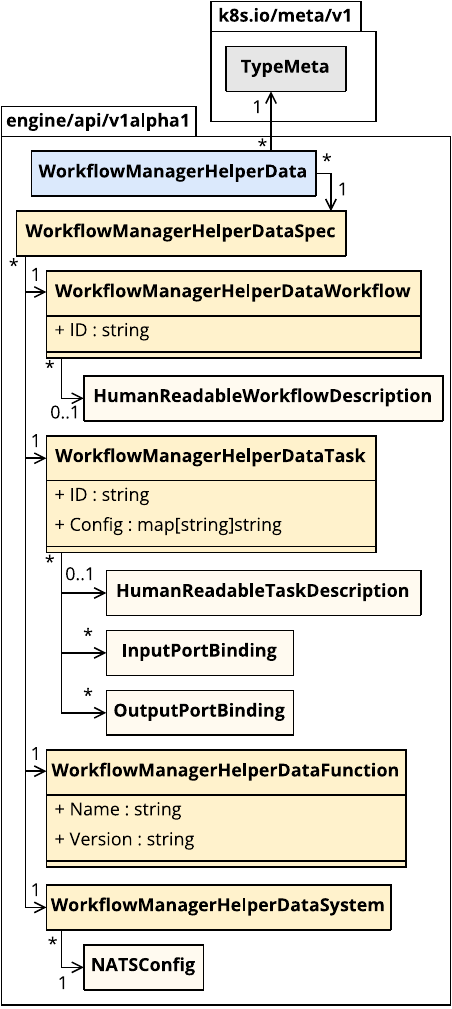}
  \caption{UML class diagram of \texttt{WorkflowManagerHelperData} related types.}
  \label{fig:nme-api-data-workflowmanagerhelper-uml-class}
  \Description{UML class diagram of \texttt{WorkflowManagerHelperData} related types}
\end{figure}

The Workflow Manager Helper bundles two controllers that run concurrently. Both are generally configured through a configuration file. Additionally, a data file provided through a Kubernetes \texttt{Secret} is specific to a particular task and determines how it is instantiated (cf.~Section~\ref{subsec:task-controller}). Figure~\ref{fig:nme-api-config-workflowmanagerhelper-uml-class} depicts the types for the configuration file while Figure~\ref{fig:nme-api-data-workflowmanagerhelper-uml-class} shows the types for the data file. Again, both follow a structure similar to Kubernetes resources.

The configuration for the Reports Controller only contains options for the HTTP server (see Section~\ref{subsec:http-server}). The configuration for the Task Controller governs the use of the NBMP task API client. First, the \texttt{TaskAPI} field sets the URL to the NBMP task API endpoint. The default value uses the loopback network interface and is set to \texttt{http://127.0.0.1:8888/v2/tasks}. The \texttt{Create-}, \texttt{Retrieve-}, \texttt{Update-} and \texttt{DeleteRequestTimeout} options set timeouts for the corresponding API operations. Moreover, \texttt{ObservePeriod} determines how often the status of a running task is probed. For this, the \texttt{MaxFailedProbes} sets a threshold for how many probes can fail in a row until the task is considered unsuccessful.

The data file contains aggregated fields from the \texttt{Workflow}, \texttt{Task} and \texttt{(Cluster)Function} resources that are necessary for the instantiation of the task. Most importantly, the input and output port bindings with the description of the media stream as well as the task configuration are included. Note that the URLs of the media streams are already transformed by the Task Controller of the Workflow Manager. In particular, our custom URL schema described in Section~\ref{subsubsec:common-types} is resolved to the underlying URL beforehand. The Workflow Manager Helper can therefore directly use the URLs to instantiate the task. In addition to the task specification, the data file contains the configuration for connecting to a NATS server. Where possible, existing types that where detailed in Section~\ref{subsec:custom-kubernetes-resources} have been reused.

\subsection{Task Controller}
\label{subsec:workflow-manager-helper-task-controller}

\begin{figure}[h]
  \centering
  \includegraphics[width=0.9\columnwidth]{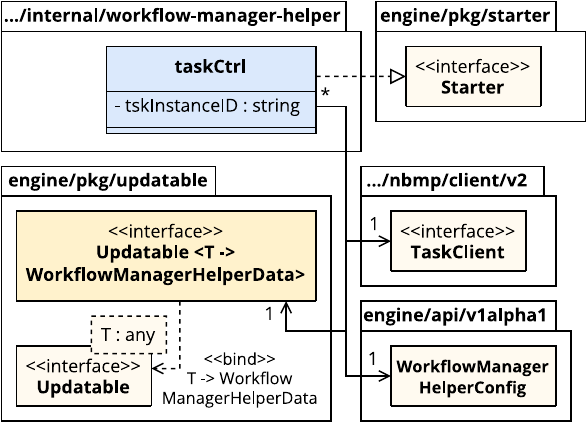}
  \caption{UML class diagram of Task Controller related types.}
  \label{fig:nme-cmd-workflow-manager-helper-task-controller-uml-class}
  \Description{UML class diagram of Task Controller related types}
\end{figure}

The Task Controller implements the standardized interactions with the task over the NBMP task API. It thus oversees the task lifecycle. Figure~\ref{fig:nme-cmd-workflow-manager-helper-task-controller-uml-class} depicts the \texttt{taskCtrl} type together with related types.

\texttt{taskCtrl} implements the \texttt{Starter} interface to execute in its own Goroutine (see Section~\ref{subsec:starter-and-manager}). Based on the configuration, a \texttt{TaskClient} is constructed (see Section~\ref{subsec:nbmp}). While the configuration is static, the data file might be changed after updates to the corresponding \texttt{Secret} resource in the worker cluster. We use the \texttt{Updatable} interface described in Section~\ref{subsec:updatable} to subscribe and react to file updates. The execution of the multimedia function might fail. Ultimately, this results in the termination of the function container with an error code. Depending on the \texttt{Job} specification, Kubernetes might subsequently restart the function container. We call each function execution a ``task instance'' that has a corresponding ``task instance ID''. The Workflow Manager Helper only handles one task instance. If it observes an error or a restarted container, it will also terminate to be restarted by Kubernetes (fail-fast property). The task instance ID is stored in the \texttt{tskInstanceID} field.

The Task Controller is structured in the three phases ``create'', ``observe'' and ``delete''. In the first phase, a \texttt{Converter} is used to transform the parsed data file into a TDD (see Section~\ref{subsec:data-transformations}). The TDD is further extended with a \texttt{Reporting} descriptor that points to the report API endpoint of the Workflow Manager Helper. Finally, a create request with the TDD is sent to the NBMP task API. We generally use retries with exponential backoff for all API requests but limit the combined duration of all tried requests to one minute. If that is exceeded, the task is considered unsuccessful, which will lead to container restarts by Kubernetes. A successful request yields a response with the assigned NBMP task ID, which is used as task instance ID. This concludes the ``create'' phase and the Task Controller transitions to the next phase.

While the function is actively executing, the Task Controller is in the ``observe'' phase. Here, various actions are performed. The primary objective is to monitor the state of the task. For that purpose, retrieve requests are sent to the NBMP task API in regular intervals. By default, these probes happen every two seconds. Various outcomes are possible. If there is a general error during the request, a counter is increased until \texttt{MaxFailedProbes} is reached. The Task Controller will then transition to the next phase with an error. The same applies when the request is successful, but the NBMP \texttt{State} property indicates the task is in an ``error'' state. Otherwise, if the \texttt{State} property is set to ``destroyed'', i.e.~the task terminated normally, the Task Controller transitions to the next phase without recording an error. Lastly, the \texttt{State} property could indicate that the task is still active. Consequently, the Task Controller resets the failed requests counter and remains in the ``observe'' phase.

Next to monitoring the state, the Task Controller also sends events that were previously reported (potentially by other task instances) to specific metadata ports of the function if such ports are defined in the task input bindings of the data file. This way, restarted tasks could recover the state by replaying previously emitted events. This feature is specific to \texttt{nagare media engine} and not standardized by NBMP. However, it is implemented using NBMP constructs. In previous work, we demonstrated how this can be used for task error recovery~\cite{neugebauer_nagaremediaengine_2024}. Events are read from the NATS JetStream defined in the data file. The Reports Controller (see Section~\ref{subsec:reports-controller}) persists events in the NATS JetStream beyond the lifetime of a task instance. Recorded events can thus be replayed easily.

During the ``observe'' phase, the Task Controller furthermore handles data file updates. It subscribes to the \texttt{Updatable} instance and is thus notified about changes through a Go channel. When receiving a new version, the data file is again converted to a TDD and an update request is sent to the NBMP task API. A configuration change of the task, for instance, is hence first requested as a workflow update by the NBMP source through the NBMP Gateway resulting in an updated \texttt{Task} resource which subsequently triggers a reconciliation by the Workflow Manager that updates the \texttt{Secret} resource which is noticed by the Workflow Manager Helper and leads to update requests to the NBMP task API. Usually this chain reaction is relatively quick also due to some forced updates we implemented, yet \texttt{nagare media engine} still follows an eventual consistency model where the workflow state might momentarily be inconsistent. Developers of multimedia functions therefore need to implement robust handling of, e.g.~unexpected input streams.

When the Task Controller reaches the ``delete'' phase, the task either succeeded or the monitoring failed. In any case, the Task Controller tries to send a delete request to the NBMP task API to properly terminate the task. It is then expected that the function container stops. The success or failure status of the task is consequently represented in the exit codes of the Workflow Manager Helper and the function containers. Kubernetes handles the terminated \texttt{Job} accordingly.

\subsection{Reports Controller}
\label{subsec:reports-controller}

\begin{figure}[h]
  \vspace{1pt}
  \centering
  \includegraphics[width=\columnwidth]{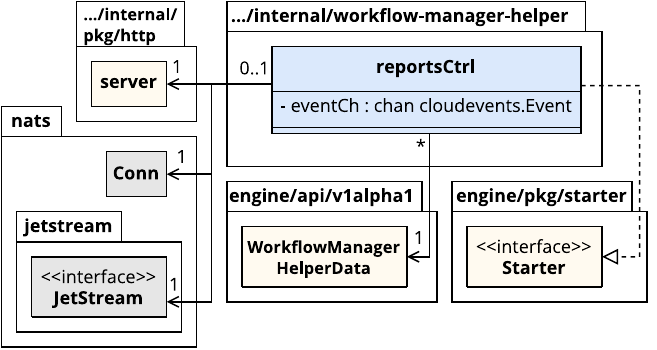}
  \caption{UML class diagram of Reports Controller related types.}
  \label{fig:nme-cmd-workflow-manager-helper-reports-controller-uml-class}
  \Description{UML class diagram of Reports Controller related types}
  \vspace{1pt}
\end{figure}

The Reports Controller provides the report API for events emitted by the multimedia function. The \texttt{reportsCtrl} and related types are depicted in Figure~\ref{fig:nme-cmd-workflow-manager-helper-reports-controller-uml-class}.

\texttt{reportsCtrl} again implements the \texttt{Starter} interface to run concurrently to other controllers. It instantiates an HTTP server and mounts the Health API (see Section~\ref{subsec:health-api}) as well as the Event API (see Section~\ref{subsec:event-api}). The \texttt{eventCh} channel is passed to the Event API instance. The Reports Controller is thus structured as an event loop that handles events sent to \texttt{eventCh}. Received events are then published to the NATS JetStream configured in the data file. We make sure that the Reports Controller terminates after the Task Controller. Hence, it can record termination events sent by the task.

\section{Task Shim}
\label{sec:task-shim}

This section discusses the Task Shim component. We once more start with a general introduction in Section~\ref{subsec:task-shim-overview}. After that, Section~\ref{subsec:task-service} discusses the implementation of the task service logic for the NBMP task API. Finally, Section~\ref{subsec:actions} outlines available actions that administrators can configure to adapt multimedia functions to the NBMP standard.

\subsection{Overview}
\label{subsec:task-shim-overview}

Once a task has been scheduled onto an MPE, the NBMP standard prescribes the use of the NBMP task API to instantiate the multimedia function. Moreover, the NBMP task API is used to monitor and update the task during its execution. Lastly, the task is deleted when it is no longer needed, e.g.~it finished processing the input streams. We already detailed the client-side implementation with the Workflow Manager Helper component in Section~\ref{sec:workflow-manager-helper}. For compatibility with NBMP, the server-side would need to be implemented by all multimedia functions. Consequently, already existing multimedia functions not implementing the NBMP task API would be considered incompatible although being perfectly able to process input streams and provide an output stream. An adapter is therefore needed to mediate between NBMP and the function. Furthermore, even if multimedia function developers started implementing the NBMP task API, having multiple implementations would still not be ideal. Despite NBMP being a standard, variances, e.g. due to bugs or different interpretations of NBMP, are still likely in this scenario. Additionally, a single implementation can provide common logic such as sending general events to the report API. We therefore propose Task Shim as a general adapter and server-side implementation of the NBMP task API. Each API operation leads to a configurable list of actions performed by the Task Shim component. Ultimately, the adapted multimedia function is spawned as a subprocess. In \texttt{nagare media engine}, these two processes run within the same container. Note that Task Shim is also used for multimedia functions we provide. The container image hence includes the Task Shim binary next to the multimedia function. Administrators thus get a demonstration of how they could adapt multimedia functions they are interested~in.

\begin{figure}[h]
  \centering
  \includegraphics[width=\columnwidth]{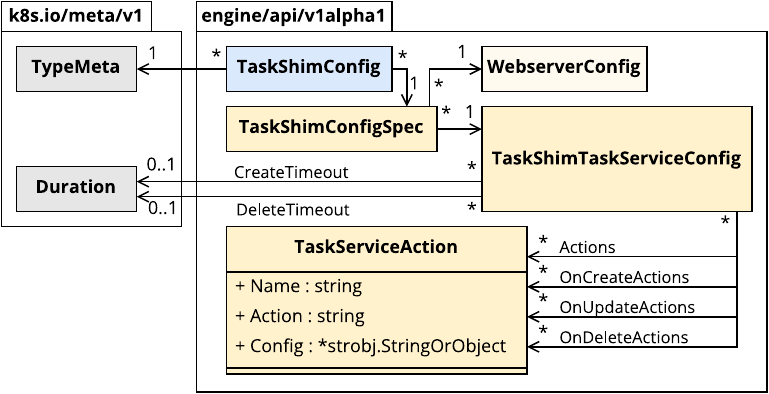}
  \caption{UML class diagram of \texttt{TaskShimConfig} related types.}
  \label{fig:nme-api-config-taskshim-uml-class}
  \Description{UML class diagram of \texttt{TaskShimConfig} related types}
\end{figure}

The adaptation as well as the general behavior is governed by a configuration file, again with a structure similar to Kubernetes resources. Figure~\ref{fig:nme-api-config-taskshim-uml-class} depicts the necessary types.

Similar to other components, we include the \texttt{WebserverConfig} type for the general configuration of the HTTP server (see Section~\ref{subsec:http-server}). Next, administrators can configure four lists of actions. The \texttt{OnCreate-}, \texttt{OnUpdate-} and \texttt{OnDeleteActions} fields determine what actions are performed when create, update and delete requests are handled, respectively. Additionally, the \texttt{Actions} field lists the necessary actions to spawn the multimedia function. In simple scenarios, administrators only need to configure the \texttt{Actions} field out of these four. If unconfigured, the other fields default to starting, restarting and stopping the subprocess, respectively. Each action has a \texttt{Name} that describes its purpose and is used in logging. The \texttt{Action} field's value is interpreted as an identifier and determines what action is performed. Section~\ref{subsec:actions} discusses the available actions. Lastly, the optional \texttt{Config} field further configures the action. The configuration file additionally contains the \texttt{Create-} and \texttt{DeleteTimeout} fields. The former sets a timeout for receiving a create request, while the latter does the same for the delete request that is sent after the multimedia function terminated successfully. If the Task Shim does not receive these requests within the configured durations, it considers the client-side of the NBMP task API as faulty and terminates with an error. By default, it expects a request one minute after the startup and termination of the multimedia function, respectively.

\subsection{Task Service}
\label{subsec:task-service}

\begin{figure}[h]
  \centering
  \includegraphics[width=0.75\columnwidth]{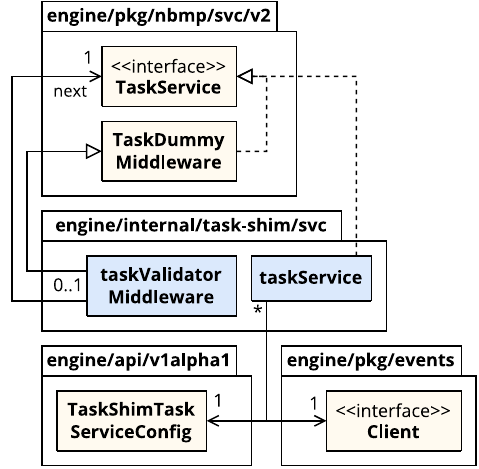}
  \caption{UML class diagram of the types in the \texttt{.../task-shim/svc} package.}
  \label{fig:nme-cmd-task-shim-svc-uml-class}
  \Description{UML class diagram of the types in the \texttt{.../task-shim/svc} package}
\end{figure}

The way the HTTP server is constructed is similar to the NBMP Gateway component (see Section~\ref{sec:nbmp-gateway}). The health and NBMP task APIs (see Section~\ref{subsec:health-api} and~\ref{subsec:nbmp}) are mounted to the HTTP server and the telemetry middleware (see Section~\ref{subsec:telemetry-middleware}) is added for NBMP task API requests. The general task service middleware types \texttt{task\linebreak{}DefaulterSpecMiddleware} and \texttt{taskValidatorSpecLaxMiddle\linebreak{}ware} (see Section~\ref{subsec:nbmp}) are instantiated as is the \texttt{taskValidator\linebreak{}Middleware} type that is specific to Task Shim. Finally, the HTTP server construction is completed with the \texttt{taskService} type that implements the \texttt{TaskService} interface. Figure~\ref{fig:nme-cmd-task-shim-svc-uml-class} depicts a UML class diagram of types relevant for the Task Service.

\begin{figure*}[b]
  \centering
  \includegraphics[width=0.65\textwidth]{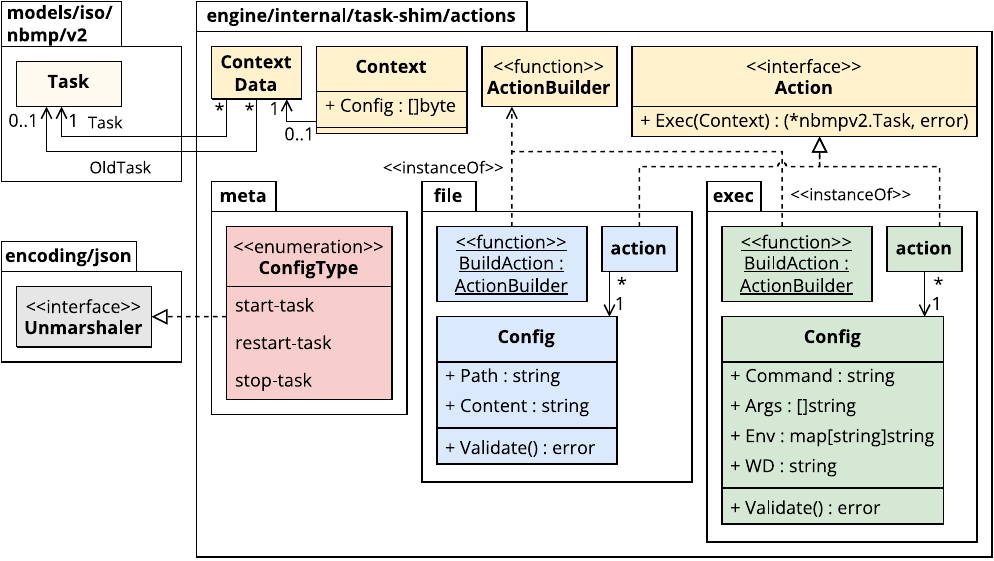}
  \caption{UML class/object diagram of the types/objects in the \texttt{.../task-shim/actions} package.}
  \label{fig:nme-cmd-task-shim-actions-uml-class}
  \Description{UML class/object diagram of the types/objects in the \texttt{.../task-shim/actions} package}
\end{figure*}

The \texttt{taskService} has access to the configuration. When receiving an NBMP task API request, it thus performs the configured actions. Moreover, it maintains general constraints, e.g.~the task can only be created once or the task needs to be created first before an update operation is possible. While we do not expect concurrent requests from the Workflow Manager Helper component, we nonetheless subject all request handlers to a read-write mutex for a strict serializability of the operations. If the TDD contains a \texttt{Reporting} descriptor, an event \texttt{Client} is instantiated and used to report general task events (created, updated, deleted, started, stopped, succeeded and failed).

\subsection{Actions}
\label{subsec:actions}

This section additionally refers to types depicted in Figure~\ref{fig:nme-cmd-task-shim-actions-uml-class}. In order to perform a configured action, the Task Shim first looks into an in-memory registry based on the configured action identifier. The registry contains references to \texttt{ActionBuilder} instances, i.e.~Go functions that construct new \texttt{Action} instances. After finding the \texttt{ActionBuilder} instance, the configured action \texttt{Config} field is processed next. We use the \texttt{StringOrObject} type as outlined in Section~\ref{subsubsec:common-types}. If \texttt{Config} is an object, i.e.~the configuration contains nested structures, it is used as is. Otherwise, if \texttt{Config} is a string, the Task Shim regards it as a template that needs further processing. We use Go's built-in templating library to parse the string as a template and render it. Administrators can thus make use of Go's templating statements and expressions. Moreover, we expose the helper functions implemented in the \texttt{github.com/Masterminds/sprig/v3}\footnote{\url{https://github.com/Masterminds/sprig}} library next to some custom functions. A \texttt{ContextData} instance is passed as input to the template execution. Hence, depending on the NBMP operation, templates can access the current and, for updates, the old TDD. This way, administrators can configure actions based on descriptors and properties of the TDD. The executed template should result in a valid YAML or JSON document. Regardless of whether the \texttt{Config} field was an object or a string, the action configuration is passed to the \texttt{ActionBuilder} as a byte array with the help of the \texttt{Context} type. The \texttt{ActionBuilder} can then decode the configuration, e.g.~into a Go structure. Once an \texttt{Action} instance is constructed by the \texttt{ActionBuilder}, the \texttt{Exec} method is called to perform the action.

We currently implement the three actions \texttt{meta}, \texttt{file} and \texttt{exec}. The first is a special case and is directly handled by the Task Shim without using an \texttt{ActionBuilder}. The configuration for the \texttt{meta} action is decoded as \texttt{ConfigType} enumeration that can take the following values. The \texttt{start-task} \texttt{meta} action spawns the multimedia function as a subprocess by performing the actions configured by the administrator in a new Goroutine. \texttt{stop-task}, on the other hand, makes sure the subprocess has terminated by canceling its Go \texttt{Context}. Lastly, \texttt{restart-task} combines these two by first stopping and then starting the multimedia function.

The \texttt{file} action can be used to create files on the filesystem. Adapting existing multimedia functions could require generating configuration files before execution. Using templating, administrators are able to derive the file's content from the TDD. This action's \texttt{Config} type has a \texttt{Path} field that determines where in the filesystem the file is created. The \texttt{Content} field then defines its content. Future work might add additional common fields such as file permissions or ownership.

Finally, the \texttt{exec} action executes a program. This can be the multimedia function or a program that runs beforehand and prepares its execution. Administrators can configure the \texttt{Command} to execute as well as its arguments using the \texttt{Args} field. Moreover, the \texttt{Env} field allows setting environment variables and \texttt{WD} defines the working directory. Again, templating can make these fields dynamic based on the TDD.

In future work, we might extend this set of actions. For example, an \texttt{event} action could be useful to report custom events if this is unsupported by the multimedia function.

\section{Functions}
\label{sec:functions}

This section covers multimedia functions as the last component in the \texttt{nagare media engine} system. We first give an overview in Section~\ref{subsec:functions-overview}. After that, we go into more detail on how the data flow is handled in Section~\ref{subsec:data-flow}. Lastly, Section~\ref{subsec:included-functions} gives an overview of the functions that are currently included in \texttt{nagare media engine}.

\subsection{Overview}
\label{subsec:functions-overview}

We provide prototypical implementations of various multimedia functions. A common approach is followed in their design and implementation. This way, not only the development is simplified, but the build process and ultimately the usage as well. All functions are bundled in a multi-call binary, i.e.~depending on the name of the binary, different functions are executed. For example, if the filename of the binary is \texttt{media-encode} the included function with the same name is executed (cf.~Section~\ref{subsubsec:media-encode}). Alternatively, if the default name \texttt{functions} is used, the first argument determines the selected function. This approach allows us to easily add new functions without changes to the build process just by extending the Go code. Moreover, the way the binary is executed is standardized as well, thus simplifying the usage. Listing~\ref{lst:task-shim-config} shows the Task Shim configuration for \texttt{media-encode} as an example (cf.~Section~\ref{sec:task-shim}). First, the full TDD is written as a file to the filesystem. For this, the \texttt{toJson} helper function is used in the template to convert the in-memory representation to a JSON document. The path to this file is then passed as an argument to the \texttt{media-encode} binary.
\begin{listing}[t]
\begin{minted}{yaml}
apiVersion: engine.nagare.media/v1alpha1
kind: TaskShimConfig
task:
  actions:
    - name: write task description document
      action: task-shim.engine.nagare.media/file
      config: |
        path: /tmp/nbmp.tdd
        content: |
          {{ toJson .Task }}
    - name: execute function
      action: task-shim.engine.nagare.media/exec
      config:
        command: /bin/media-encode
        args:
          - /tmp/nbmp.tdd
\end{minted}
  \caption{Example Task Shim configuration for the \texttt{media-encode} multimedia function.}
  \label{lst:task-shim-config}
  \Description{Example Task Shim configuration for the \texttt{media-encode} multimedia function.}
\end{listing}

\begin{figure}[h]
  \centering
  \includegraphics[width=0.7\columnwidth]{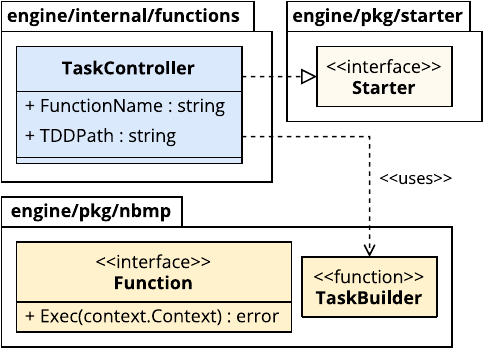}
  \caption{UML class diagram of \texttt{TaskController} related types.}
  \label{fig:nme-cmd-functions-task-controller-uml-class}
  \Description{UML class diagram of \texttt{TaskController} related types}
\end{figure}

After parsing the arguments, the function name and the path to the TDD thus have been determined. With that, an instance of the \texttt{TaskController} type is created and started. Figure~\ref{fig:nme-cmd-functions-task-controller-uml-class} depicts the relevant types in a UML class diagram.

The \texttt{TaskController} then proceeds to parse the TDD file and to look up the function name in an in-memory registry of \texttt{Task\linebreak{}Builder} instances (see Section~\ref{subsec:nbmp}). These are Go functions that instantiate a \texttt{Function} implementation with a given TDD. Thus, with a successfully constructed \texttt{Function} instance, the \texttt{Exec} method is~called.

\subsection{Data Flow}
\label{subsec:data-flow}

NBMP describes the flow of data through the notions of in- and output streams bound to function ports. We model and extend these notions in a custom function IO library. Section~\ref{subsubsec:data-flow-overview} gives a first overview and introduces the types and interfaces. In Section~\ref{subsubsec:http-server-and-port}, we then discuss an implementation that uses HTTP as the underlying network protocol. Lastly, Section~\ref{subsubsec:buffered-port} details a buffered port implementation.

\subsubsection{Overview}
\label{subsubsec:data-flow-overview}

\begin{figure*}[t]
  \centering
  \includegraphics[width=\textwidth]{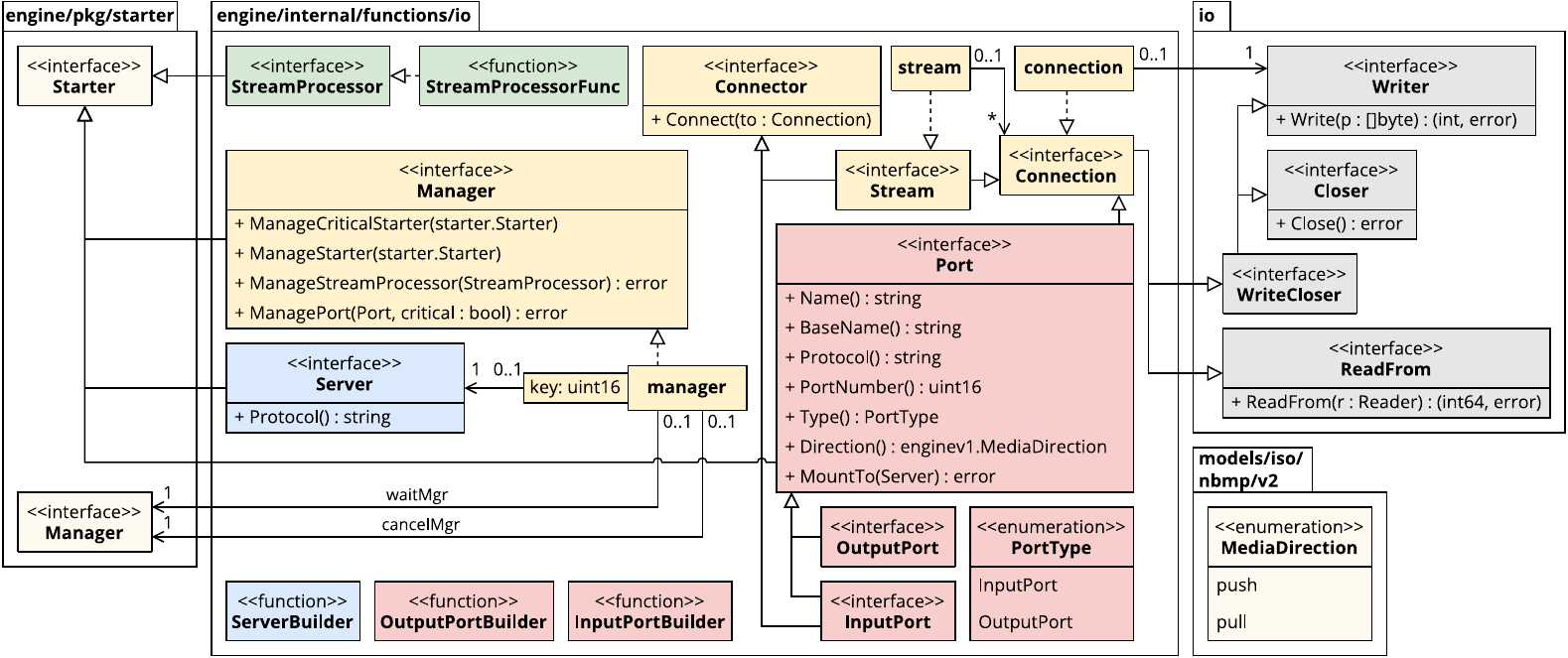}
  \caption{UML class diagram of the types in the \texttt{.../functions/io} package.}
  \label{fig:nme-cmd-functions-io-uml-class}
  \Description{UML class diagram of the types in the \texttt{.../functions/io} package}
\end{figure*}

The main goal of the function IO library is to provide building blocks that are based on the NBMP notions while at the same time offer compatibility to how Go handles IO in general. Figure~\ref{fig:nme-cmd-functions-io-uml-class} visualizes the types we introduced.

Go's \texttt{io} package contains various interfaces to describe types that perform IO operations. Many interfaces only prescribe a single method and are then combined into larger interfaces. For instance, \texttt{WriteCloser} describes a type that implements the single-method interfaces \texttt{Writer} and \texttt{Closer}. The former describes the most basic type that supports writing data, while the latter is for types that need to be closed after completing IO operations. We then defined the \texttt{Connection} interface that combines \texttt{WriteCloser} and \texttt{ReadFrom}. The \texttt{ReadFrom} interface describes a writer type that supports writing all data it reads from a given reader. The \texttt{Connection} interface describes the base capabilities for all our IO types. We thus describe the data flow in terms of interconnected writer types. Accordingly, a \texttt{Connector} is a type that can be connected to, i.e.~it outputs to, a \texttt{Connection}. For compatibility, we implemented a \texttt{connection} type that adapts a general Go writer.

The \texttt{Stream} interface is the next building block. It outputs to multiple \texttt{Connection} types and thus implements a fan-out, but it is also used for single output scenarios. A special consumer is the \texttt{StreamProcessor} implementation. As an implementation of the \texttt{Starter} interface, it describes a type that runs in its own Goroutine and processes the data in the \texttt{Stream}. It thus models the core business logic of a function. The \texttt{StreamProcessorFunc} interface allows using Go functions to define a \texttt{StreamProcessor}.

Next, we define the \texttt{Port} interface together with the specializations \texttt{InputPort} and \texttt{OutputPort}. As the names suggest, these interfaces model the NBMP function ports. The \texttt{InputPort} connects to one \texttt{Connection} while the \texttt{OutputPort} is the final element in the chain. Concrete implementations of the \texttt{OutputPort} could, of course, forward the written data internally to further writers such as a network connection. \texttt{Port} types are active components, i.e.~they implement the \texttt{Starter} interface and run concurrently in their own Goroutine. Moreover, they can be push- or pull-based as already discussed in Section~\ref{subsec:network-based-media-processing}. An \texttt{InputPort} in a pull-based or an \texttt{OutputPort} in a push-based configuration act as a (network) client. They initiate the transfer of data between tasks. An \texttt{InputPort} in a push-based or an \texttt{OutputPort} in a pull-based configuration, on the other hand, actively listen for transfer requests and thus act as a (network) server. To allow the bundling of multiple \texttt{Port} instances, we introduced the \texttt{Server} interface. \texttt{Port} instances are hence mounted to a \texttt{Server} instance. Both need to be compatible, i.e.~their network protocol must be identical. The \texttt{Port} will inform the \texttt{Server} which underlying network port should be used.

Each \texttt{Port} has a name that corresponds to the identifier in the TDD. In an NBMP workflow, an output port can only be connected to a single input port or act as output of the whole workflow. Still, being able to process the output of one task in multiple subsequent tasks is desirable. In addition to a \texttt{data-copy} function (see Section~\ref{subsubsec:data-copy}), we therefore introduced a special naming convention for \texttt{Port} names in \texttt{nagare media engine}. Functions that output to a port with a certain base name will thus duplicate the output to configured ports that start with this base name followed by a dot and a port instance name. For example, if the TDD describes two output ports named ``out.0'' and ``out.1'', then also two \texttt{OutputPort} instances are created, but they both receive the same data. In this way, the multimedia functions provided by \texttt{nagare media engine} already implement a fan-out capability alleviating the necessity to introduce additional fan-out tasks in the workflow.

The NBMP standard is indifferent regarding the network protocols. Multiple protocols can be used side-by-side within the same workflow as defined in the WDD and TDD. We therefore introduced the \texttt{InputPort-}, \texttt{OutputPort-} and \texttt{ServerBuilder} interfaces that construct instances based on NBMP descriptors.

We implemented various building blocks that run concurrently in their own Goroutines. To easily control all active components, we introduced a special \texttt{Manager} interface together with the \texttt{manager} implementation based on the types outlined in Section~\ref{subsec:starter-and-manager}. All IO building blocks can thus be started simultaneously. The \texttt{manager} will further make sure that no two \texttt{Server} instances use the same underlying networking port. Furthermore, we differentiate between critical and non-critical \texttt{Starter} types. The termination of a critical \texttt{Starter} has the termination of all others as the consequence. For example, if a critical \texttt{InputPort} instance stops, the \texttt{Stream\linebreak{}Processor} and \texttt{OutputPort} instances should ultimately also terminate. If a \texttt{Starter} is non-critical, on the other hand, its termination will not trigger further actions.

\subsubsection{HTTP Server and Port}
\label{subsubsec:http-server-and-port}

\begin{figure*}[t]
  \centering
  \includegraphics[width=0.8\textwidth]{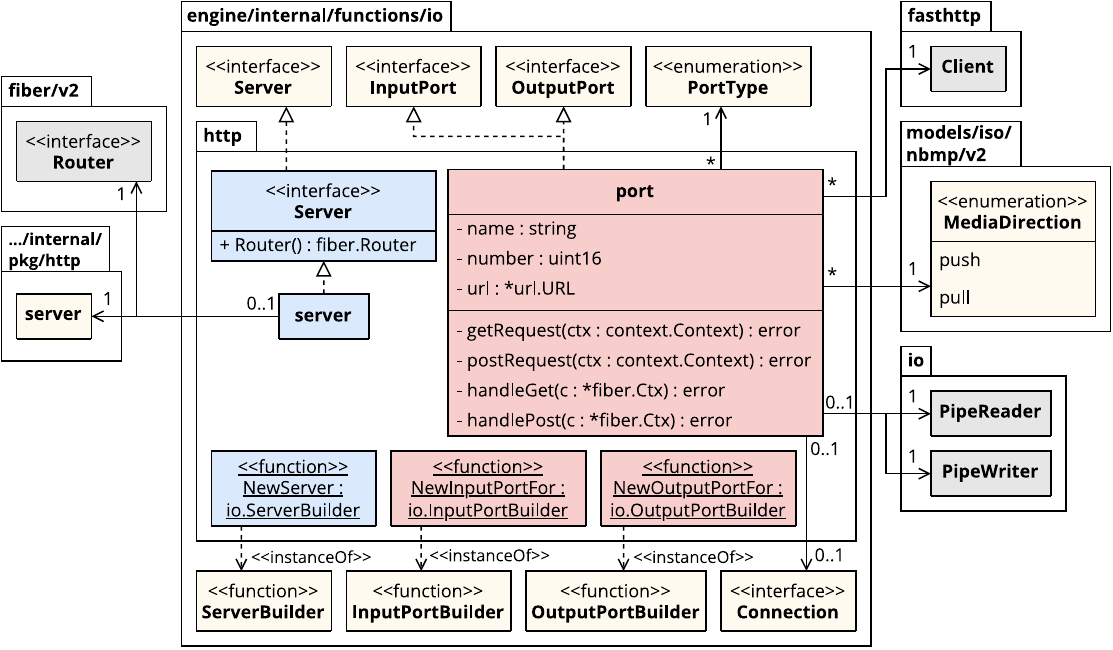}
  \caption{UML class/object diagram of the types/objects in the \texttt{.../functions/io/http} package.}
  \label{fig:nme-cmd-functions-io-http-uml-class}
  \Description{UML class/object diagram of the types/objects in the \texttt{.../functions/io/http} package}
\end{figure*}
Beyond the general interface implementations already provided by the function IO library, the \texttt{Server} and \texttt{Port} interfaces need to be implemented for a specific network protocol. Various protocols exist to transport data in multimedia systems. For our prototype, we chose HTTP, in particular HTTP/1.1 with chunked transfer encoding~(CTE)~\cite{fielding_http11_2022}. Because HTTP is used universally, including in multimedia systems, we consider it a good first candidate. Future work could expand support for additional protocols such as MOQT that may be more suited to transport multimedia data internally between tasks. As can be seen in Figure~\ref{fig:nme-cmd-functions-io-http-uml-class}, we implemented \texttt{Server} and \texttt{Port} types.

First, we extended the \texttt{Server} interface with a \texttt{Router} method. When mounting a compatible \texttt{Port}, it can be used to register handler code for specific request paths. The \texttt{server} type then implements the HTTP server. We base this implementation on our general HTTP server type discussed in Section~\ref{subsec:http-server}. Next, the \texttt{port} type implements both the \texttt{InputPort} and \texttt{OutputPort} interfaces. Based on its configuration, it thus acts as an input or output port. Accordingly, we implemented the necessary builder functions for the \texttt{server} and the \texttt{port} types.

Depending on the configuration, up to two Goroutines are involved in the writing process of a \texttt{port}. Internally, we coordinate the writing process between the different Goroutines through Go's \texttt{PipeReader} and \texttt{PipeWriter} types. These types operate as in-memory analogy to a UNIX pipe, i.e.~the same data that is written to a \texttt{PipeWriter} is read from the \texttt{PipeReader}. Write and read operations must run concurrently or are otherwise blocked. In the following, we go over the different modes \texttt{port} can be used in.

\begin{figure}[h]
  \centering
  \includegraphics[width=0.9\columnwidth]{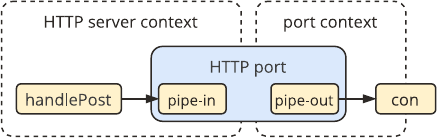}
  \caption{Logical view of the data flow through an HTTP input port in push configuration.}
  \label{fig:nme-cmd-functions-io-http-input-push}
  \Description{Logical view of the data flow through an HTTP input port in push configuration}
\end{figure}

First, configured as a push-based input port, it waits for an HTTP \texttt{POST} request to arrive at the path configured in the TDD. For this path, the \texttt{handlePost} method is registered as handler. Once the request arrives, this method is executed by the Fiber HTTP framework within a new Goroutine, i.e.~the ``HTTP server context''. This Goroutine then writes the request body to the pipe. Concurrently, the \texttt{port} Goroutine reads from the pipe and writes it to the associated \texttt{Connection}. Figure~\ref{fig:nme-cmd-functions-io-http-input-push} visualizes this scenario.

\begin{figure}[h]
  \centering
  \includegraphics[width=0.9\columnwidth]{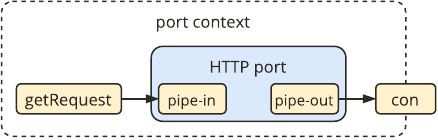}
  \caption{Logical view of the data flow through an HTTP input port in pull configuration.}
  \label{fig:nme-cmd-functions-io-http-input-pull}
  \Description{Logical view of the data flow through an HTTP input port in pull configuration}
\end{figure}

Alternatively, the input port could be in pull configuration. Here, all IO operations are performed within Goroutines started by the \texttt{port}. First, an HTTP \texttt{GET} request is sent to the configured URL. The response body is then written to the pipe. A second Goroutine then reads from the pipe and writes it to the \texttt{Connection}. Figure~\ref{fig:nme-cmd-functions-io-http-input-pull} depicts this configuration.

\begin{figure}[h]
  \centering
  \includegraphics[width=\columnwidth]{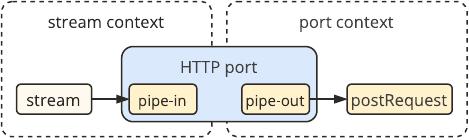}
  \caption{Logical view of the data flow through an HTTP output port in push configuration.}
  \label{fig:nme-cmd-functions-io-http-output-push}
  \Description{Logical view of the data flow through an HTTP output port in push configuration}
\end{figure}

In case the \texttt{port} is configured as a push-based output, the data flows as shown in Figure~\ref{fig:nme-cmd-functions-io-http-output-push}. Writing to an output port always occurs in the Goroutine of the previous building block, here the ``stream context''. Again, the written data is sent to the internal pipe. In the \texttt{port} Goroutine, the data read from the pipe is used as the request body for an HTTP \texttt{POST} request to the configured URL. CTE is used to stream the data in chunks.

\begin{figure}[b]
  \centering
  \includegraphics[width=\columnwidth]{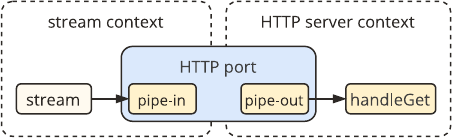}
  \caption{Logical view of the data flow through an HTTP output port in pull configuration.}
  \label{fig:nme-cmd-functions-io-http-output-pull}
  \Description{Logical view of the data flow through an HTTP output port in pull configuration}
\end{figure}
Lastly, as an output port in pull configuration, the data flow occurs in the following way. On the receiving side, the data is again written to the \texttt{port} within the ``stream context''. However, the write operation will block until the \texttt{server} receives an HTTP \texttt{GET} request to the configured path and executes the \texttt{handleGet} method. We use CTE in the response to transfer the data read from the pipe. Note that in this case the \texttt{port} Goroutine only facilitates the data flow by constructing the connections, but it is otherwise inactive. Figure~\ref{fig:nme-cmd-functions-io-http-output-pull} depicts this scenario.

\subsubsection{Buffered Port}
\label{subsubsec:buffered-port}

\begin{figure}[h]
  \centering
  \includegraphics[width=\columnwidth]{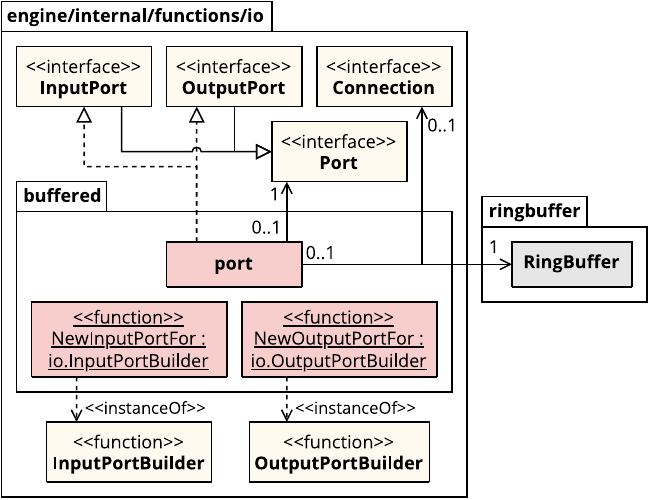}
  \caption{UML class/object diagram of the types/objects in the \texttt{.../functions/io/buffered} package.}
  \label{fig:nme-cmd-functions-io-buffered-uml-class}
  \Description{UML class/object diagram of the types/objects in the \texttt{.../functions/io/buffered} package}
\end{figure}

Our \texttt{Port} implementation for HTTP uses an internal pipe that blocks if writer and reader do not operate at the same speed. This can be undesirable. The bit rate of encoded multimedia streams is typically variable if it is not artificially padded. This can consequently result in a variable duration required for processing the stream. Input and output buffers are thus commonly employed to smooth out this fluctuation. We therefore implemented a buffered \texttt{Port} implementation that wraps another \texttt{Port} for buffered IO. Compared to the HTTP implementation, we replaced the pipe with a ring buffer that only blocks if the buffer is full (write operations) or empty (read operations). The ring buffer implementation is provided by the \texttt{github.com/smallnest/ringbuffer}\footnote{\url{https://github.com/smallnest/ringbuffer}} library. Figure~\ref{fig:nme-cmd-functions-io-buffered-uml-class} depicts the relevant types.

\begin{figure}[h]
  \centering
  \includegraphics[width=0.8\columnwidth]{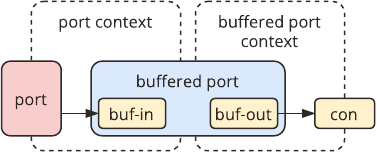}
  \caption{Logical view of the data flow through a buffered input port.}
  \label{fig:nme-cmd-functions-io-buffered-input}
  \Description{Logical view of the data flow through a buffered input port}
\end{figure}

In the following, we go over the possible configuration options of the buffered \texttt{port}. This time, only two scenarios need to be considered. Figure~\ref{fig:nme-cmd-functions-io-buffered-input} depicts the buffered \texttt{port} as input. The wrapped \texttt{Port} is connected to the buffered \texttt{port} and thus writes to the internal ring buffer. Within the buffered \texttt{port} Goroutine, the data is read from the buffer and sent to the associated \texttt{Connection}.

\begin{figure}[h]
  \centering
  \includegraphics[width=0.8\columnwidth]{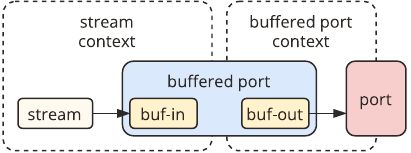}
  \caption{Logical view of the data flow through a buffered output port.}
  \label{fig:nme-cmd-functions-io-buffered-output}
  \Description{Logical view of the data flow through a buffered output port}
\end{figure}

Configured as output, the buffered \texttt{port} is situated before the wrapped \texttt{Port}. The received data is thus written to the ring buffer within the ``stream context''. The buffered \texttt{port} Goroutine then reads from the ring buffer and forwards it to the wrapped \texttt{Port}. This scenario is illustrated in Figure~\ref{fig:nme-cmd-functions-io-buffered-output}.

The buffered \texttt{port} is created if the stream URL configured in the TDD uses ``\texttt{buffered}'' as protocol, e.g.~\texttt{buffered://127.0.0.1/\linebreak{}input?nme-buffered-protocol=http}. The query parameter \texttt{nme-\linebreak{}buffered-protocol} then determines what type of \texttt{Port} is wrapped. The buffered \texttt{port} build functions thus also constructed the wrapped port based on a transformed URL, e.g.~\texttt{http://127.0.0.1/input}. By default, the ring buffer will allocate 10~megabytes.

\subsection{Included Functions}
\label{subsec:included-functions}

In the following, we will give an overview of the multimedia functions that are currently included in \texttt{nagare media engine}. Sections~\ref{subsubsec:generic-noop} to~\ref{subsubsec:script-lua} discuss the \texttt{generic-noop}, \texttt{generic-\linebreak{}sleep}, \texttt{data-discard}, \texttt{data-copy}, \texttt{media-generate-test\linebreak{}pattern}, \texttt{media-encode} and \texttt{script-lua} functions, respectively.

\subsubsection{\texttt{generic-noop}}
\label{subsubsec:generic-noop}

\begin{figure}[h]
  \centering
  \includegraphics[width=0.7\columnwidth]{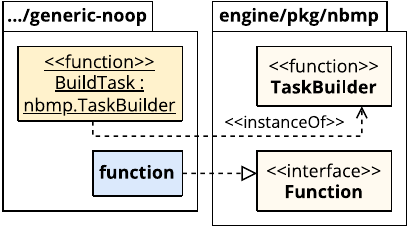}
  \caption{UML class/object diagram of the types/objects in the \texttt{.../generic-noop} package.}
  \label{fig:nme-cmd-functions-functions-generic-noop-uml-class}
  \Description{UML class/object diagram of the types/objects in the \texttt{.../generic-noop} package}
\end{figure}

\texttt{generic-noop} is a dummy function for testing purposes that does nothing and directly terminates. It ignores all configured inputs and outputs. It further has no configuration options. Figure~\ref{fig:nme-cmd-functions-functions-generic-noop-uml-class} depicts the implementation.

\subsubsection{\texttt{generic-sleep}}
\label{subsubsec:generic-sleep}

\begin{figure}[h]
  \centering
  \includegraphics[width=0.7\columnwidth]{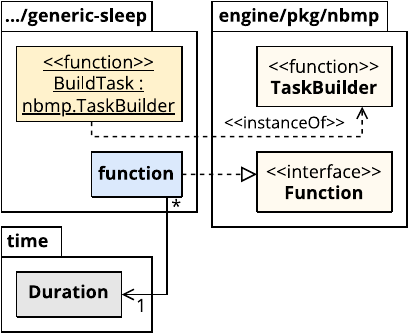}
  \caption{UML class/object diagram of the types/objects in the \texttt{.../generic-sleep} package.}
  \label{fig:nme-cmd-functions-functions-generic-sleep-uml-class}
  \Description{UML class/object diagram of the types/objects in the \texttt{.../generic-sleep} package}
\end{figure}

The \texttt{generic-sleep} function sleeps for a configured amount of time and then terminates. We initially implemented this function for testing purposes. Unlike \texttt{generic-noop}, a task does not terminate immediately and we can simulate that a certain process is running. We do not expect \texttt{generic-sleep} to be used in production workflows. It ignores all configured inputs and outputs. Users can set the sleeping duration through the function configuration. The implementation is illustrated in Figure~\ref{fig:nme-cmd-functions-functions-generic-sleep-uml-class}.

\subsubsection{\texttt{data-discard}}
\label{subsubsec:data-discard}

\begin{figure}[h]
  \centering
  \includegraphics[width=0.55\columnwidth]{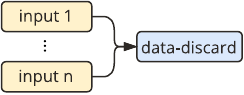}
  \caption{Logical view of the data flow through the \texttt{data-discard} function.}
  \label{fig:nme-cmd-functions-functions-data-discard}
  \Description{Logical view of the data flow through the \texttt{data-discard} function}
\end{figure}

The \texttt{data-discard} function discards the data it receives from its input ports. We implemented this function to test workflows that have no NBMP media sink, i.e.~no consumer of the workflow output. Figure~\ref{fig:nme-cmd-functions-functions-data-discard} illustrates the logical data flow through this function.

\begin{figure}[b]
  \centering
  \includegraphics[width=0.7\columnwidth]{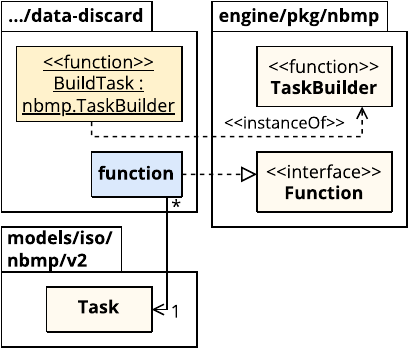}
  \caption{UML class/object diagram of the types/objects in the \texttt{.../data-discard} package.}
  \label{fig:nme-cmd-functions-functions-data-discard-uml-class}
  \Description{UML class/object diagram of the types/objects in the \texttt{.../data-discard} package}
\end{figure}
The \texttt{data-discard} function has no configuration options, but it creates \texttt{InputPort} instances for all ports described in the TDD. We then connect the \texttt{InputPort} instances to Go's \texttt{Discard} writer that is wrapped by a \texttt{connection} instance. The input streams are thus actually read to completion. Figure~\ref{fig:nme-cmd-functions-functions-data-discard-uml-class} depicts the involved types.

\subsubsection{\texttt{data-copy}}
\label{subsubsec:data-copy}

\begin{figure}[h]
  \centering
  \includegraphics[width=0.8\columnwidth]{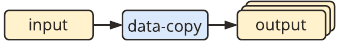}
  \caption{Logical view of the data flow through the \texttt{data-copy} function.}
  \label{fig:nme-cmd-functions-functions-data-copy}
  \Description{Logical view of the data flow through the \texttt{data-copy} function}
\end{figure}

In NBMP, a port can only be bound to one stream. Nonetheless, workflow authors might want to connect the output of a task to the inputs of multiple tasks. In \texttt{nagare media engine}, we hence implemented a fan-out mechanism internal to all functions based on a naming convention (see Section~\ref{subsubsec:data-flow-overview}). In general, however, the fan-out must be accomplished through a task within the workflow. We therefore implemented the \texttt{data-copy} function. It takes a single input and copies the incoming data to multiple outputs as can be seen in Figure~\ref{fig:nme-cmd-functions-functions-data-copy}. Additionally, if buffered ports are described in the TDD, the \texttt{data-copy} function can operate as a buffer between tasks.

\begin{figure}[h]
  \centering
  \includegraphics[width=0.7\columnwidth]{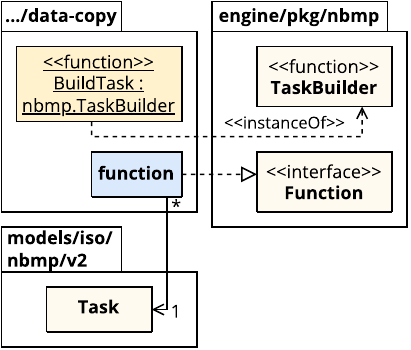}
  \caption{UML class/object diagram of the types/objects in the \texttt{.../data-copy} package.}
  \label{fig:nme-cmd-functions-functions-data-copy-uml-class}
  \Description{UML class/object diagram of the types/objects in the \texttt{.../data-copy} package}
\end{figure}

\texttt{InputPort} and \texttt{OutputPort} instances are created based on the TDD. They are connected by a \texttt{Stream} instance that handles the fan-out. This function does not have any configuration options. Future work could add support for more than one input. Each input could then be mapped to multiple outputs through the function configuration. Figure~\ref{fig:nme-cmd-functions-functions-data-copy-uml-class} visualizes our implementation in a UML diagram.

\subsubsection{\texttt{media-generate-testpattern}}
\label{subsubsec:media-generate-testpattern}

\begin{figure}[h]
  \centering
  \includegraphics[width=0.75\columnwidth]{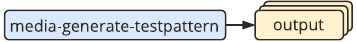}
  \caption{Logical view of the data flow through the \texttt{media-generate-testpattern} function.}
  \label{fig:nme-cmd-functions-functions-media-generate-testpattern}
  \Description{Logical view of the data flow through the \texttt{media-generate-testpattern} function}
\end{figure}

The \texttt{media-generate-testpattern} function allows generating a multimedia stream that shows a test pattern. Accordingly, this function has no inputs but multiple outputs, as illustrated in Figure~\ref{fig:nme-cmd-functions-functions-media-generate-testpattern}.

\begin{figure}[h]
  \centering
  \includegraphics[width=0.7\columnwidth]{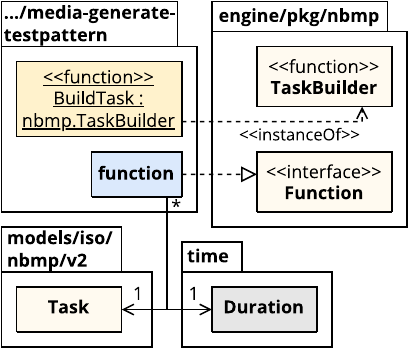}
  \caption{UML class/object diagram of the types/objects in the \texttt{.../media-generate-testpattern} package.}
  \label{fig:nme-cmd-functions-functions-media-generate-testpattern-uml-class}
  \Description{UML class/object diagram of the types/objects in the \texttt{.../media-generate-testpattern} package}
\end{figure}

We use FFmpeg running as a subprocess to generate the test pattern. FFmpeg writes the encoded bitstream to its standard output that we connect to a \texttt{Stream} instance. Additionally, log messages from FFmpeg that are written to standard error are included in the function's log. The execution of FFmpeg is implemented as concurrently running \texttt{StreamProcessor}. The \texttt{Stream} instance then connects to \texttt{OutputPort} instances as described by the TDD. Users can optionally configure a duration after which the \texttt{media-generate-\linebreak{}testpattern} function stops generating the multimedia stream. Figure~\ref{fig:nme-cmd-functions-functions-media-generate-testpattern-uml-class} visualizes our implementation.

\subsubsection{\texttt{media-encode}}
\label{subsubsec:media-encode}

\begin{figure}[h]
  \centering
  \includegraphics[width=0.8\columnwidth]{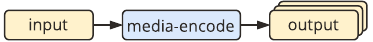}
  \caption{Logical view of the data flow through the \texttt{media-encode} function.}
  \label{fig:nme-cmd-functions-functions-media-encode}
  \Description{Logical view of the data flow through the \texttt{media-encode} function}
\end{figure}

The \texttt{media-encode} function takes one input, encodes the multimedia stream and sends it to multiple outputs. Figure~\ref{fig:nme-cmd-functions-functions-media-encode} illustrates this data flow.

FFmpeg is used to implement the media encoding process. By default, an H.264/AVC~\cite{isoiec_isoiec1449610_2022} bitstream is created and packaged in an MPEG transport stream~(MPEG-TS) container~\cite{isoiec_recituth2220_2023}. Users can control the output resolution and bit rate through configuration options. Alternatively, users have the option to directly set FFmpeg input and output flags and thus use arbitrary video codecs.

\begin{figure}[h]
  \centering
  \includegraphics[width=0.75\columnwidth]{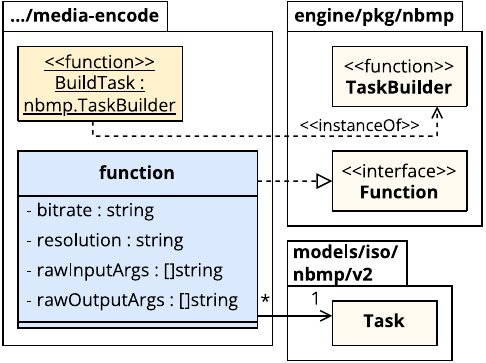}
  \caption{UML class/object diagram of the types/objects in the \texttt{.../media-encode} package.}
  \label{fig:nme-cmd-functions-functions-media-encode-uml-class}
  \Description{UML class/object diagram of the types/objects in the \texttt{.../media-encode} package}
\end{figure}

We execute FFmpeg as a subprocess within a \texttt{StreamProcessor} Goroutine. The created \texttt{InputPort} instance is connected to FFmpeg's standard input. Next, a \texttt{Stream} instance is created and connected to FFmpeg's standard output. We also forward log messages written to standard error to the log output of the \texttt{media-encode} function. Finally, the \texttt{Stream} is connected to \texttt{OutputPort} instances as defined in the TDD. The involved types can be seen in Figure~\ref{fig:nme-cmd-functions-functions-media-encode-uml-class}.
\pagebreak

\subsubsection{\texttt{script-lua}}
\label{subsubsec:script-lua}

\addtocounter{figure}{1}
\begin{figure*}[b]
  \centering
  \includegraphics[width=\textwidth]{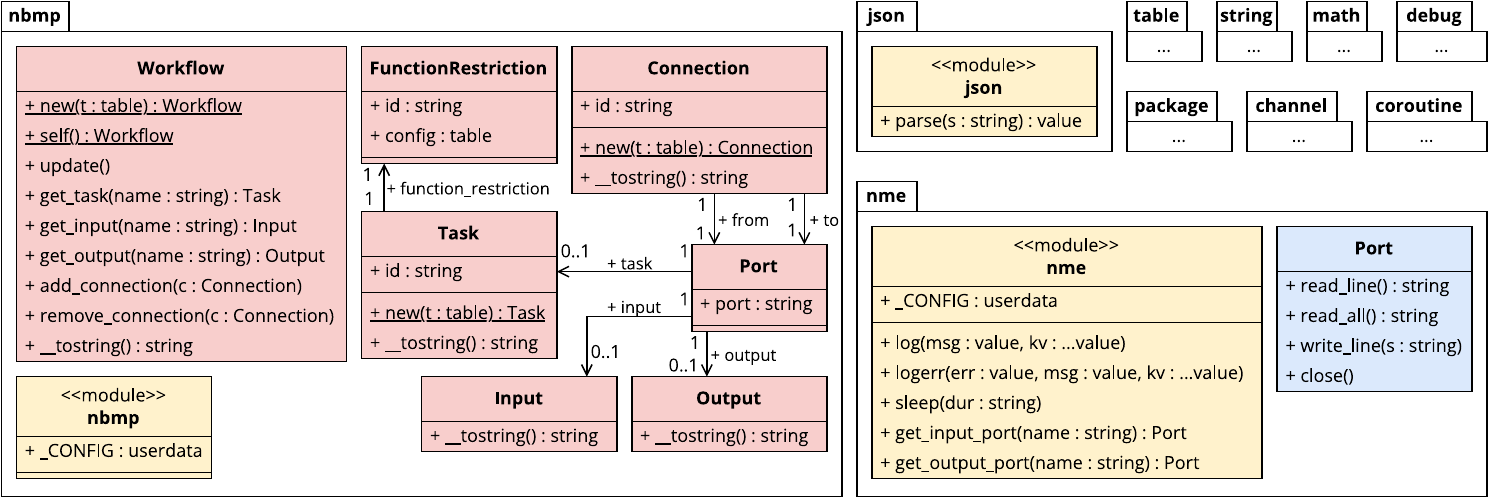}
  \caption{UML class diagram of the modules and types available in Lua scripts within the \texttt{script-lua} function.}
  \label{fig:nme-cmd-functions-functions-script-lua-modules-uml-class}
  \Description{UML class diagram of the modules and types available in Lua scripts within the \texttt{script-lua} function}
\end{figure*}
\addtocounter{figure}{-2}

In NBMP, the workflow is defined in a declarative manner through the WDD. The NBMP workflow manager interprets the WDD and creates a corresponding DAG of tasks. Theoretically, the NBMP workflow manager itself could adapt this DAG when it finds a better structure, e.g.~after changed circumstances. However, we expect that NBMP workflow manager implementations only construct the DAG once. In~\cite{neugebauer_nagaremediaengine_2025}, we therefore wanted to give workflow authors the option to develop workflows with inherent self-adaptability that are compatible with NBMP. We implemented the \texttt{script-lua} function that executes a user-given script with the self-adaptation logic. This way, NBMP workflows can optimize for higher-level objectives, e.g.~dynamically start and stop encoding tasks for specific resolutions based on user requests in order to minimize electricity and computing resource usage.

\begin{figure}[h]
  \centering
  \includegraphics[width=0.8\columnwidth]{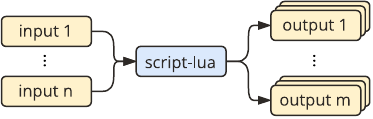}
  \caption{Logical view of the data flow through the \texttt{script-lua} function.}
  \label{fig:nme-cmd-functions-functions-script-lua}
  \Description{Logical view of the data flow through the \texttt{script-lua} function}
\end{figure}

The script is passed to the \texttt{script-lua} function as a configuration parameter. The WDD and, in turn, the TDD thus fully define the self-adaptation logic without further components. The function performs the adaptations by altering the WDD and sending an update request to the NBMP workflow API. Users therefore also have to set the URL to the NBMP workflow API as a configuration option. Moreover, the script can access the input and output ports defined in the TDD. We expect that mostly metadata streams will be consumed and created by the \texttt{script-lua} function. Figure~\ref{fig:nme-cmd-functions-functions-script-lua} illustrates the possible data flow.
\addtocounter{figure}{1}

As the function name suggests, \texttt{script-lua} executes a script that is implemented in the programming language Lua\footnote{\url{https://www.lua.org/}}. This language was especially designed to be embedded into a host application. Executing the script thus allows influencing the behavior of the host application and accessing exposed functionality. For \texttt{script-lua}, we use \texttt{github.com/yuin/gopher-lua}\footnote{\url{https://github.com/yuin/gopher-lua}}~(GopherLua), a Go implementation of Lua version~5.1. Additionally, a slightly modified version of the \texttt{github.com/layeh/gopher-luar}\footnote{\url{https://github.com/layeh/gopher-luar}} library helps with exposing Go types to Lua scripts. We thus provide an internal domain-specific language~(DSL) to Lua scripts based on NBMP notions. Figure~\ref{fig:nme-cmd-functions-functions-script-lua-modules-uml-class} shows the modules and types available to Lua scripts. Note that we depict module functions as UML classes with the \texttt{<<module>>} stereotype.

We expose selected modules from the Lua standard library~\cite{ierusalimschy_lua51reference_2006}, i.e.~the \texttt{coroutine}, \texttt{debug}, \texttt{math}, \texttt{package}, \texttt{string} and \texttt{table} modules. The standard modules \texttt{io} and \texttt{os}, on the other hand, are not available because of undesired functionality, e.g.~the \texttt{os.exit} Lua function terminates the host program forcefully. The \texttt{channel} module is specific to GopherLua and allows Lua to interact with Go channels. We then added the \texttt{json}, \texttt{nbmp} and \texttt{nme} modules. All available modules are loaded by \texttt{script-lua} before the script executes. Users hence do not need to add \texttt{require} statements to their scripts and can concentrate on the adaptation logic.

The \texttt{json} module provides a function to parse a JSON string to a Lua-native value.

The \texttt{nme} module provides functionality to interact with \texttt{nagare media engine} as the host system. The \texttt{log} and \texttt{logerr} functions allow writing messages to the \texttt{script-lua} log output. Next, the \texttt{sleep} function halts the execution of the script for the given duration. The duration is passed as a string to include a time unit, e.g.~``2m'' for two minutes. Finally, the \texttt{get\_input\_port} and \texttt{get\_output\_port} functions return the \texttt{Port} instance with the given name. These correspond to the input and output ports of the \texttt{script-lua} function. Users thus are able to read and write to streams from within Lua.

Lastly, the \texttt{nbmp} module provides functionality to adapt an NBMP workflow. The \texttt{Workflow} type represents a WDD. Through the \texttt{self} constructor, an instance of the workflow the script is running in can be retrieved. Then, the \texttt{get\_task}, \texttt{get\_input} and \texttt{get\_output} methods return instances of the corresponding types. Users adapt the WDD by adding or removing connections between tasks or connections between tasks and inputs or outputs. This is modelled with the \texttt{Connection}, \texttt{Task}, \texttt{Port}, \texttt{Input} and \texttt{Output} types. The \texttt{add\_connection} and \texttt{remove\_connection} methods of a \texttt{Workflow} then change the in-memory representation of the workflow. To persist the changes and thus adapt the workflow, the \texttt{update} method must be called. Hence, it is possible to perform multiple simultaneous adaptations. The host then sends an update request to the NBMP workflow API. Note that we currently do not expose the full NBMP data model in Lua. Future work could improve upon this and provide a complete set of NBMP descriptors and properties.

\begin{figure}[b]
  \centering
  \includegraphics[width=0.9\columnwidth]{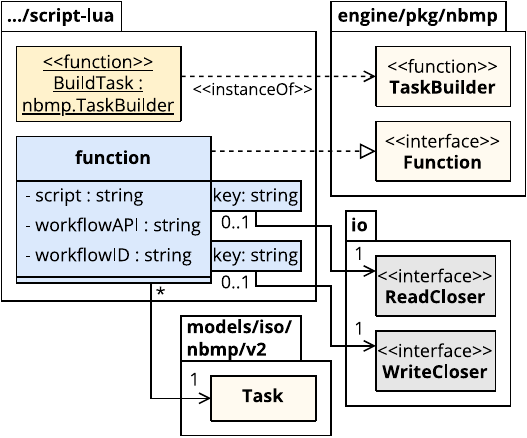}
  \caption{UML class/object diagram of the types/objects in the \texttt{.../script-lua} package.}
  \label{fig:nme-cmd-functions-functions-script-lua-uml-class}
  \Description{UML class/object diagram of the types/objects in the \texttt{.../script-lua} package}
\end{figure}
Figure~\ref{fig:nme-cmd-functions-functions-script-lua-uml-class} depicts the types of the host, i.e.~the \texttt{script-lua} function. Next to the \texttt{scrip} and \texttt{workflowAPI} attributes, the \texttt{function} type keeps a map to inputs (\texttt{ReadCloser} types) and outputs (\texttt{Write\linebreak{}Closer} type). We use these Go interfaces directly to implement the read and write operations in Lua.

\section{Conclusion}
\label{sec:conclusion}

This section concludes the technical report. We first give a summary in Section~\ref{subsec:summary} and then outline potentials for future work in Section~\ref{subsec:future-work}

\subsection{Summary}
\label{subsec:summary}

This technical report outlined the open source software \texttt{nagare media engine}. It implements a system for cloud- and edge-native network-based multimedia workflows. Following the NBMP standard, a workflow in \texttt{nagare media engine} logically consists of tasks that are connected via a network. Tasks are instantiated multimedia functions with input and output streams that are deployed onto an MPE. The NBMP reference architecture consists of multiple components that interact via REST APIs. \texttt{nagare media engine} implements the reference architecture in a series of components. Kubernetes is used as the platform for control plane (NBMP source, workflow manager) as well as media plane components (MPE, function, task). The management Kubernetes cluster that hosts the NBMP control plane is extended with custom resources that form the \texttt{nagare media engine} data model. The NBMP Gateway component runs in the management cluster and translates between the NBMP and the \texttt{nagare media engine} data models. It implements the NBMP workflow API for managing workflows and thus performs the requested operations on Kubernetes resources. This, in turn, triggers the execution of a reconciliation implemented by corresponding custom Kubernetes controllers. \texttt{nagare media engine} bundles all controllers in the Workflow Manager component. First, the MPE Controller establishes connections to MPEs, i.e.~Kubernetes clusters. The Workflow and Task Controllers reconcile the \texttt{Workflow} and \texttt{Task} resources, respectively, and thus manage their lifecycle. Finally, the Job Controller observes the execution of NBMP tasks that are running in the form of Kubernetes \texttt{Job}s. The Workflow Manager Helper component runs as an extension of the Workflow Manager next to each multimedia function as an additional container within the same \texttt{Job}. It uses the NBMP task API to initialize and control the multimedia function. Furthermore, it provides a reporting API that multimedia functions can use to report relevant events. NATS JetStream, an event sourcing system, is used to persist all reported events. Later on, functions may thus replay reported events, e.g.~to recover after a restart. With the Task Shim component, \texttt{nagare media engine} provides a common implementation of the NBMP task API. Hence, developers of multimedia functions can focus on the core processing logic. Moreover, administrators can easily adapt existing functions that do not implement the NBMP task API. \texttt{nagare media engine} already includes various multimedia functions for testing purposes and for encoding multimedia streams. Additionally, the \texttt{script-lua} function allows describing self-adapting workflows through a Lua script.

\subsection{Future Work}
\label{subsec:future-work}

\texttt{nagare media engine} is a prototypical implementation of the NBMP standard. As such, not all aspects of NBMP are already implemented. Various descriptors and properties of the NBMP data model are not yet supported. Furthermore, support could be added for the NBMP function API as well as the new NBMP MDD API that was introduced in the second edition of the standard. Finally, the full NBMP data model could be made available to the Lua script of the \texttt{script-lua} function.

Future work could also explore how an uninterrupted multimedia stream can be maintained. Kubernetes is a dynamic system where workloads could intentionally be disrupted, e.g.~because workloads have been rescheduled to other nodes. For \texttt{nagare media engine} tasks, this currently results in an interruption of the multimedia stream. Typically, Kubernetes workloads run replicated on different nodes to combat disruptions, but this is not easily adoptable for stateful multimedia functions that would then need to support replicated input streams. Recently, MPEG standardized redundant encoding and packaging for segmented live media~(REaP) as \mbox{ISO/IEC~23009-9}~\cite{isoiec_isoiec230099_2025} that may guide this work.

There exist a multitude of storage systems, streaming protocols and functions for multimedia systems. Therefore, \texttt{nagare media engine} could be extended in various ways. In particular, support for the standardized multimedia functions defined in the second edition of NBMP could be added. Moreover, MOQT seems to be a good candidate for an efficient transport protocol of multimedia streams both between tasks and for workflow inputs and outputs.

To improve the production readiness, the monitoring could be improved. Support for a standard tracing format, in particular, could improve the visibility of the workflow execution. How NBMP workflows could be presented in user interfaces is generally an open question and could be subject to further research. Lastly, the packaging of \texttt{nagare media engine} components in Helm charts would ease the installation process.

\bibliographystyle{ACM-Reference-Format}
\bibliography{references,references-zotero}

% \appendix
% \input{appendix}

\end{document}